
%
\input amstex
%
%
%
%
%
%


\magnification=1200
\hsize=31pc
\vsize=55 truepc
\hfuzz=2pt
\vfuzz=4pt
\pretolerance=500
\tolerance=500
\parskip=0pt plus 1pt
\parindent=16pt
%

%
%
\font\fourteenrm=cmr10 scaled \magstep2
\font\fourteeni=cmmi10 scaled \magstep2
\font\fourteenbf=cmbx10 scaled \magstep2
\font\fourteenit=cmti10 scaled \magstep2
\font\fourteensy=cmsy10 scaled \magstep2

%
\font\large=cmbx10 scaled \magstep1

%
\font\sans=cmssbx10

%

%

%
\font\eightrm=cmr8
\font\eighti=cmmi8
\font\eightbf=cmbx8
\font\eightit=cmti8

\font\eightsy=cmsy8
\font\sixrm=cmr6
\font\sixi=cmmi6
\font\sixsy=cmsy6

%
\def\tenpoint{\def\rm{\fam0\tenrm}%
  \textfont0=\tenrm \scriptfont0=\sevenrm
                      \scriptscriptfont0=\fiverm
  \textfont1=\teni  \scriptfont1=\seveni
                      \scriptscriptfont1=\fivei
  \textfont2=\tensy \scriptfont2=\sevensy
                      \scriptscriptfont2=\fivesy
  \textfont3=\tenex   \scriptfont3=\tenex
                      \scriptscriptfont3=\tenex
  \textfont\itfam=\tenit  \def\it{\fam\itfam\tenit}%
  \textfont\slfam=\tensl  \def\sl{\fam\slfam\tensl}%
  \textfont\bffam=\tenbf  \scriptfont\bffam=\sevenbf
                            \scriptscriptfont\bffam=\fivebf
                            \def\bf{\fam\bffam\tenbf}%
  \normalbaselineskip=20 truept
  \setbox\strutbox=\hbox{\vrule height14pt depth6pt
width0pt}%
  \let\sc=\eightrm \normalbaselines\rm}
\def\eightpoint{\def\rm{\fam0\eightrm}%
  \textfont0=\eightrm \scriptfont0=\sixrm
                      \scriptscriptfont0=\fiverm
  \textfont1=\eighti  \scriptfont1=\sixi
                      \scriptscriptfont1=\fivei
  \textfont2=\eightsy \scriptfont2=\sixsy
                      \scriptscriptfont2=\fivesy
  \textfont3=\tenex   \scriptfont3=\tenex
                      \scriptscriptfont3=\tenex
  \textfont\itfam=\eightit  \def\it{\fam\itfam\eightit}%
  \textfont\bffam=\eightbf  \def\bf{\fam\bffam\eightbf}%
  \normalbaselineskip=16 truept
  \setbox\strutbox=\hbox{\vrule height11pt depth5pt width0pt}}
\def\fourteenpoint{\def\rm{\fam0\fourteenrm}%
  \textfont0=\fourteenrm \scriptfont0=\tenrm
                      \scriptscriptfont0=\eightrm
  \textfont1=\fourteeni  \scriptfont1=\teni
                      \scriptscriptfont1=\eighti
  \textfont2=\fourteensy \scriptfont2=\tensy
                      \scriptscriptfont2=\eightsy
  \textfont3=\tenex   \scriptfont3=\tenex
                      \scriptscriptfont3=\tenex
  \textfont\itfam=\fourteenit  \def\it{\fam\itfam\fourteenit}%
  \textfont\bffam=\fourteenbf  \scriptfont\bffam=\tenbf
                             \scriptscriptfont\bffam=\eightbf
                             \def\bf{\fam\bffam\fourteenbf}%
  \normalbaselineskip=24 truept
  \setbox\strutbox=\hbox{\vrule height17pt depth7pt width0pt}%
  \let\sc=\tenrm \normalbaselines\rm}
\def\today{\number\day\ \ifcase\month\or
  January\or February\or March\or April\or May\or June\or
  July\or August\or September\or October\or November\or
December\fi
  \space \number\year}
\def\monthyear{\ifcase\month\or
  January\or February\or March\or April\or May\or June\or
  July\or August\or September\or October\or November\or
December\fi
  \space \number\year}

%
\newcount\secno      
\newcount\subno      
\newcount\subsubno   
\newcount\appno      
\newcount\tableno    
\newcount\figureno   
%

%
\normalbaselineskip=20 truept
\baselineskip=20 truept

%
%
\def\title#1
   {\vglue1truein
   {\baselineskip=24 truept
    \pretolerance=10000
    \raggedright
    \noindent \fourteenpoint\bf #1\par}
    \vskip1truein minus36pt}
%

%
\def\author#1
  {{\pretolerance=10000
    \raggedright
    \noindent {\large #1}\par}}

%
\def\address#1
   {\bigskip
    \noindent \rm #1\par}

%
\def\shorttitle#1
   {\vfill
    \noindent \rm Short title: {\sl #1}\par
    \medskip}

%
\def\pacs#1
   {\noindent \rm PACS number(s): #1\par
    \medskip}

%
\def\jnl#1
   {\noindent \rm Submitted to: {\sl #1}\par
    \medskip}

%
\def\date
   {\noindent Date: \today\par
    \medskip}

%

%
\def\keyword#1
   {\bigskip
    \noindent {\bf Keyword abstract: }\rm#1}

%

%
%

%
\def\entry#1#2#3
   {\noindent
    \hangindent=20pt
    \hangafter=1
    \hbox to20pt{#1 \hss}#2\hfill #3\par}

%
\def\subentry#1#2#3
   {\noindent
    \hangindent=40pt
    \hangafter=1
    \hskip20pt\hbox to20pt{#1 \hss}#2\hfill #3\par}
\def\checkforsub{\futurelet\nexttok\decide}
\def\ssf{\relax}
\def\decide{\if\nexttok\ssf\let\endspace=\nospace
                \else\let\endspace=\extraspace\fi\endspace}
\def\nospace{\nobreak\par\nobreak}
%
%
\def\section#1{%
    \goodbreak
    \vskip24pt plus12pt minus12pt
    \nobreak
    \gdef\extraspace{\nobreak\bigskip\noindent\ignorespaces}%
    \noindent
    \subno=0 \subsubno=0
    \global\advance\secno by 1
    \noindent {\bf \the\secno. #1}\par\checkforsub}

%
\def\subsection#1{%
     \goodbreak
     \vskip24pt plus12pt minus6pt
     \nobreak
     \gdef\extraspace{\nobreak\medskip\noindent\ignorespaces}%
     \noindent
     \subsubno=0
     \global\advance\subno by 1
     \noindent {\sl \the\secno.\the\subno. #1\par}\checkforsub}

%
\def\subsubsection#1{%
     \goodbreak
     \vskip15pt plus6pt minus6pt
     \nobreak\noindent
     \global\advance\subsubno by 1
     \noindent {\sl \the\secno.\the\subno.\the\subsubno. #1}\null.
     \ignorespaces}

%
\def\appendix#1
   {\vskip0pt plus.1\vsize\penalty-250
    \vskip0pt plus-.1\vsize\vskip24pt plus12pt minus6pt
    \subno=0
    \global\advance\appno by 1
    \noindent {\bf Appendix \the\appno. #1\par}
    \bigskip
    \noindent}

%
\def\subappendix#1
   {\vskip-\lastskip
    \vskip36pt plus12pt minus12pt
    \bigbreak
    \global\advance\subno by 1
    \noindent {\sl \the\appno.\the\subno. #1\par}
    \nobreak
    \medskip
    \noindent}

%
\def\ack
   {\vskip-\lastskip
    \vskip36pt plus12pt minus12pt
    \bigbreak
    \noindent{\bf Acknowledgments\par}
    \nobreak
    \bigskip
    \noindent}


%

%
\def\tabcaption#1
   {\global\advance\tableno by 1
    \noindent {\bf Table \the\tableno.} \rm#1\par
    \bigskip}

%

%

%

%

%

%
\def\figcaption#1
   {\global\advance\figureno by 1
    \noindent {\bf Figure \the\figureno.} \rm#1\par
    \bigskip}

%

%

%
\def\refjl#1#2#3#4
   {\hangindent=16pt
    \hangafter=1
    \rm #1
   {\frenchspacing\sl #2
    \bf #3}
    #4\par}

%
\def\refbk#1#2#3
   {\hangindent=16pt
    \hangafter=1
    \rm #1
   {\frenchspacing\sl #2}
    #3\par}

%
\def\numrefjl#1#2#3#4#5
   {\parindent=40pt
    \hang
    \noindent
    \rm {\hbox to 30truept{\hss #1\quad}}#2
   {\frenchspacing\sl #3\/
    \bf #4}
    #5\par\parindent=16pt}

%
\def\numrefbk#1#2#3#4
   {\parindent=40pt
    \hang
    \noindent
    \rm {\hbox to 30truept{\hss #1\quad}}#2
   {\frenchspacing\sl #3\/}
    #4\par\parindent=16pt}

%

\def\ref#1{\par\noindent \hbox to 21pt{\hss
#1\quad}\frenchspacing\ignorespaces}

%
\def\frac#1#2{{#1 \over #2}}

%

%
\def\d{\hbox{\rm d}}

%
\def\e{\operatorname{e}}


\def\i{\operatorname{i}}
\chardef\ii="10

%

%

%
\def\footnoterule{}
\catcode`\@=11
\def\vfootnote#1{\insert\footins\bgroup
    \interlinepenalty=\interfootnotelinepenalty
    \splittopskip=\ht\strutbox 
    \splitmaxdepth=\dp\strutbox \floatingpenalty=20000
    \leftskip=0pt \rightskip=0pt \spaceskip=0pt \xspaceskip=0pt
    \noindent\eightpoint\rm #1\ \ignorespaces\footstrut\futurelet\next\fo@t}

%
%
\def\eq(#1){\hfill\llap{(#1)}}
\catcode`\@=12
%
%



%
%





%
%

%
%

%
%

%

%

%

%
\def\gap{\;\lower3pt\hbox{$\buildrel > \over \sim$}\;}
%
%
\def\lap{\;\lower3pt\hbox{$\buildrel < \over \sim$}\;}
\def\tqs{\hbox to 25pt{\hfil}}


%
%
%
%


{\obeylines\gdef\startdisplay#1
  {\catcode`\^^M=5$$#1\halign\bgroup\indent##\hfil&&\qquad##\hfil\cr}}
\outer\def\enddisplay{\crcr\egroup$$}

\chardef\other=12
\def\ttverbatim{\begingroup \catcode`\\=\other \catcode`\{=\other
  \catcode`\}=\other \catcode`\$=\other \catcode`\&=\other
  \catcode`\#=\other \catcode`\%=\other \catcode`\~=\other
  \catcode`\_=\other \catcode`\^=\other
  \obeyspaces \obeylines \tt}
{\obeyspaces\gdef {\ }}  

\outer\def\begintt{$$\let\par=\endgraf \ttverbatim \parskip=0pt
  \catcode`\|=0 \rightskip=-5pc \ttfinish}
{\catcode`\|=0 |catcode`|\=\other 
  |obeylines 
  |gdef|ttfinish#1^^M#2\endtt{#1|vbox{#2}|endgroup$$}}

\catcode`\|=\active
{\obeylines\gdef|{\ttverbatim\spaceskip=.5em plus.25em minus.15em
                                            \let^^M=\ \let|=\endgroup}}%

\TagsOnRight
\hsize=16.0truecm
\vsize=24.0truecm
\hfuzz=3pt

\tracingstats=1    

\font\twelverm=cmr10 scaled 1200

\normalbaselineskip=12pt
\baselineskip=12pt
%
\def\AA{{\hbox{\rm A}\NUM}}
\def\CA{{\Cal A}}
\def\CB{{\Cal B}}
\def\CD{{\Cal D}}
\def\CL{{\Cal L}}
\def\CM{{\Cal M}}
\def\CP{{\Cal P}}
\def\CQ{{\Cal Q}}
\def\cn{\operatorname{cn}}
\def\dc{\operatorname{dc}}
\def\diag{\operatorname{diag}}
\def\dn{\operatorname{dn}}
\def\me{\operatorname{me}}
\def\Me{\operatorname{Me}}
\def\nc{\operatorname{nc}}
\def\ps{\operatorname{ps}}
\def\sphpsi{\operatorname{psi}}
\def\sc{\operatorname{sc}}
\def\Si{\operatorname{Si}}
\def\sn{\operatorname{sn}}
\def\SO{\operatorname{SO}}
\def\PSL{\operatorname{PSL}}
\def\OO{\operatorname{O}}
\def\SU{\operatorname{SU}}
\font\sans=cmssbx10
\def\sf{\sans}
\def\bbbc{{\mathchoice {\setbox0=\hbox{\rm C}\hbox{\hbox
to0pt{\kern0.4\wd0\vrule height0.9\ht0\hss}\box0}}
{\setbox0=\hbox{$\textstyle\hbox{\rm C}$}\hbox{\hbox
to0pt{\kern0.4\wd0\vrule height0.9\ht0\hss}\box0}}
{\setbox0=\hbox{$\scriptstyle\hbox{\rm C}$}\hbox{\hbox
to0pt{\kern0.4\wd0\vrule height0.9\ht0\hss}\box0}}
{\setbox0=\hbox{$\scriptscriptstyle\hbox{\rm C}$}\hbox{\hbox
to0pt{\kern0.4\wd0\vrule height0.9\ht0\hss}\box0}}}} 
\def\bbbr{\operatorname{{I\!R}}}                     
\def\bbbn{\operatorname{{I\!N}}}                     
\def\bbbz{{\mathchoice {\hbox{$\sf\textstyle Z\kern-0.4em Z$}}
{\hbox{$\sf\textstyle Z\kern-0.4em Z$}}
{\hbox{$\sf\scriptstyle Z\kern-0.3em Z$}}
{\hbox{$\sf\scriptscriptstyle Z\kern-0.2em Z$}}}}    
\def\bbbone{\mathchoice {\operatorname{1\mskip-4mu l}}
{\operatorname{1\mskip-4mu l}}
{\operatorname{1\mskip-4.5mu l}} {\operatorname{1\mskip-5mu l}}}
\def\vec#1{{\textfont1=\tenbf\scriptfont1=\sevenbf
\textfont0=\tenbf\scriptfont0=\sevenbf
\mathchoice{\hbox{$\displaystyle#1$}}{\hbox{$\textstyle#1$}}
{\hbox{$\scriptstyle#1$}}{\hbox{$\scriptscriptstyle#1$}}}}
\def\ih{{\i\over\hbar}}
\def\hi{{\hbar\over\i}}

\def\hbarm{{\hbar^2\over2m}}
\def\viert{{1\over4}}
\def\half{{1\over2}}
\def\bhalf{\hbox{$\half$}}
\def\bviert{\hbox{$\viert$}}
\def\myalign{\allowdisplaybreaks\align}
\def\dfrac{\dsize\frac}
\def\Norm{\left({m\over2\pi\i\epsilon\hbar}\right)}
\def\bqj{ {\bar{\vec q}}_j }

\def\Energysdrei{\exp\bigg[-{\i\hbar T\over2m}l(l+2)\bigg]}
\def\Energylzwei{
     \exp\bigg[-{\i\hbar T\over2m}\bigg(p^2+\viert\bigg)\bigg]}
\def\Energyldrei{\exp\bigg[-{\i\hbar T\over2m}(p^2+1)\bigg]}
\def\energylzwei{\e^{-\i\hbar T(p^2+\viert)/2m}}
\def\energyldrei{\e^{-\i\hbar T(p^2+1)/2m}}

\def\footnoterule{\kern-3pt\hrule width 2true cm\kern2.6pt}
\newcount\footcount \footcount=0
\def\advftncnt{\advance\footcount by1\global\footcount=\footcount}
\def\fonote#1{\advftncnt$^{\the\footcount}$\begingroup\eightpoint
\eightrm
\parfillskip=0pt plus 1fil
\def\textindent##1{\hangindent0.5\oldparindent\noindent\hbox
to0.5\oldparindent{##1\hss}\ignorespaces}%
\vfootnote{$^{\the\footcount}$}{#1\vskip-9.69pt}\endgroup}
\def\ezwei{{E^{(2)}}}
\def\edrei{{E^{(3)}}}
\def\eD{{E^{(D)}}}
\def\szwei{{S^{(2)}}}
\def\sdrei{{S^{(3)}}}
\def\sD{{S^{(D-1)}}}
\def\lzwei{{\Lambda^{(2)}}}
\def\ldrei{{\Lambda^{(3)}}}
\def\lD{{\Lambda^{(D-1)}}}

\newcount\Chapno
\def\PLUS{\advance\Chapno by 1}
\def\NUM{\the\Chapno}
\Chapno=0

\newcount\glno
\def\plus{\advance\glno by 1}
\def\minus{\advance\glno by -1}
\def\num{\the\glno}

\newcount\Refno
\def\add{\advance\Refno by 1}
\Refno=1

\edef\ABS{\the\Refno}\add
\edef\ABHK{\the\Refno}\add
\edef\AHK{\the\Refno}\add
\edef\ART{\the\Refno}\add
\edef\AMSS{\the\Refno}\add
\edef\ASS{\the\Refno}\add
\edef\AST{\the\Refno}\add
\edef\BV{\the\Refno}\add
\edef\BIJ{\the\Refno}\add
\edef\BJb{\the\Refno}\add
\edef\BJc{\the\Refno}\add
\edef\BJd{\the\Refno}\add
\edef\BUC{\the\Refno}\add
\edef\CAST{\the\Refno}\add
\edef\CHe{\the\Refno}\add
\edef\DAR{\the\Refno}\add
\edef\POGOb{\the\Refno}\add
\edef\DEW{\the\Refno}\add
\edef\REUT{\the\Refno}\add
\edef\DOTO{\the\Refno}\add
\edef\DURb{\the\Refno}\add
\edef\DURd{\the\Refno}\add
\edef\DKa{\the\Refno}\add
\edef\DKb{\the\Refno}\add
\edef\EG{\the\Refno}\add
\edef\EIS{\the\Refno}\add
\edef\EGM{\the\Refno}\add
\edef\EMOTa{\the\Refno}\add
\edef\EMOTb{\the\Refno}\add
\edef\EWA{\the\Refno}\add
\edef\FEY{\the\Refno}\add
\edef\FH{\the\Refno}\add
\edef\FLM{\the\Refno}\add
\edef\FMSUW{\the\Refno}\add
\edef\GY{\the\Refno}\add
\edef\GJ{\the\Refno}\add
\edef\GOOb{\the\Refno}\add
\edef\GEGR{\the\Refno}\add
\edef\GGV{\the\Refno}\add
\edef\GLJA{\the\Refno}\add
\edef\GRA{\the\Refno}\add
\edef\GSW{\the\Refno}\add
\edef\GROa{\the\Refno}\add
\edef\GROb{\the\Refno}\add
\edef\GROc{\the\Refno}\add
\edef\GROe{\the\Refno}\add
\edef\GROg{\the\Refno}\add
\edef\GROf{\the\Refno}\add
\edef\GROj{\the\Refno}\add
\edef\GROm{\the\Refno}\add
\edef\GROn{\the\Refno}\add
\edef\GROq{\the\Refno}\add
\edef\GROw{\the\Refno}\add
\edef\GRSa{\the\Refno}\add
\edef\GRSb{\the\Refno}\add
\edef\GRSc{\the\Refno}\add
\edef\GRSf{\the\Refno}\add
\edef\GRSg{\the\Refno}\add
\edef\GRSh{\the\Refno}\add
\edef\GUTc{\the\Refno}\add
\edef\GUTd{\the\Refno}\add
\edef\HEJ{\the\Refno}\add
\edef\HIG{\the\Refno}\add
\edef\INOa{\the\Refno}\add
\edef\INOb{\the\Refno}\add
\edef\INOWI{\the\Refno}\add
\edef\JUNc{\the\Refno}\add
\edef\BJa{\the\Refno}\add
\edef\KAL{\the\Refno}\add
\edef\KAMIa{\the\Refno}\add
\edef\KAMIb{\the\Refno}\add
\edef\KLEh{\the\Refno}\add
\edef\KLE{\the\Refno}\add
\edef\KOKCAS{\the\Refno}\add
\edef\KLEMUS{\the\Refno}\add
\edef\KUB{\the\Refno}\add
\edef\KUZ{\the\Refno}\add
\edef\LL{\the\Refno}\add
\edef\POGOa{\the\Refno}\add
\edef\MCSCH{\the\Refno}\add
\edef\MSVW{\the\Refno}\add
\edef\MASA{\the\Refno}\add
\edef\MADO{\the\Refno}\add
\edef\MESCH{\the\Refno}\add
\edef\MIZa{\the\Refno}\add
\edef\MDEW{\the\Refno}\add
\edef\MF{\the\Refno}\add
\edef\NEL{\the\Refno}\add
\edef\OMO{\the\Refno}\add
\edef\OLE{\the\Refno}\add
\edef\PAKSc{\the\Refno}\add
\edef\PI{\the\Refno}\add
\edef\PELST{\the\Refno}\add
\edef\PODO{\the\Refno}\add
\edef\POGOc{\the\Refno}\add
\edef\POL{\the\Refno}\add
\edef\QUES{\the\Refno}\add
\edef\ROEP{\the\Refno}\add
\edef\SCHUa{\the\Refno}\add
\edef\SCHU{\the\Refno}\add
\edef\SCHW{\the\Refno}\add
\edef\SEL{\the\Refno}\add
\edef\SIMON{\the\Refno}\add
\edef\STEc{\the\Refno}\add
\edef\STEP{\the\Refno}\add
\edef\STORCH{\the\Refno}\add
\edef\VENa{\the\Refno}\add
\edef\VENb{\the\Refno}\add
\edef\VENc{\the\Refno}\add
\edef\VIL{\the\Refno}\add
\edef\VISM{\the\Refno}\add
\edef\WIE{\the\Refno}\add
\edef\YODEWM{\the\Refno}\add


{\nopagenumbers
\pageno=0
\centerline{DESY 93 - 141 \hfill ISSN 0418 - 9833}
\centerline{October 1993\hfill}
\centerline{\hfill hep-th/9311001}
\vskip1cm
\centerline{\fourteenpoint PATH INTEGRATION AND SEPARATION OF }
\bigskip
\centerline{\fourteenpoint VARIABLES IN SPACES OF CONSTANT CURVATURE}
\bigskip
\centerline{\fourteenpoint IN TWO AND THREE DIMENSIONS}
\vskip1cm
\centerline{\twelverm CHRISTIAN GROSCHE$^*$}
\bigskip
\centerline{\it II.\ Institut f\"ur Theoretische Physik}
\centerline{\it Universit\"at Hamburg, Luruper Chaussee 149}
\centerline{\it 22761 Hamburg, Germany}
\vfill
\midinsert
\narrower
\noindent
{\bf Abstract.}
In this paper path integration in two- and three-dimensional spaces of
constant curvature is discussed: i.e.\ the flat spaces $\bbbr^2$ and
$\bbbr^3$, the two- and three-dimensional sphere and the two- and three
dimensional pseudosphere. The Laplace operator in these spaces admits
separation of variables in various coordinate systems. In all these
coordinate systems the path integral formulation will be stated,
however in most of them an explicit solution in terms of the spectral
expansion can be given only on a formal level. What can be stated in
all cases, are the propagator and the corresponding Green function,
respectively, depending on the invariant distance which is a
coordinate independent quantity. This property gives rise to numerous
identities connecting the corresponding path integral representations
and propagators in various coordinate systems with each other.
\endinsert

\bigskip\noindent
\centerline{\vrule height0.25pt depth0.25pt width4cm\hfill}
\noindent
{\eightpoint\eightrm
 $^*$ Supported by Deutsche Forschungsgemeinschaft under contract
 number GR 1031/2--1.}
\eject\pageno=0\centerline{\ }\vfill
\eject}
\pageno=1


\PLUS\glno=0                      
\section{Introduction}
The invention of the path integral by Feynman [\FEY] is one of the major
achievements of theoretical physics. In its now 50 years history it has
become an indispensable tool in field theory, cosmology, molecular
physics, condensed matter physics and string theory as well [\GSW,
\POL].

Originally developed as a ``space-time approach to non-relativistic
quantum mechanics'' [\FEY] with the famous solution of the harmonic
oscillator, it became soon of paramount importance in quantum
electrodynamics (QED), especially in the development of the nowadays
so-called ``Feynman rules''. It did not take long and Feynman succeeded
in discussing problems not only in QED but also in the theory of
super-fluidity.
\fonote{For this kind of general overview we do not want to
cite all the relevant references in detail, but would instead invite
the interested reader to consult some of the textbooks on the Feynman
path integral, in particular Feynman and Hibbs [\FH], respectively the
collection of Schwinger [\SCHW] and the early review paper of Gel'fand
and Yaglom [\GY].}
However, it took a considerably long time before it
was generally accepted by most physicists as a powerful tool to analyse
a physical system, giving non-perturbative global information, instead
of only perturbative, respectively local information, as in an operator
approach.

Eventually, a satisfying theory should be based on a field theory
formulation, let it be the second quantization of the Schr\"odinger,
respectively the Dirac equation, let it be a field theory path integral.
However, the path integral is quite a formidable and difficult
functional-analytic object. The very early field theory formulations by
a path integral, first by Feynman and in the following by e.g.\ Matthews
and Salam [\GY, \MASA] remained only on a formal level, however with
well-described rules to extract the relevant information, say, for
Feynman diagrams. Indeed, in field theory the path integral is cursed
by several pathologies which cause people now and then to state
that ``the path integral does not exist``.

What does exists, however, is the very originally Feynman path integral,
i.e.\ Feynman's ``space-time approach to non-relativistic quantum
mechanics''. Thanks to the work of many mathematicians and physicists as
well, the theory of the Feynman path integral can be considered as
quite comprehensively developed. Actually, the theory of the
``Wiener-integral'' [\GY] existed some twenty years right before
Feynman published his ideas, and was developed in the theory of
diffusion processes. The Wiener integral itself represents an
``imaginary time'' version of the ``real time'' Feynman path integral.
This particular feature of the Wiener integral makes it a not too
complicated and convenient tool in functional analysis, mostly because
convergence properties are easily shown. These convergence properties
are absent in the Feynman path integral and the emerging challenge
attracted many mathematicians and mathematical physicists, c.f.\ the
references given in [\GY]. Let us in addition mention Nelson [\NEL]
concerning the Feynman path integral in cartesian coordinates, DeWitt
concerning curvilinear coordinates [\DEW], Morette-DeWitt et al.\
[\MDEW] and Albeverio et al.\ (e.g.~[\ABHK, \AHK]) who developed a
theory of ``pseudomeasures'' appropriate to the interference of
probability amplitudes in the path integral. This interference of
probabilities, in particular in the lattice formulation of the path
integral (see below in Section~2) leads to the very interpretation of
the Feynman path integral. One encounters for finite lattice spacing,
i.e.\ finite $N$, a complex number $\Phi(\vec q_1,\hdots,\vec q_{N-1})$
which is a function of the variables $\vec q_j$ defining a $\vec q(t)$,
the path integral can be interpreted as a ``sum over all paths'' or a
``sum over all histories''
\plus$$
  K(\vec q'',\vec q';T)
   =\sum_{\scriptstyle\text{\eightpoint over all paths}
    \atop
    \scriptstyle\text{\eightpoint from $\vec q'$ to $\vec q''$}}
  \Phi[\vec q(t)]
   =\sum_{\scriptstyle\text{\eightpoint over all paths}
    \atop
    \scriptstyle\text{\eightpoint from $\vec q'$ to $\vec q''$}}
   \e^{\i S[\vec q(t)]/\hbar}\enspace.
  \tag\NUM.\num$$
The path integral then gives a prescription how to compute the important
quantity $\Phi$ for each path: ``The paths contribute equally in
magnitude, but the phase of their contribution is the classical action
(in units of $\hbar$). \dots That is to say, the contribution $\Phi[\vec
q(t)]$ from a given path $\vec q(t)$ is proportional to $\exp\big(\ih
S[\vec q(t)]\big)$, where the {\it action\/} is the time integral of
the classical Lagrangian taking along the path in question'' [\FEY]. All
possible paths enter and interfere which each other in the convolution
of the probability amplitudes. In fact, the nowhere differentiable
paths span the continuum in the set of all paths, the differentiable
ones are being a set of measure zero (the quantity $\Delta\vec q_j/
\Delta t_j$ does not exist, whereas $(\Delta\vec q_j)^2/\Delta t_j$
does).

A more comprehensive discussion will be given in a forthcoming
publication [\GRSg], including a table of exactly solvable path
integrals, in our lecture notes [\GRSh], and alternatively, we
refer to the several already existing textbooks on path integrals
Albeverio et at.\ [\ABHK, \AHK], Dittrich and Reuter [\REUT], Feynman
and Hibbs [\FH], Glimm and Jaffe [\GLJA], Kleinert [\KLE], Roepstorff
[\ROEP], Schulman [\SCHU], Simon [\SIMON], and Wiegel [\WIE]. For a
short reference we refer to [\GRSf].

The subject of this paper will be the path integral formulations in
(spatial) two and three-dimensional spaces of constant curvature. These
are the most important needed in physics. It is often necessary to
consider a given problem from various points of views, i.e.\ coordinate
system realizations, say, and moreover find the relevant Fourier
expansions needed for the harmonic analysis and for the transition
from one coordinate system to another one, respectively. E.g.\ in the
space $\ldrei$ (the three-dimensional  pseudosphere) this task was
undertaken by Vilenkin and Smorodinsky [\VISM] for five coordinate
systems on $\ldrei$. Historically, Lam\'e started the project of
finding all trirectangular systems in Euclidean space which admit
separation of variables of the corresponding Hamilton-Jacobi and
Laplace equations. Further, the works of Darboux [\DAR], Eisenhart
[\EIS] and Stepanov [\STEP] were set out to solve this problem. The
result is that there are four real coordinate systems in two dimensions
and eleven real coordinates systems in three dimensions of the sought
type (which e.g.\ can be found in Morse and Feshbach [\MF]). The
corresponding problem for spaces of (non-vanishing) constant curvature,
i.e.\ on spheres and pseudospheres was solved by Olevski\v\ii\ [\OLE],
and a systematic approach for the D-dimensional generalization is due
to Kalnins et al.~[\KAL, \KAMIa] and references therein. I will return
to this classification in the next Section.

Some of the coordinate  systems - cartesian, polar, parabolic - are
very familar, others much less so. They are all obtained as
degenerations of the confocal ellipsoidal coordinates (c.f.~[\MESCH,
\MF] for such discussions in the three-dimensional Euclidean plane).
It turns out that the  pseudosphere $\lD$ for a fixed dimension D has
the richest structure of all of them. This is not too surprising. The
uniformization theorem for Riemann surfaces states that the fundamental
domain of $\lzwei/\Gamma$, where $\Gamma$ is a discrete fixed-point
free subgroups of the automorphisms on $\lzwei$, i.e.\ $\Gamma$ is a
Fuchsian group, tesselates the entire hyperbolic plane. The Riemannian
surfaces may have any genus $g\geq2$ and may be arbitrarily shaped
according to the corresponding Teichm\"uller space. This rich structure
makes $\lzwei$ interesting in the Polyakov approach to string theory,
respectively in $1+1$-dimensional quantum gravity [\GSW]: in the
perturbative expansion of the Polyakov path integral one is left with a
summation over all topologies of world sheets a string can sweep out,
and an integral over the moduli space of Riemann surfaces. In contrast,
the sphere $\szwei$ can represent only a genus zero Riemann surface,
i.e.\ a sphere.

The  pseudosphere $\ldrei$ is on the one hand of particular
importance, because the manifolds of constant $u^2$, $u$ being the
four-velocity or the four-momentum, represent the physical domain of
the variables for particles with real mass (on the mass shell).
On the other, a similar harmonic analysis as in $\lzwei$ can be
studied, c.f.~[\EGM, \VENa], giving a tessalation of $\ldrei$ in the
form of three-manifolds. This structure makes $\ldrei$ interesting in
$2+1$-dimensional quantum gravity.

I am now going to study and compute explicitly as far as possible,
all the path integral formulations in spaces of constant curvature in
two and three dimensions in terms of all possible coordinate systems
allowing the complete separation of variables. Not all our path integral
representations will be new. However, we put them all, i) into the
context of exactly solvable examples of the path integral in a given
space of constant curvature. ii) We establish in this way numerous
identities in terms of a specific coordinate formulation of a path
integral and its corresponding spectral expansion on the one hand, and
its {\it explicitly known\/} form of the propagator and the Green
function in terms of the invariant distance (norm) in the space
expressed in these coordinates on the other. Spectral theory of
functional analysis guarantees the equivalence of the identities.
In the majority of the cases, especially for $\ldrei$, where no explicit
path integral evaluations are possible, this equivalence sets out
further progress in finding exactly solvable examples for the Feynman
path integral by connecting the various examples with each other.

Whereas in this paper we focus on coordinate systems and their
corresponding path integral representations, one can also look on the
matter from a group theoretical point of view [\BJb]: For flat space we
have the Euclidean group, for the sphere we have $\sD\cong\SO(D)/\SO
(D-1)$, in particular $\sdrei\cong\SO(4)/\SO(3)$ and $\szwei \cong\SO(3)
/\SO(2)$ and for the  pseudosphere $\lD\cong\SO_0(D-1,1)/\SO(D-1)$.

In order to look at such a path integral formulation we consider the
Lagrangian $\CL(\vec x,{\dot{\vec x}})-V(\vec x)$ ($\vec x\in\bbbr^{p+q}
$) as formulated, say, in a not-necessarily positive definite space with
signature
\plus$$(g_{ab})=\diag\big(\underbrace{+1,\hdots,+1}_{p\ times},
         \underbrace{-1,\hdots,-1}_{q\ times}\big)\enspace.
  \tag\NUM.\num$$
One introduces polar coordinates
\plus$$x_\nu=re_\nu(\theta_1,\hdots,\theta_{p+q-1})
  \enspace,\qquad \nu=1,\hdots,p+q\enspace,
  \tag\NUM.\num$$
where the $\vec e$ are unit vectors in some suitable chosen
(timelike, spacelike or lightlike) set [\BJb]. One then expresses the
Lagrangian in terms of these polar coordinates and seeks for an
expansion of the quantity $\e^{z(\vec e_1\cdot\vec e_2)}$ expressed in
terms of group elements $g_1\,,g_2$. If this is possible one can
re-express the path integration of the coordinates $\vec x$ into a path
integration over group elements $g$ yielding [\BJb]
$$\myalign
  \int\limits_{\vec x(t')=\vec x'}^{\vec x(t'')=\vec x''}\CD\vec x(t)
  \exp\left[\ih\int_{t'}^{t''}\CL(\vec x,{\dot{\vec x}})dt\right]
       &
  \mapsto
  \int dE_\lambda d_\lambda\sum_{mn}
  {\hat f}_{mn}^\lambda D_{mn}^\lambda({g'}^{-1}g'')
  \tag\NUM.\num\\   \global\plus
       &
  =\int dE_\lambda d_\lambda\sum_{mn}{\hat f}_{mn}^\lambda
   \sum_kD_{m,k}^{\lambda\,*}(g')D_{m,k}^\lambda(g'')\enspace.
  \tag\NUM.\num\endalign$$
Here ${\hat f}^\lambda_{mn}$ is defined via the Fourier transformation
\plus$$f(g)=\int dE_\lambda d_\lambda
       \sum_{mn}\hat f^\lambda_{mn}D^\lambda_{mn}(g)
  \enspace,\qquad
  \hat f^\lambda_{mn}=\int_G f(g)D^{\lambda\,*}_{mn}(g^{-1})dg\enspace,
  \tag\NUM.\num$$
and $dg$ is the invariant group (Haar) measure. $\int dE_\lambda$
stands for a Lebesque-Stieltjes integral to include discrete $(\int
dE_\lambda\to\sum_\lambda$) as well as continuous representations.
$\int dE_\lambda$ is to be taken over the complete set $\{\lambda\}$ of
class one representations. $d_\lambda$ denotes (in the compact case)
the dimension of the representation and we take
\plus$$d_\lambda\int_G D^\lambda_{mn}(g)
       D^{\lambda'\,*}_{m'n'}(g)dg=\delta(\lambda,\lambda')
  \delta_{m,m'}\delta_{n,n'}
  \tag\NUM.\num$$
as a definition for $d_\lambda$. $\delta(\lambda,\lambda')$ can denote
a Kronecker delta, respectively, a $\delta$-function, depending whether
the quantity $\lambda$ is a discrete or continuous parameter. We have
furthermore used the group (composition) law
\plus$$D^\lambda_{mn}(g_a^{-1}g_b)
  =\sum_k D^{\lambda\,*}_{kn}(g_a) D^\lambda_{km}(g_b)\enspace.
  \tag\NUM.\num$$
Choosing a basis $\{\vec b\}$ in the relevant Hilbert space fixes the
matrix elements $D_{mn}^\lambda$ through $D^\lambda_{mn}=(D^\lambda(g)
b_m ,b_n)$ of the representation $D^\lambda(g)$ of the group. In
particular the $D^\lambda_{0m}$ are called associated spherical
harmonics, and the $D^\lambda_{00}$ are the zonal harmonics. These
spherical functions are eigen-functions of the corresponding
Laplace-Beltrami operator on a, say, homogeneous space, and the entire
Hilbert space is spanned by a complete set of associated spherical
functions $D^\lambda_{0m}$ [\GEGR, \GGV, \VIL].

``All what remains'' is to look at a convenient representation, i.e.\
a coordinates system. One may say, that this is the very subject of this
paper.

In our case of the free motion things are not too difficult, and
the group path integral technique can be fruitfully exploited. This
kind of approach is always possible in cases where a model has a known
underlying dynamical group structure (e.g.~[\INOb]). Here the most
famous example is the hydrogen atom with its $\OO(4)$ symmetry.
Actually, this property enabled Duru and Kleinert to apply the
so-called Kustaanheimo-Stiefel transformation to the path integral
problem of the hydrogen atom [\DKa, \DKb]. This dynamical group
structure is also important in order to discuss the so-called
``super-integrable'' potentials [\EWA] (where the Coulomb potential and
the harmonic oscillator in $\bbbr^3$ are but two examples), and of
course their path integral representations; we return shortly to this
topic in the discussion in Section~6.

The further content will be as follows: In the second Section we outline
some basic information concerning the path integral in curved spaces.
We sketch the relevant (transformation) techniques along the lines
as presented in our earlier work [\GROa, \GROj, \GRSb]. Next we sketch
the classification of separating variables for Riemannian spaces of
constant curvature. Here some basic information as found in Kalnins
[\KAL] is given. This includes the development of separability of
variables in the language of path integrals.

Sections~4 and~5 then deal with the enumeration of the path integral
representations, spectral expansions, and path integral evaluations in
two and three dimensions, respectively. These two Sections represent the
principal part of the paper and because we also list some already known
results, some parts will have a review character.

The last Section contains a summary and a discussion of the results.
In the three appendices some additional information is given as needed
in Sections~4 and ~5: In Appendix~1 some basic path integrals are cited,
in Appendix~2 a particular dispersion relation is discussed, and in
Appendix~3 the path integral solution on $\lD$ in a particular
coordinate system is given.


\PLUS\glno=0                      
\section{Formulation of the Path Integral}
In order to set up our notation we proceed in the canonical way for
path integrals on curved spaces (DeWitt [\DEW], D'Olivio and Torres
[\DOTO], Feynman [\FEY], Gervais and Jevicki [\GJ], [\GROa, \GRSb],
McLaughlin and Schulman [\MCSCH], Mayes and Dowker [\MADO], Mizrahi
[\MIZa], and Omote [\OMO]). In the following $\vec x$ denotes a
D-dimensional cartesian coordinate, $\vec q$ a D-dimensional arbitrary
coordinate, and $x,y,z$ etc.\ one-dimensional coordinates. A quantity
$\b o$ denotes an operator. We start by considering the classical
Lagrangian corresponding to the line element $ds^2=g_{ab}dq^adq^b$ of
the classical motion in some $D$-dimensional Riemannian space
\plus$$\CL_{Cl}(\vec q,\dot{\vec q})
   ={m\over2}\bigg({ds\over dt}\bigg)^2-V(\vec q)
                    ={m\over2}g_{ab}\dot q^a\dot q^b-V(\vec q).
  \tag\NUM.\num$$
The quantum Hamiltonian is {\it constructed} by means of the
Laplace-Beltrami operator
\plus$$\b H=-\hbarm\Delta_{LB}+V(\vec q)
   =-\hbarm
     {1\over\sqrt{g}}{\partial\over\partial q_a}g^{ab}\sqrt{g}
     {\partial\over\partial q_b}+V(\vec q)
  \tag\NUM.\num$$
\edef\numbh{\NUM.\num}%
as a {\it definition} of the quantum theory on a curved space [\PODO].
Here $g=\det{(g_{ab})}$ and $(g^{ab})=(g_{ab})^{-1}$. The scalar
product for wave-functions on the manifold reads $(f,g)=\int d\vec
q\sqrt{g}\,f^*(\vec q)g(\vec q)$, and the momentum operators which are
hermitean with respect to this scalar product are given by
\plus$$p_a=\hi\bigg({\partial\over\partial q^a}
                             +{\Gamma_a\over2}\bigg)\enspace,
  \qquad\Gamma_a={\partial\ln\sqrt{g}\over\partial q^a}\enspace.
  \tag\NUM.\num$$
\edef\numbg{\NUM.\num}%
In terms of the momentum operators (\numbg) we can rewrite $H$ by using
the Weyl-ordering prescription
([\GRSb, \MIZa], $W$={\it W\/}eyl):
\plus$$\b H(\vec p,\vec q)={1\over8m}(\b g^{ab}\,\b p_a
       \b p_b+2\b p_a\,\b g^{ab}\,\b p_b+\b p_a \b p_b\,\b g^{ab})
       +V(\vec q)+\Delta V_W(\vec q)\enspace.
  \tag\NUM.\num$$
Here a well-defined quantum correction appears which is given by
[\GRSb, \MIZa, \OMO]:
\plus$$\Delta
V_W={\hbar^2\over8m}(g^{ab}\Gamma^d_{ac}\Gamma^c_{bd}-R)
      ={1\over8m}\Big[g^{ab}\Gamma_a\Gamma_b
       +2(g^{ab}\Gamma_a)_{,b}+g^{ab}_{\ \ ,ab}\Big]
  \tag\NUM.\num$$
The corresponding {\it Lagrangian path
integral\/} reads ($MP$ = {\it M\/}id-{\it P\/}oint):
\plus$$\myalign
  K(\vec q'',\vec q';T)
       &
  =[g(\vec q')g(\vec q'')]^{-1/4}
   \int\limits_{\vec q(t')=\vec q'}^{\vec q(t'')=\vec q''}
  \sqrt{g}\,\CD_{MP}\vec q(t)
  \exp\bigg[\ih\int_{t'}^{t''}\CL_{eff}(\vec q,\dot{\vec q})dt\bigg]
  \\  &
  \equiv[g(\vec q')g(\vec q'')]^{-1/4}\lim_{N\to\infty}
  \Norm^{ND\over2}\left(\prod_{j=1}^{N-1}\int d\vec q_j\right)
  \prod_{j=1}^N\sqrt{g(\bqj)}
  \\   &\qquad\times
  \exp\left\{\ih\bigg[{m\over2\epsilon}g_{ab}(\bqj)
  \Delta q^a_j\Delta q^b_j-\epsilon V(\bqj)
  -\epsilon\Delta V_W(\bqj)\bigg]\right\}\enspace.
  \tag\NUM.\num\endalign$$
Here we have used the abbreviations $\epsilon=(t''-t')/N\equiv T/N$,
$\Delta\vec q_j=\vec q_j-\vec q_{j-1}$, $\bqj=\half(\vec q_j+\vec
q_{j-1})$ for $\vec q_j=\vec q(t'+j\epsilon)$ $(t_j=t'+\epsilon j,\
j=0,\dots,N)$ and we interpret the limit $N\to \infty$ as equivalent to
$\epsilon\to0$, $T$ fixed. The lattice representation can be
obtained by exploiting the composition law of the time-evolution
operator $\b U=\exp(-\i\b HT/\hbar)$, respectively its semi-group
property. The Weyl-ordering prescription is the most discussed ordering
prescription in the literature.

In an alternative approach the metric tensor is  rewritten
as a product according to $g_{ab}=h_{ac}h_{cb}$ [\GROa].
Then we obtain for the Hamiltonian (\numbh)
\plus$$\b H=-\hbarm\Delta_{LB}+V(\vec q)
  ={1\over2m}\b h^{ac}\b p_a\b p_b\b h^{cb}
     +\Delta V_{PF}(\vec q)+V(\vec q)
  \tag\NUM.\num$$
and for the path integral (PF - {\it P\/}roduct-{\it F\/}orm)
\plus$$\myalign
  &K(\vec q'',\vec q';T)
  \\   &
  =\int\limits_{\vec q(t')=\vec q'}^{\vec q(t'')=\vec q''}\sqrt{g}\,
  \CD_{PF}\vec q(t)
  \exp\bigg\{\ih\int_{t'}^{t''}
      \bigg[{m\over2}h_{ac}(\vec q)h_{cb}(\vec q)\dot q^a\dot q^b
   -V(\vec q)-\Delta V(\vec q)\bigg]dt\bigg\}
  \\   &\equiv\lim_{N\to\infty}
  \Norm^{ND/2}\prod_{j=1}^{N-1}\int d\vec q_j\sqrt{g(\vec q_j)}
  \\   &\qquad\times
  \exp\bigg\{\ih\sum_{j=1}^N\bigg[{m\over2\epsilon}
  h_{bc}(\vec q_j)h_{ac}(\vec q_{j-1})\Delta q_j^a\Delta q_j^b
  -\epsilon V(\vec q_j)-\epsilon\Delta V_{PF}(\vec q_j)\bigg]\bigg\}
  \enspace.
  \tag\NUM.\num\endalign$$
\edef\numba{\NUM.\num}%
$\Delta V_{PF}$ denotes the well-defined quantum potential
\plus$$\Delta V_{PF}={\hbar^2\over8m}
  \Big[g^{ab}\Gamma_a\Gamma_b+2(g^{ab}\Gamma_b)_{,b}
   +{g^{ab}}_{,ab}\Big]
  +{\hbar^2\over8m}
  \Big(2h^{ac}{h^{bc}}_{,ab}-{h^{ac}}_{,a}{h^{bc}}_{,b}
                -{h^{ac}}_{,b}{h^{bc}}_{,a}\Big)
  \tag\NUM.\num$$
\edef\numbc{\NUM.\num}%
arising from the specific lattice formulation for the path integral,
respectively the ordering prescription for position and momentum
operators in the quantum Hamiltonian. We only use the lattice
formulation of (\numba) in this paper unless otherwise (and explicitly)
stated.

Indispensable tools in path integral techniques are transformation
rules. In order to avoid cumbersome notation, we restrict ourselves to
the one-di\-men\-sio\-nal case. For the general case we refer to DeWitt
[\DEW], Fischer, Leschke and M\"uller [\FLM], Gervais and Jevicki
[\GJ], [\GROm, \GROw, \GRSb, \GRSf], Junker [\JUNc], Kleinert [\KLEh,
\KLE], Pak and S\"okmen [\PAKSc], Steiner [\STEc] and Storchak
[\STORCH], and references therein. We consider the one-dimensional
path integral
\plus$$K(x'',x';T)
  =\int\limits_{x(t')=x'}^{x(t'')=x''}\CD x(t)
  \exp\bigg[\ih\int_{t'}^{t''}\bigg({m\over2}\dot x^2
  -V(x)\bigg)dt\bigg]
  \tag\NUM.\num$$
\edef\numbb{\NUM.\num}%
and perform the coordinate transformation $x=F(q)$. Implementing this
transformation, one has to keep all terms of $O(\epsilon)$ in (\numbb).
Expanding about midpoints, the result is
\plus$$\multline
  \!\!\!\!\!\!
  K\big(F(q''),F(q');T\big)\!=\!\Big[F'(q'')F'(q')\Big]^{-1/2}
  \!\lim_{N\to\infty}\bigg({m\over2\pi\i\epsilon\hbar}\bigg)^{1/2}
  \prod_{j=1}^{N-1}\int dq_j\cdot\prod_{j=1}^N F'(\bar q_j)
  \hfill\\   \times
  \exp\left\{\ih\sum_{j=1}^N\left[{m\over2\epsilon}
       {F'}^2(\bar q_j)(\Delta q_j)^2
       -\epsilon V(F(\bar q_j))-{\epsilon\hbar^2\over8m}
        {{F''}^2(\bar q_j)\over{F'}^4(\bar q_j)}\right]\right\}
  \enspace.
  \endmultline
  \tag\NUM.\num$$
\edef\numbd{\NUM.\num}%

It is obvious that the path integral representation (\numbd) is not
completely satisfactory. Whereas the transformed potential $V(F(q))$
may have a convenient form when expressed in the new coordinate $q$,
the kinetic term ${m\over2}{F'}^2\dot q^2$ is in general nasty. Here
the so-called ``time-transformation'' comes into play which leads in
combination with the ``space-transformation'' already carried out to
general ``space-time transformations'' in path integrals. The
time-transformation is implemented [\DKa, \DKb, \STEc, \STORCH] by
introducing a new ``pseudo-time'' $s''$ by means of $s''=\int_{t'}^{t''}
ds/{F'}^2(q(s))$. A rigorous lattice derivation is far from being
trivial and has been discussed by many authors. Recent attempts to put
it on a sound footing can be found in Refs.~[\CAST, \FLM, \YODEWM]. A
convenient way to derive the corresponding transformation formul\ae\
uses the energy dependent Green's function $G(E)$ of the kernel $K(T)$
defined by
\plus$$G(x'',x';E)=\bigg<q''
  \bigg\vert {1\over\b H-E-\i\epsilon}\bigg\vert q'\bigg>
  =\ih\int_0^\infty dT \e^{\i(E+\i\epsilon)T/\hbar}
    K(x'',x';T)\enspace.
  \tag\NUM.\num$$
where a small positive imaginary part $(\epsilon>0)$ has been added to
the energy $E$. (Usually we not explicitly write the $\i\epsilon$, but
will tacitly assume that the various expressions are regularized
according to this rule). For the path integral (\numbb) one obtains the
following transformation formul\ae
$$\myalign
  K(x'',x';T)&=\int_{\bbbr}{dE\over2\pi\i}
               \e^{-\i ET/\hbar}G(q'',q';E)\enspace,
  \tag\NUM.\num\\   \global\plus
  G(q'',q';E)&=\ih\Big[F'(q'')F'(q')\Big]^{1/2}
  \int_0^\infty ds''\widehat K(q'',q';s'')\enspace,
  \tag\NUM.\num\endalign$$
\minus
\edef\numbf{\NUM.\num}\plus%
with the transformed path integral $\widehat K(s'')$ given by
\plus$$\multline
  \widehat K(q'',q';s'')
  =\lim_{N\to\infty}\bigg({m\over2\pi\i\epsilon\hbar}\bigg)^{1/2}
   \prod_{j=1}^{N-1}\int dq_j
  \\   \qquad\times
  \exp\Bigg\{\ih\sum_{j=1}^N\Bigg[{m\over2\epsilon}(\Delta q_j)^2
       -\epsilon {F'}^2(\bar q_j)\Big(V(F(\bar q_j))-E\Big)
  \hfill\\
       -{\epsilon\hbar^2\over8m}\Bigg(
        3{{F''}^2(\bar q_j)\over{F'}^2(\bar q_j)}
         -2{F'''(\bar q_j)\over F'(\bar q_j)}
        \Bigg)\Bigg]\Bigg\}\enspace.
  \endmultline
  \tag\NUM.\num$$
\edef\numbj{\NUM.\num}%
Further refinements are possible and general formul\ae\ of practical
interest and importance can be derived. Let us note that also an
explicitly time-dependent ``space-time transformation'' $x=F(q,t)$
can be formulated similarly to the formul\ae\ (\numbf-\numbj), c.f.\
Refs.~[\GROw, \GRSg, \GRSh, \KLE, \PELST, \STORCH].

Finally we consider a pure time transformation in a path
integral. We consider
\plus$$G(\vec q'',\vec q';E)=\sqrt{f(\vec q')f(\vec q'')}
  \ih\int_0^\infty ds''
  \Big<\vec q''\Big\vert\exp\Big(-\i s''\sqrt{f}\,(\b
               H-E)\sqrt{f}\,/\hbar\Big)\Big\vert\vec q'\Big>\enspace,
  \tag\NUM.\num$$
which corresponds to the introduction of the ``pseudo-time''
$s''=\int_{t'}^{t''}ds/f(\vec q(s))$ and we assume that the
Hamiltonian $H$ is product ordered. Then
\plus$$G(\vec q'',\vec q';E)=\ih(f'f'')^{\half(1-D/2)}
  \int_0^\infty\,\widetilde K(\vec q'',\vec q';s'')\,ds''
  \tag\NUM.\num$$
with the path integral
\plus$$\multline
  \widetilde K(\vec q'',\vec q';s'')
  =\int\limits_{\vec q(t')=\vec q'}^{\vec q(t'')=\vec q''}
   \sqrt{\tilde g}\,\CD_{PF}\vec q(t)
      \\   \times
  \exp\left\{\ih\int_0^{s''}\bigg[{m\over2}
  \tilde h_{ac}\tilde h_{cb}\dot q^a\dot q^b
  -f\Big(V(\vec q)+\Delta V_{PF}(\vec q)-E\Big)\bigg]ds\right\}\enspace.
  \endmultline
  \tag\NUM.\num$$
\edef\numBde{\NUM.\num}%
Here are $\tilde h_{ac}=h_{ac}/\sqrt{f}$, $\sqrt{\tilde g}=\det
(\tilde h_{ac})$ and (\numBde) is of the canonical product form.

The third ingredient in our calculations will be the technique of
separation of variables in path integrals [\GROb]. Let us sketch the
most important features of this technique. We assume that a potential
problem $V(\vec x)$ has an exact solution according to
\plus$$
  \int\limits_{\vec x(t')=\vec x'}^{\vec x(t'')=\vec x''}\CD\vec x(t)
  \exp\bigg[
 \ih\int_{t'}^{t''}\bigg({m\over2}\dot{\vec x}^2-V(\vec x)\bigg)dt\bigg]
  =\int dE_\lambda\,e^{-\i E_\lambda T/\hbar}
  \Psi_\lambda^*(\vec x')\Psi_\lambda(\vec x'')\enspace,
  \tag\NUM.\num$$
and $\vec x$ is a set of variables of dimension $d$. Now we consider
the path integral
\plus$$\myalign
  &K(\vec z'',\vec z',\vec x'',\vec x';T)
  =\int\limits_{\vec z(t')=\vec z'}^{\vec z(t'')=\vec z''}
  f^d(\vec z)G(\vec z)\CD\vec z(t)
  \int\limits_{\vec x(t')=\vec x'}^{\vec x(t'')=\vec x''}\CD\vec x(t)
  \\   &\qquad\times
  \exp\left\{\ih\int_{t'}^{t''}\bigg[{m\over2}
  \Big((\vec g\cdot{\dot{\vec z}})^2+f^2{\dot{\vec x}}^2
  -\bigg({V(\vec x)\over f^2(\vec z)}
  +V(\vec z)+\Delta\tilde V(\vec z)\bigg)\bigg]dt\right\}\enspace.
  \tag\NUM.\num\endalign$$
\edef\numbk{\NUM.\num}%
Here, $\vec z$ denotes a $d'$-dimensional coordinate with $d+d'=D$,
$g_i$ and $f$ the elements of the $D$-dimensional metric tensor
$g_{ab}=\diag[g_1^2,\dots, g_{d'}^2,f^2,\dots,f^2]$, $\Delta\tilde V$
the quantum potential of (\numbc), and $\det(g_{ab})=f^{2d}\prod_i g_i^2
\equiv f^{2d}G(z)$. As shown in Ref.~[\GROb] by performing a
time-transformation (see Duru and Kleinert [\DKb] and Kleinert [\KLE])
forth and back in the path integral (\numbk) we can separate the $\vec
x$ from the $\vec z$ variables yielding
\plus$$\multline
  K(\vec z'',\vec z',\vec x'',\vec x';T)
  =[f(\vec z')f(\vec z'')]^{-d/2}
   \int dE_\lambda\Psi_\lambda^*(\vec x')\Psi_\lambda(\vec x'')
  \hfill\\  \times
  \int\sqrt{G(\vec z)}\,\CD\vec z(t)\exp\left\{\ih\int_{t'}^{t''}
  \bigg[{m\over2}\big(\vec g\cdot{\dot{\vec z}}\big)^2
   -V(\vec z)-\Delta\tilde V(\vec z)
   -{E_\lambda\over f^2(\vec z)}\bigg]dt\right\}.
  \endmultline
  \tag\NUM.\num$$
\edef\numbl{\NUM.\num}%
Of course, also $\vec x$-depended metric terms can be included (with
their corresponding quantum potentials) in the separated $\vec x$-path
integration without changing the general feature.


\PLUS\glno=0                      
\section{Separable Coordinate Systems on Spaces of Constant Curvature}
\ssf
\subsection{The Coordinates Systems}
A systematic approach for the D-dimensional on the problem on the
classification of separable coordinate systems in spaces of constant
curvature (positive, zero, negative) is due to Kalnins et al.~[\KAL,
\KAMIa]. By the notion ``separable coordinate system'' we man any
coordinate system which separate the classical Hamilton-Jacobi
equations, respectively the Schr\"odinger equation. The classification
goes at follows [\KAL]:

\subsubsection{The Sphere $\sD$}
We denote the coordinates on the sphere $\sD$ by the vector
$\vec s=(s_0,\dots,s_{D-1})$. The basic building blocks of separable
coordinates systems on $\sD$ are the (D-1)-sphere {\it elliptic\/}
coordinates
\plus$$
  s_j^2=\dsize\frac{\prod_{i=1}^{D-1}(\rho_i-e_j)}
                   {\prod_{j\not=i}(e_i-e_j)}\enspace,\quad
  (j=0,\dots,D-1)\enspace,\quad
  \sum_{j=0}^{D-1}s_j^2=1\enspace,
  \tag\NUM.\num$$
corresponding to a metric
\plus$$
  ds^2=-{1\over4k}\sum_{i=1}^{D-1}{1\over P_D(\rho_i)}
  \bigg[\prod_{j\not=i}(\rho_i-\rho_j)\bigg](d\rho_i)^2\enspace,
  \qquad
  P_D(\rho)=\prod_{i=0}^D(\rho-e_i)
  \tag\NUM.\num$$
\edef\numcb{\NUM.\num}%
($k>0$ curvature). In order to find the possible explicit coordinate
systems one must pay attention to the requirements that, (i) the metric
must be positive definite, (ii) the variables $\{\rho\}_{i=1}^{D-1}$
should vary in such a way that they correspond to a coordinates patch
which is compact. There is a unique solution to these requirements
given by
\plus$$
  e_0<\rho_1<e_1<\dots<e_{D-1}<\rho_D<e_D\enspace.
  \tag\NUM.\num$$

\subsubsection{The Euclidean Space $\eD$}
In D-dimensional Euclidean space we have first the coordinate system
corresponding to the D-sphere elliptic (\numcb)
\plus$$
  x_j^2=c^2\dsize\frac{\prod_{i=1}^D(\rho_i-e_j)}
                   {\prod_{j\not=i}(e_i-e_j)}\enspace,\qquad
  (j=1,\dots,D)
  \tag\NUM.\num$$
($c^2$ constant). In addition there is a second class of coordinate
systems, namely the parabolic coordinates
\plus$$\aligned
  x_1^2&={c\over2}(\rho_1+\dots+\rho_D+e_1+\dots+e_{D-1})\enspace,
  \\
  x_j^2&=-c^2\dsize\frac{\prod_{i=1}^D(\rho_i-e_j)}
                   {\prod_{j\not=i}(e_i-e_j)}\enspace,\quad
  (j=2,\dots,D)\enspace.
  \endaligned
  \tag\NUM.\num$$

\subsubsection{The Pseudo-Sphere $\lD$}
On the  pseudosphere $\lD$ the complexity increases considerably.
One starts by considering the line element
\plus$$
  ds^2=-{1\over4k}\sum_{i=1}^{D-1}{1\over P_D(\rho_i)}
  \bigg[\prod_{j\not=i}(\rho_i-\rho_j)\bigg](d\rho_i)^2
  \tag\NUM.\num$$
($k<0$ curvature), and one must require that $ds^2>0$. It turns out that
there are four classes of solutions determined by the character of the
solutions of the characteristic equation $P_D(\rho)=0$.

\medskip\noindent{\sl \the\secno.\the\subno.\the\subsubno.A.}
The first class is characterised by $e_i\not=e_j$ for $i,j=0,\dots,D-1$.
If $D-1=n=2p+1$ is odd then
\plus$$
  \dots\rho_{i-2}<e_{i-2}<\rho_{i-1}<e_{i-1}<e_i<e_{i+1}<
  \rho_i<e_{i+2}<\dots<e_{2p+2}<e_{2p+1}\enspace,
  \tag\NUM.\num$$
($i=0,\dots,p$) with the convention that $e_j,\rho_j=0$ for $j$ a
non-positive integer which give $p+1$ distinct possibilities. Using
$E_i^{(j)}= e_{i+j+1}$ ($i=1,\dots,2p+2$, $j=1,\dots,p+1$) and
$e_r=e_s$ for $r=s\mod(n+1)$, the coordinates on $\lD$ are written in
the following way
\plus$$
  u_0^2=\dsize\frac{\prod_{i=1}^n(\rho_i-E_1^{(j)})}
                   {\prod_{k\not=1}(E_k^{(j)}-E_1^{(j)})}
  \enspace,\qquad
  u_l^2=\dsize\frac{\prod_{i=1}^n(\rho_i-E_{l+1}^{(j)})}
                   {\prod_{k\not=l+1}(E_k^{(j)}-E_{l+1}^{(j)})}
  \enspace.
  \tag\NUM.\num$$
Similarly if $D-1=n=2p+2$ is even ($i=0,\dots,p$)
\plus$$
  \dots\rho_{i-2}<e_{i-2}<\rho_{i-1}<e_{i-1}<e_i<e_{i+1}<
  \rho_i<\dots<e_{2p+1}<e_{2p}\enspace.
  \tag\NUM.\num$$

\medskip\noindent{\sl \the\secno.\the\subno.\the\subsubno.B.}
The second class is characterised by the fact that there can be two
complex conjugate zeros of $P_D(\rho)=0$ denoted by $e_1=\alpha+\i
\beta$, $e_2=\alpha-\i\beta$ ($\alpha,\beta\in\bbbr$), respectively.
Together with the convention $e_{i+1}\equiv f_{i-1}$ for all other
$e_j$  there is the one possibility
\plus$$
  \rho_1<f_1<\rho_2<f_2<\dots<\rho_{n-1}<f_{n-1}<\rho_n\enspace.
  \tag\NUM.\num$$
A suitable choice of coordinates is ($j=2,\dots,n$)
\plus$$
  (u_0+\i u_1)^2
  ={\i\over\beta}\dsize\frac{\prod_{i=1}^n(\rho_i-\alpha-\i\beta)}
                   {\prod_{i=1}^{n-1}(f_i-\alpha-\i\beta)}
  \enspace,\quad
  u_j^2=\dsize\frac{-\prod_{i=1}^n(\rho_i-f_{j-1})}
        {[(\alpha-f_{j-1})^2+\beta^2]\prod_{i\not=j-1}(f_i-f_{j-1})}
  \enspace.
  \tag\NUM.\num$$

\medskip\noindent{\sl \the\secno.\the\subno.\the\subsubno.C.}
In the third class we have the two-fold root $e_1=e_2=a$, say. Let us
denote  $G_j^{(i)}=g_{j+1}$ ($j=1,\dots,n-1, i=0,\dots,p$), where
$e_j=g_{j-2}$ ($j=3,\dots,n+1$) with $g_k\not=g_l$ if $k\not=l$ and
$g_k\not=a$ for any $k$, $g_r=g_s$ for $r=s\mod(n+1)$, $n=2p+1$ for
$n$ odd, and $n=2p$ for $n$ even, respectively. This case divides into
two families with coordinates varying in the ranges ($i=0,\dots,p$)
\plus$$\gather
  \dots\rho_{i-1}<g_{i-1}<\rho_i<g_i<a<\rho_{i+1}<g_{i+1}<\dots
       g_{n-1}<\rho_n\enspace,
  \tag\NUM.\num a\\
  \dots\rho_{i-1}<g_{i-1}<\rho_i<g_i<\rho_{i+1}<a<g_{i+1}<\dots
       g_{n-1}<\rho_n\enspace,
  \tag\NUM.\num b\endgather$$
\edef\numcc{\NUM.\num}%
and in either case of $n$ there are $p+1$ distinguishable cases to
consider. A suitable choice of coordinates is
\plus$$\left.\aligned
  (u_0-u_1)^2
  &=\epsilon\dsize\frac{\prod_{i=1}^n(\rho_i-a)}
                   {\prod_{k=1}^{n-1}(G_j^{(i)}-a)}\enspace,
  \\
  (u_0^2-u_1^2)
  &={\partial\over\partial a}
           \dsize\frac{\prod_{i=1}^n(\rho_i-a)}
                   {\prod_{k=1}^{n-1}(G_j^{(i)}-a)}\enspace,
  \\
  u_j^2&=-\dsize\frac{\prod_{i=1}^n(\rho_i-G_{j-1}^{(i)})}
        {(a-G_{j-1}^{(i)})^2\prod_{l\not=j-1}(G_l^{(i)}-G_{j-1}^{(i)})}
  \enspace,
  \endaligned\qquad\qquad\right\}
  \tag\NUM.\num$$
($j=2,\dots,n$) and $\epsilon=+1$ in (\numcc a), $\epsilon=-1$ in
(\numcc b).

\medskip\noindent{\sl \the\secno.\the\subno.\the\subsubno.D.}
The forth case is characterised by $e_1=e_2=e_3=b$. We set
$e_j=h_{j-3}$ ($j=4,\dots,n+1$) with $h_k\not=k_l$ for $k\not=l$ and
$h_k\not=b$ for any $k$. Then
\plus$$
  \dots<\rho_{i-1}<h_{i-1}<\rho_i<\rho_{i+1}<b<\rho_{i+2}<h_{i+1}
  <\dots<\rho_{n-1}<h_{n-2}<\rho_n\enspace,
  \tag\NUM.\num$$
($i=0,\dots,p$) and there are $p+1$ distinct cases. A suitable choice of
coordinates is
\plus$$\left.\aligned
  (u_0-u_1)^2
  &=-\dsize\frac{\prod_{i=1}^n(\rho_i-b)}
                   {\prod_{k=1}^{n-2}(H_j^{(i)}-b)}\enspace,
  \\
  2u_2(u_0-u_1)
  &=-{\partial\over\partial b}
           \dsize\frac{\prod_{i=1}^n(\rho_i-b)}
                   {\prod_{k=1}^{n-2}(H_j^{(i)}-b)}\enspace,
  \\
  (u_0^2-u_1^2-u_2^2)
  &=-\half{\partial^2\over\partial b^2}
           \dsize\frac{\prod_{i=1}^n(\rho_i-b)}
                   {\prod_{k=1}^{n-2}(H_j^{(i)}-b)}
  \\
  u_j^2&=-\dsize\frac{\prod_{i=1}^n(\rho_i-H_{j-2}^{(i)})}
        {\prod_{l\not=j-2}(H_l^{(i)}-H_{j-2}^{(i)})(b-H_{j-2}^{(i)})^3 }
  \enspace,
  \endaligned\qquad\qquad\right\}
  \tag\NUM.\num$$
($j=3,\dots,n$) and $H_j^{(i)}=h_{j+1}$ ($j=1,\dots,n-2$,
$i=0,\dots,p$), and $h_r=h_s\mod(n-2)$.

\noindent
For the lowest two case we obtain the following small table
of distinct coordinate systems
$$\aligned
\vbox{\offinterlineskip
\hrule
\halign{&\vrule#&
  \strut\quad\hfil#\quad\hfil\quad\cr
height2pt&\omit&&\omit&&\omit&\cr
&              &&$D=2$
               &&$D=3$                                      &\cr
height2pt&\omit&&\omit&&\omit&\cr
\noalign{\hrule}
\noalign{\hrule}
height2pt&\omit&&\omit&&\omit&\cr
&$\sD$         &&2
               &&6                                         &\cr
height2pt&\omit&&\omit&&\omit&\cr
\noalign{\hrule}
height2pt&\omit&&\omit&&\omit&\cr
&$\eD$         &&4
               &&11                                        &\cr
height2pt&\omit&&\omit&&\omit&\cr
\noalign{\hrule}
height2pt&\omit&&\omit&&\omit&\cr
&$\lD$         &&9
               &&34                                        &\cr
height2pt&\omit&&\omit&&\omit&\cr}\hrule}
  \endaligned$$
There seems to be no obvious closed recursions for D a natural number
[\KUZ].

\subsection{Separability of Variables in the Path Integral}
Separation of variables is always of great importance in many
calculations. The separation formula (\numbl) provides us already with
a prescription in the case where we know a separating coordinate system.
However, the situation is often somewhat more complicated and one has
to look first for the coordinate systems which separate the relevant
partial differential equations, i.e.\ the Hamiltonian, and, more
important from our point of view, the path integral. In order to develop
such a theory we consider according to [\MF] the Lagrangian $\CL={m\over
2}\sum_{i=1}^Dh_i^2\dot x_i^2$ and the Laplacian $\Delta_{LB}$,
respectively, in the following way (where only orthogonal coordinate
systems are taken into account)
\plus$$\myalign
   \Delta_{LB}
   &=\sum_{i=1}^D{1\over\prod_{j=1}^Dh_j(\{\xi\})}
   {\partial\over\partial\xi_i}\left(
   {\prod_{k=1}^Dh_k(\{\xi\})\over h_i^2(\{\xi\})}
     {\partial\over\partial \xi_i}\right)
  \\   &
  =:\sum_{i=1}^D{1\over\prod_{j=1}^Dh_j(\{\xi\})}
   {\partial\over\partial\xi_i}\left(
   g_i(\xi_1,\dots,\xi_{i-1},\xi_{i+1},\dots,\xi_D)f(\xi_i)
   {\partial\over\partial\xi_i}\right)\enspace,
  \tag\NUM.\num\endalign$$
where $\{\xi\}$ denotes the set of variables $(\xi_1,\dots,\xi_D)$, and
the existence of the functions $f_i, g_i$ is necessary for the
separation [\KAL, \MF]. Note that the factorizing in terms of the $h_i$
is perfectly adopted to the product form prescription. We introduce the
St\"ackel-determinant [\KAL, \MF, \OLE]
\plus$$
  S(\{\xi\})=\det(\Phi_{ij})=\prod_{i=1}^D{h_i(\{\xi\})\over f_i(\xi_i)}
  \ ,\
  M_i(\xi_1,\dots,\xi_{i-1},\xi_{i+1},\dots,\xi_D)
  ={\partial S\over\partial\Phi_{i1}}
  ={S(\{\xi\})\over h_i^2(\{\xi\})}\enspace,
  \tag\NUM.\num$$
and abbreviate  $\Gamma_i=f_i'/f_i$. Then
\plus$$
  g_i(\xi_1,\dots,\xi_{i-1},\xi_{i+1},\dots,\xi_D)
  =M_i(\xi_1,\dots,\xi_{i-1},\xi_{i+1},\dots,\xi_D)
  \prod_{\scriptstyle j=1 \atop\scriptstyle i\not=j}^D
  f_j(\xi_j)\enspace,
  \tag\NUM.\num$$
which fixes the functions $g_i$. Introducing the (new) momentum
operators
\plus$$ P_{\xi_i}=\hi
   \bigg({\partial\over\partial\xi_i}+\half{f_i'\over f_i}\bigg)
   \enspace,
  \tag\NUM.\num$$
we write the Legendre transformed Hamiltonian [\GRSa] as follows
\plus$$\myalign
  0&=H-E=-\hbarm\Delta_{LB}-E
  \\   &
   =-\hbarm\sum_{i=1}^D{1\over\prod_{j=1}^Dh_j}
   {\partial\over\partial\xi_i}\left(
   {\prod_{k=1}^Dk_k\over h_i^2}{\partial\over\partial\xi_i}\right)-E
  =-\hbarm{1\over S}\sum_{i=1}^D\bigg[{1\over f_i}
   {\partial\over\partial\xi_i}
   \bigg(f_i{\partial\over\partial\xi_i}\bigg)\bigg]-E
  \\   &
  =-\hbarm{1\over S}\sum_{i=1}^DM_i
   \bigg({\partial^2\over\partial\xi_i^2}
   +\Gamma_i{\partial\over\partial\xi_i}\bigg)-E
  \\   &
  ={1\over S}\left[{1\over2m}\sum_{i=1}^DM_iP_i^2-ES+{\hbar^2\over8m}
       \sum_{i=1}^DM_i\Big(\Gamma_i^2+2\Gamma_i'\Big)\right]\enspace.
  \tag\NUM.\num\endalign$$
Therefore we obtain according to the general theory the following
identity in the path integral by means of the space-time transformation
technique
\plus$$\myalign
  &\prod_{i=1}^D\int\limits_{\xi_i(t')=\xi_i'}^{\xi_i(t'')=\xi_i''}
   h_i\CD\xi_i(t)
   \exp\left\{\ih\int_{t'}^{t''}\left[{m\over2}\sum_{i=1}^D
      h_i^2\dot\xi_i^2-\Delta V_{PF}(\{\xi\})\right]dt\right\}
  \\   &
  =\prod_{i=1}^D\int\limits_{\xi_i(t')=\xi_i'}^{\xi_i(t'')=\xi_i''}
   \sqrt{S\over M_i}\,\CD\xi_i(t)
   \exp\left\{\ih\int_{t'}^{t''}\left[{m\over2}S\sum_{i=1}^D
   {\dot\xi_i^2\over M_i}-\Delta V_{PF}(\{\xi\})\right]dt\right\}
  \\   &
  =(S'S'')^{\half(1-D/2)}\int_{\bbbr}{dE\over2\pi\hbar}
  \e^{-\i ET/\hbar}\int_0^\infty ds''
   \prod_{i=1}^D\int\limits_{\xi_i(0)=\xi_i'}^{\xi_i(s'')=\xi_i''}
   M_i^{-1/2}\CD\xi_i(s)
  \\   &\qquad\qquad\times
   \exp\left\{\ih\int_{0}^{s''}\left[{m\over2}\sum_{i=1}^D
   {\dot\xi_i^2\over M_i}+ES
    -{\hbar^2\over8m}\sum_{i=1}^DM_i\Big(\Gamma_i^2+2\Gamma_i'\Big)
    \right]ds\right\}\enspace.
  \tag\NUM.\num\endalign$$
\edef\numca{\NUM.\num}%


\PLUS\glno=0                      
\section{Separation of Variables in Two Dimensions}
The subject of the next two Sections will be the path integral
formulation in the three two-dimensional spaces of constant curvature,
the flat space $\ezwei$, the sphere $\szwei$ and the  pseudosphere
$\lzwei$, respectively the path integral formulation in the three
three-dimensional spaces of constant curvature, the flat space
$\edrei$, the sphere $\sdrei$, and the  pseudosphere $\ldrei$.
For notation, in the two-dimensional case we denote by $\vec x=(x,y)
=(x_1,x_2)$ the flat space coordinates, by $\vec s=(s_0,s_1,s_2)$
coordinates on the sphere, and by $\vec u=(u_0,u_1,u_2)$ coordinates on
the pseudosphere, and in the three-dimensional case $\vec x=(x,y,z)$
the flat space coordinates and $\vec s$ and $\vec u$ in obvious
generalization from the two-dimensional case. In each case, we first
state the general form of the propagator and the Green function
in $D=2$ and $D=3$, respectively, in terms of the invariant distance
(norm) $d_\eD(\vec q'', \vec q')= \vec x'\cdot\vec x''$ in $\eD$,
$\cos\psi_\sD(\vec q'', \vec q')=\vec s'\cdot\vec s''/R^2$ in $\sD$,
and $\cosh d_{\lD}(\vec q'',\vec q')=\vec u'\cdot\vec u''/R^2$ in
 $\lD$. The sphere $\sD$ is described by the constraint $\vec s^2=R^2$,
and $\lD$ by $\vec u^2= u_0^2-\sum_{i=1}^{D-1}u_i^2=R^2$ which fixes the
signature of the metric in $\vec u$-space. The metric is determined via
the classical Lagrangians, say,
\plus$$
  \CL_{Cl}^\eD(\vec x,{\dot{\vec x}})={m\over2}{\dot{\vec x}}^2
  \enspace,\qquad
  \CL_{Cl}^\sD(\vec s,{\dot{\vec s}})={m\over2}{\dot{\vec s}}^2
  \enspace,\qquad
  \CL_{Cl}^\lD(\vec u,{\dot{\vec u}})
  =-{m\over2}{\dot{\vec u}}^2\enspace.
  \tag\NUM.\num$$
We put $R^2=1$, since $R$ is just a scaling factor of the systems.
Consequently, the Laplacians are determined by
\plus$$
  \Delta_\eD=\sum_{i=1}^D{\partial^2\over\partial x_i^2}
  \enspace,\qquad
  \Delta_\sD=\sum_{i=0}^{D-1}{\partial^2\over\partial s_i^2}
  \enspace,\qquad
  \Delta_\lD={\partial^2\over\partial u_0^2}
  -\sum_{i=1}^{D-1}{\partial^2\over\partial u_i^2}
  \enspace.
  \tag\NUM.\num$$
We would like to point out that the various path integral evaluations
are straightforward by applying the basic path integral solutions of
Appendix~1. For completeness we present in this Section some examples
somewhat more comprehensive to help understanding; in the next Section
it is then sufficient just to take reference to the former ones. It is
obvious that in two dimensions, the more easy path integral solutions
are already known and the emphasize of this Section will be more on
the collection of the various identities, instead on the path integral
evaluations. On the other, where the theory of special functions is not
well developed, such path integral evaluations cannot be explicitly
done, therefore only an indirect reasoning is possible. Especially in
these cases we state the general form of the propagator, respectively
the Green function. We do not discus coordinate systems which are
equivalent with a presented one and which can be obtained by a simple
reparameterization of the variables. Compare e.g.\ the two different
definitions of the parabolic coordinates [\EWA, \MF] in $\ezwei,
\edrei$, and of the equidistant coordinates [\GROc, \OLE, \VISM] for
$\lzwei$, respectively.

\subsection{The Flat Space $\ezwei$}
\subsubsection{General Form of the Propagator and the Green Function}
First of all we discuss the general form of the propagator and the
Green function in the two-dimensional flat space $\ezwei$, which are
easiest calculated in cartesian coordinates (see below). One finds that
in $\ezwei$ they are four different coordinate systems [\MF]. In terms
of the norm $d_\ezwei(\vec q'',\vec q')$ (where $\vec q$ denotes any
coordinate system) they are given by
$$\myalign
  K^\ezwei\Big(d_{\ezwei}(\vec q'',\vec q');T\Big)
  &={m\over2\pi\i\hbar T}\exp\bigg[{\i m\over2\hbar T}
             d^2_{\ezwei}(\vec q'',\vec q')\bigg]\enspace,
  \tag\NUM.\num\\   \global\plus
  G^\ezwei\Big(d_{\ezwei}(\vec q'',\vec q');E\Big)
  &={m\over\pi\hbar^2}K_0\bigg({d_\ezwei(\vec q'',\vec q')\over\hbar}
    \sqrt{-2mE}\,\bigg)\enspace.
  \tag\NUM.\num\endalign$$
The norm in $\ezwei$ is given by (for the definition of the coordinates
see below)
\plus$$\myalign
  &\hbox{Cartesian Coordinates:}
  \\   &
  d_\ezwei^2(\vec q'',\vec q')
        =\vert\vec x''-\vec x'\vert^2\enspace,
  \tag\NUM.\num a
  \\   &\hbox{Cylindrical Coordinates:}
  \\   &\phantom{d_\ezwei^2}
  ={r'}^2+{r''}^2-2r'r''\cos(\phi''-\phi')\enspace,
  \tag\NUM.\num b
  \\   &\hbox{Parabolic Coordinates:}
  \\   &\phantom{d_\ezwei^2}
  =\viert\Big[({\eta''}^2+{\xi''}^2)^2+({\eta'}^2+{\xi'}^2)^2
   -2({\eta'}^2-{\xi'}^2)({\eta''}^2-{\xi''}^2)
   -8\eta'\eta''\xi'\xi''\Big]\enspace,\qquad\qquad
  \tag\NUM.\num c
  \\   &\hbox{Elliptic Coordinates:}
  \\   &\phantom{d_\ezwei^2}
  =d^2(\cosh\mu'\cosh\mu''\cos\nu'\cos\nu''
   +\sinh\mu'\sinh\mu''\sin\nu'\sin\nu'')\enspace.
  \tag\NUM.\num d\endalign$$

\subsubsection{Cartesian Coordinates}
We consider the usual cartesian coordinates $(x,y)=\vec x\in\bbbr^2$.
The metric is given by $(g_{ab})=\bbbone_2$ and for the momentum
operators we have $p_x=-\i\hbar\partial_x$, $p_y=-\i\hbar\partial_y$.
Therefore:
\plus$$
  -\hbarm\Delta_\ezwei
  =-\hbarm\bigg({\partial^2\over\partial x^2}
               +{\partial^2\over\partial y^2}\bigg)
  ={1\over2m}(p_x^2+p_y^2)\enspace.
  \tag\NUM.\num$$
The path integral formulation is well-known and given by [\FEY, \FH]
$$\myalign
  \int\limits_{\vec x(t')=\vec x'}^{\vec x(t'')=\vec x''}\CD\vec x(t)
  \exp\left({\i m\over2\hbar}\int_{t'}^{t''}{\dot{\vec x}}^2dt\right)
  &={m\over2\pi\i\hbar T}\exp\bigg({\i m\over2\hbar T}
         \vert\vec x''-\vec x'\vert^2\bigg)
  \tag\NUM.\num\\   \global\plus
      &
  =\int_{\bbbr^2}{d\vec p\over4\pi^2}
    \exp\bigg[-{\i\hbar T\over2m}\vec p^2
        +\i\vec p\cdot(\vec x''-\vec x')\bigg]\enspace.\qquad
  \tag\NUM.\num\endalign$$
Note the formulation via a path integration over the Euclidean group
[\BJc].

\subsubsection{Cylindrical Coordinates}
We consider two-dimensional polar coordinates
\plus$$\alignedat 3
  x&=r\cos\phi\enspace,
  &\qquad
  &r>0\enspace,
  \\
  y&=r\sin\phi\enspace,
  &\qquad
  &0\leq\phi<2\pi\enspace.
  \endalignedat
  \tag\NUM.\num$$
The metric is given by $(g_{ab})=\diag(1,r^2)$, and the momentum
operators have the form
\plus$$
  p_r=\hi\bigg({\partial\over\partial r}+{1\over2r}\bigg)
  \enspace,\qquad
  p_\phi=\hi{\partial\over\partial\phi}\enspace.
  \tag\NUM.\num$$
\edef\numdf{\NUM.\num}%
This gives for the Hamiltonian
\plus$$
  -\hbarm\Delta_\ezwei
  =-\hbarm\bigg(
   {\partial^2\over\partial r^2}+{1\over r}{\partial\over\partial r}
   +{1\over r^2}{\partial^2\over\partial\phi^2}\bigg)
  ={1\over2m}\bigg(p_r^2+{1\over r^2}p_\phi^2\bigg)
   -{\hbar^2\over8mr^2}\enspace.
  \tag\NUM.\num$$
We therefore obtain the path integral identity
[\ART, \EG, \GRSa, \PI, \STEc]
\minus
$$\myalign
  &\int\limits_{r(t')=r'}^{r(t'')=r''}r\CD r(t)
  \int\limits_{\phi(t')=\phi'}^{\phi(t'')=\phi''}\CD\phi(t)
  \exp\left\{\ih\int_{t'}^{t''}\bigg[{m\over2}\Big(\dot r^2
      +r^2\dot\phi^2\Big)+{\hbar^2\over8mr^2}\bigg]dt\right\}
  \\   &
  =(r'r'')^{-1/2}\sum_{l\in\bbbz}{\e^{\i l(\phi''-\phi')}\over2\pi}
  \int\limits_{r(t')=r'}^{r(t'')=r''}\CD r(t)
  \exp\left[\ih\int_{t'}^{t''}\bigg({m\over2}\dot r^2
      -\hbar^2{l^2-\viert\over2mr^2}\bigg)dt\right]
  \\   &
  =(r'r'')^{-1/2}\sum_{l\in\bbbz}{\e^{\i l(\phi''-\phi')}\over2\pi}
  \int\limits_{r(t')=r'}^{r(t'')=r''}\CD r(t)\mu_l[r^2]
  \exp\left({\i m\over2\hbar}\int_{t'}^{t''}\dot r^2dt\right)
  \\   &
  ={m\over2\pi\i\hbar T}\exp\bigg[{\i m\over2\hbar T}
     ({r'}^2+{r''}^2)\bigg]
   \sum_{l\in\bbbz}\e^{\i l(\phi''-\phi')}
   I_l\bigg({mr'r''\over\i\hbar T}\bigg)
  \tag\NUM.\num\\   \global\plus
       &
  ={m\over2\pi\i\hbar T}\exp\bigg[{\i m\over2\hbar T}
     \Big({r'}^2+{r''}^2-2r'r''\cos(\phi''-\phi')\Big)\bigg]
  \tag\NUM.\num\\   \global\plus
       &
  =\sum_{l\in\bbbz}{\e^{\i l(\phi''-\phi')}\over2\pi}
   \int_0^\infty pdp\,J_l(pr')J_l(pr'')\e^{-\i\hbar p^2T/2m}\enspace.
  \tag\NUM.\num\endalign$$

\subsubsection{Elliptic Coordinates}
We consider the coordinate system
\plus$$\alignedat 3
  x&=d\cosh\mu\cos\nu\enspace,
   &\qquad
   &\mu>0\enspace,
  \\
  y&=d\sinh\mu\sin\nu\enspace,
   &\qquad
   &-\pi<\nu\leq\pi\enspace.
  \endalignedat
  \tag\NUM.\num$$
The metric is $(g_{ab})=d^2(\sinh^2\mu+\sin^2\nu)\bbbone_2$ and we
obtain for the momentum operators
\plus$$
  p_\mu=\hi\bigg({\partial\over\partial\mu}
         +{\sinh\mu\cosh\mu\over\sinh^2\mu+\sin^2\nu}\bigg)\enspace,
  \qquad
  p_\nu=\hi\bigg({\partial\over\partial\nu}
         +{\sin\nu\cos\nu\over\sinh^2\mu+\sin^2\nu}\bigg)\enspace.
  \tag\NUM.\num$$
\edef\numec{\NUM.\num}%
Consequently we have for the Hamiltonian
\plus$$\myalign
  -\hbarm\Delta_\ezwei
  &=-{\hbar^2\over2md^2(\sinh^2\mu+\sin^2\nu)}
    \bigg({\partial^2\over\partial\mu^2}
         +{\partial^2\over\partial\nu^2}\bigg)
  \\   &
  ={1\over2md^2}(\sinh^2\mu+\sin^2\nu)^{-1/2}
  (p_\mu^2+p_\nu^2)(\sinh^2\mu+\sin^2\nu)^{-1/2}\enspace.
  \tag\NUM.\num\endalign$$
The path integral {\it construction\/} is straightforward, however no
explicit path {\it integration\/} is possible. Actually, an expansion
into the corresponding wave-functions in the coordinates $\mu$ and $\nu$
yields the Mathieu functions $\me_\nu(\eta,h^2)$ and $\Me_\nu^{(1)}(\xi,
h^2)$ ($h^2=mEd^2/2\hbar^2$), respectively, as eigen-function of the
Hamiltonian, a specific class of higher transcendental functions
[\MESCH]. However, because we know on the one side the eigen-functions
of the Hamiltonian in terms of these functions [\MESCH], and on the
other the kernel in $\ezwei$ in terms of the invariant distance
$d_\ezwei$ we can state the following path integral identity (note the
implemented time-transformation)
\minus
$$\myalign
 &\!\!\!\int\limits_{\mu(t')=\mu'}^{\mu(t'')=\mu''}\!\!\!\CD\mu(t)\!\!\!
  \int\limits_{\nu(t')=\nu'}^{\nu(t'')=\nu''}\!\!\!\CD\nu(t)
  d^2(\sinh^2\mu+\sin^2\nu)
  \exp\left[{\i m\over2\hbar}d^2\!\!\int_{t'}^{t''}\!\!
   (\sinh^2\mu+\sin^2\nu)(\dot\mu^2+\dot\nu^2)dt\right]
  \\   &
  =\int_{\bbbr}{dE\over2\pi\hbar}\e^{-\i ET/\hbar}\int_0^\infty ds''
   \int\limits_{\mu(0)=\mu'}^{\mu(s'')=\mu''}\CD\mu(s)
  \int\limits_{\nu(0)=\nu'}^{\nu(s'')=\nu''}\CD\nu(s)
  \\   &\qquad\qquad\times
  \exp\left\{\ih\int_{0}^{s''}
   \bigg[{m\over2}(\dot\mu^2+\dot\nu^2)
    +Ed^2(\sinh^2\mu+\sin^2\nu)\bigg]ds\right\}
  \tag\NUM.\num\\   \global\plus
       &
  ={1\over2\pi}\sum_{\nu\in\Lambda}\int_0^\infty pdp\,
  \e^{-\i\hbar p^2T/2m}
  \\   &\qquad\qquad\times
  \me_\nu^*(\eta',\hbox{${d^2p^2\over4}$})
  \me_\nu(\eta'',\hbox{${d^2p^2\over4}$})
  \Me_\nu^{(1)\,*}(\xi',\hbox{${d^2p^2\over4}$})
  \Me_\nu^{(1)}(\xi'',\hbox{${d^2p^2\over4}$})
  \tag\NUM.\num\\   \global\plus
       &
  ={m\over2\pi\i\hbar T}\exp\bigg[{\i m\over2\hbar T}
             d^2_{\ezwei}(\vec q'',\vec q')\bigg]\enspace,
  \tag\NUM.\num\endalign$$
and $d_\ezwei(\vec q'',\vec q')$ must be taken in elliptic coordinates.
             The functions $\me
_\nu(\eta,h^2)$ and \linebreak
$\Me_\nu^{(1)}(\xi,h^2)$, are mutually determined
through the separation parameter $\lambda=\lambda_\nu(h^2)$ yielding a
countable set of numbers $\nu\in\Lambda$. In particular, the functions
$\Me_\nu^{(1)}(z,h^2)$ yield in the limit $h^2\to0$ the Bessel functions
$J_\nu$, i.e.\ $\Me_\nu^{(1)}(z/h,h^2)\simeq J_\nu(z)$ ($h\to0$), and
the functions $\me_\nu(z,h^2)\propto\e^{\i\nu z}$ ($h^2\to0$), therefore
obeying the correct boundary-conditions of our problem [\MESCH].

\subsubsection{Parabolic Coordinates}
We consider the coordinate system
\plus$$
  x=\xi\eta\enspace,
   \qquad
  y=\bhalf(\eta^2-\xi^2)\enspace,\qquad
   \xi\in\bbbr,\eta>0
  \tag\NUM.\num$$
(alternatively $\xi>0$, $\eta\in\bbbr$ [\MESCH]). We have $(g_{ab})=
(\xi^2+\eta^2)\bbbone_2$, and consequently for the momentum operators
\plus$$
  p_\xi=\hi\bigg({\partial\over\partial\xi}
          +{\xi\over\xi^2+\eta^2}\bigg)\enspace,\qquad
  p_\eta=\hi\bigg({\partial\over\partial\eta}
          +{\eta\over\xi^2+\eta^2}\bigg)\enspace.
  \tag\NUM.\num$$
\edef\numdg{\NUM.\num}%
This gives for the Hamiltonian
\plus$$
  -\hbarm\Delta_\ezwei
  =-{\hbar^2\over2m(\xi^2+\eta^2)}
   \bigg({\partial^2\over\partial\xi^2}
        +{\partial^2\over\partial\eta^2}\bigg)
  ={1\over2m}(\xi^2+\eta^2)^{-1/2}
             (p_\xi^2+p_\eta^2)(\xi^2+\eta^2)^{-1/2}\enspace.
  \tag\NUM.\num$$
\edef\numeb{\NUM.\num}%
Note $\Delta V_{PF}=0$ which is a peculiarity of two dimensions with
metric $\propto\bbbone$. These coordinates have been used in literature
to discuss the two-dimensional ``Coulomb-problem'', c.f.\ [\DKb, \GROq,
\INOa]. The transformation $(x,y)\mapsto(\xi,\eta)$ actually is the
two-dimensional realization of the Kustaanheimo-Stiefel
transformation. The arising path integral in these coordinates must be
treated by a time-transformation due to the metric terms in the
Lagrangian. The transformation in its continuous and lattice
implementation, respectively, has the form
\plus$$
  s(t)=\int_{t'}^t{d\sigma\over\xi^2(\sigma)+\eta^2(\sigma)}\enspace,
  \qquad
  \epsilon=\widehat{(\xi_j^2+\eta_j^2)}\Delta s_j\enspace,
  \tag\NUM.\num$$
and decouples the $\xi$- and $\eta$-path integration giving two harmonic
oscillator path integrals in $\xi$ and $\eta$ with frequency $\omega=
\sqrt{-2E/m}$, respectively. Because $E>0$ in our case, $\omega$ is
purely imaginary and we obtain two repelling oscillators. With the help
of the harmonic oscillator path integral and the relations in
Appendix~2 we obtain the path integral identity
$$\myalign
  &\int\limits_{\xi(t')=\xi'}^{\xi(t'')=\xi''}\CD\xi(t)
  \int\limits_{\eta(t')=\eta'}^{\eta(t'')=\eta''}\CD\eta(t)
  (\xi^2+\eta^2)
  \exp\left[{\i m\over2\hbar}\int_{t'}^{t''}
   (\xi^2+\eta^2)(\dot\xi^2+\dot\eta^2)dt\right]
  \\   &
  =\int_{\bbbr}{dE\over2\pi\hbar}\e^{-\i ET/\hbar}
   \int_0^\infty ds''
   \int\limits_{\xi(0)=\xi'}^{\xi(s'')=\xi''}\CD\xi(s)
  \int\limits_{\eta(0)=\eta'}^{\eta(s'')=\eta''}\CD\eta(t)
  \\   &\qquad\qquad\times
  \exp\left\{\ih\int_0^{s''}\bigg[{m\over2}(\dot\xi^2+\dot\eta^2)
  +E(\xi^2+\eta^2)\bigg]ds\right\}
  \\   &
  =\int_{\bbbr}{dE\over2\pi\hbar}\e^{-\i ET/\hbar}
   \int_0^\infty ds''
  {m\omega\over\pi\i\hbar\sin\omega s''}
  \\   &\qquad\qquad\times
  \exp\bigg[{\i m\omega\over2\hbar\sin\omega s''}
    ({\xi'}^2+{\xi''}^2+{\eta'}^2+{\eta''}^2)\cos\omega s''\bigg]
  \cosh\bigg({m\omega(\xi'\xi''+\eta'\eta'')\over\i\hbar\sin\omega s''}
       \bigg)
  \\   &
  \tag\NUM.\num\\   \global\plus
       &
  =\sum_{e,o}\int_{\bbbr} d\zeta\int_{\bbbr}dp\,
   \e^{-\i\hbar p^2T/2m}\Psi_{p,\zeta}^{(e,o)\,*}(\xi',\eta')
   \Psi_{p,\zeta}^{(e,o)}(\xi'',\eta'')\enspace,
  \tag\NUM.\num\endalign$$
\minus
\edef\numdi{\NUM.\num}\plus
\edef\numdh{\NUM.\num}%
and $\sum_{e,o}$ denotes the summation over even and odd states
respectively; the functions $\Psi_{p,\zeta}^{(e,o)}(\xi,\eta)$ are
given by
\plus$$\myalign
  &\Psi_{p,\zeta}^{(e,o)}(\xi,\eta)
  ={\e^{\pi/2ap}\over\sqrt{2}4\pi^2}
  \\   &\ \times
    \pmatrix
    \big\vert\Gamma(\viert-{\i\zeta\over2p})\big\vert^2
    E^{(0)}_{-\half+\i\zeta/p}(\e^{-\i\pi/4}\sqrt{2p}\,\xi)
    E^{(0)}_{-\half-\i\zeta/p}(\e^{-\i\pi/4}\sqrt{2p}\,\eta)
    \\
    \big\vert\Gamma({3\over4}-{\i\zeta\over2p})\big\vert^2
    E^{(1)}_{-\half+\i\zeta/p}(\e^{-\i\pi/4}\sqrt{2p}\,\xi)
    E^{(1)}_{-\half-\i\zeta/p}(\e^{-\i\pi/4}\sqrt{2p}\,\eta)
    \endpmatrix\enspace,
  \tag\NUM.\num\endalign$$
which are $\delta$-normalized according to [\POGOa]
\plus$$
  \int_0^\infty d\eta\int_{\bbbr}d\xi(\xi^2+\eta^2)
  \Psi_{p',\zeta'}^{(e,o)\,*}(\xi,\eta)
  \Psi_{p,\zeta}^{(e,o)}(\xi,\eta)
  =\delta(p'-p)\delta(\zeta'-\zeta)\enspace.
  \tag\NUM.\num$$
Note that in the evaluation of the path integral one has to take into
account that by using the harmonic oscillator solution for the $\xi$-
and $\eta$-variable, respectively, one actually uses a double covering
of the original $(x,y)\equiv(x_1,x_2)\in\bbbr^2$-plane, i.e.\ $\vec
u\equiv(\xi,\eta)\in\bbbr^2$. Furthermore we have taken into account
that our mapping is of the ``square-root'' type which gives rise to a
sign ambiguity. ``Thus, if one considers all paths in the complex
$z=x+\i y $-plane  from $z'$ to $z''$, they will be mapped into two
different classes of paths in the $\vec u$-plane: Those which go from
$\vec u'$ to $\vec u''$ and those going from $\vec u'$ to $-\vec u''$.
In the cut complex $z$-plane for the function $\vert\vec u\vert=\sqrt{
\vert z\vert}$ these are the paths passing an even or odd number of
times through the square root from $\vert z\vert=0$ and $\vert
z\vert=-\infty$. We may choose the $\vec u'$ corresponding to the
initial $z'$ to lie on the first sheet (i.e.\ in the right half
$\vec u$-plane). The final $\vec u''$ can be in the right as well as
the left half-plane and all paths on the $z$-plane go over into paths
from $\vec u'$ to $\vec u''$ and those from $\vec u'$ to $-\vec u''$
'' [\DKb]. Thus the two contributions arise in (\numdi). The last line
of (\numdh) is then best obtained by considering the Coulomb problem
$-q_1q_2/r$ in $\bbbr^2$, applying (A.2.2), respectively (A.2.4),
performing a momentum variable transformation $(p_\xi,p_\eta)\to\big(
{1\over2p}({1\over a}+\zeta),{1\over2p}({1\over a}-\zeta)\big)$
($a=\hbar^2/mq_1q_2$ is the Bohr radius) with the new variables
$(p,\zeta)$ and finally setting $q_1q_2=0$, i.e.\ $a=\infty$.

\subsection{The Sphere $\szwei$}
\subsubsection{General Form of the Propagator and the Green Function}
The sphere $\szwei$ is the first non-trivial space of constant
positive curvature. There exist two coordinate systems which admit
separation of variables on $\szwei$: polar coordinates and
sphero-conical coordinates. The invariant distance $\cos\psi_\szwei=
\vec s'\cdot\vec s''$ in these coordinates is given by (for the
definition of the coordinates see below)
\plus$$\myalign
       &\hbox{Spherical Coordinates:}
  \\   &
  \cos\psi_\szwei(\vec q'',\vec q')
  =\cos\theta''\cos\theta'+\sin\theta''\sin\theta'\cos(\phi''-\phi')
  \enspace,
  \tag\NUM.\num a
  \\   &\hbox{Sphero-Conical Coordinates:}
  \\   &\phantom{\cos\psi_\szwei}
  =\sn\mu''\sn\mu'\dn\nu''\dn\nu'
  +\cn\mu''\cn\mu'\cn\nu''\cn\nu'
  +\dn\mu''\dn\mu'\sn\nu''\sn\nu'\enspace.\qquad
  \tag\NUM.\num b\endalign$$
The propagator and the Green function on $\szwei$ are best calculated
in terms of polar coordinates. One obtains
$$\myalign
  K^\szwei\Big(\szwei(\vec q'',\vec q');T\Big)
  &=\sum_{l=0}^\infty{2l+1\over4\pi}
    P_l\big(\cos\psi_\szwei(\vec q'',\vec q')\big)
  \exp\bigg[-{\i\hbar T\over2m}l(l+1)\bigg]\enspace,\qquad
  \tag\NUM.\num\\   \global\plus
  G^\szwei\Big(\psi_\szwei(\vec q'',\vec q');E\Big)
  &={m\over2\hbar^2}
  {P_{-\half+\sqrt{2mE/\hbar^2+1/4}}
    \Big(-\cos\psi_\szwei(\vec q'',\vec q')\Big)
  \over\sin\Big[\pi\Big(\half-\sqrt{2mE/\hbar^2+{1\over4}}\,\Big)\Big]}
  \enspace.
  \tag\NUM.\num\endalign$$

\subsubsection{Spherical Coordinates}
We consider the polar coordinates
\plus$$\left.\alignedat 3
   s_0&=\sin\theta\cos\phi\enspace,
    &\qquad
    &0<\theta<\pi\enspace,
   \\
   s_1&=\cos\theta\enspace,
    &\qquad
    &0\leq\phi<2\pi\enspace,
   \\
   s_2&=\sin\theta\sin\phi\enspace.
    &\qquad
    &\qquad
   \endalignedat\qquad\qquad\right\}
   \tag\NUM.\num$$
These are the usual two-dimensional polar coordinates on the sphere.
The metric tensor is $(g_{ab})=\diag(1,\sin^2\theta)$, and the
momentum operators have the form
\plus$$
  p_\theta=\hi\bigg(
    {\partial\over\partial\theta}+\half\cot\theta\bigg)
  \enspace,\qquad
  p_\phi=\hi{\partial\over\partial\phi}\enspace.
  \tag\NUM.\num$$
\edef\numed{\NUM.\num}%
For the Hamiltonian we obtain
\plus$$\myalign
  -\hbarm\Delta_\szwei
  &=-\hbarm\bigg({\partial^2\over\partial\theta^2}
                 +\cot\theta{\partial\over\partial\theta}
    +{1\over\sin^2\theta}{\partial^2\over\partial\phi^2}\bigg)
  \\   &
  ={1\over2m}\bigg(p_\theta^2+{1\over\sin^2\theta}p_\phi^2\bigg)
    -{\hbar^2\over8m}\bigg(1+{1\over\sin^2\theta}\bigg)
  \enspace.
  \tag\NUM.\num\endalign$$
The corresponding path integral is well-known [\BJb, \GRSb, \PI] and we
have the identity
$$\myalign
  &\int\limits_{\theta(t')=\theta'}^{\theta(t'')=\theta''}
  \sin\theta\CD\theta(t)
  \int\limits_{\phi(t')=\phi'}^{\phi(t'')=\phi''}\CD\phi(t)
  \\   &\qquad\qquad\times
  \exp\left\{\ih\int_{t'}^{t''}\bigg[{m\over2}\Big(
      \dot\theta^2+r^2\sin^2\theta\dot\phi^2\Big)
  +{\hbar^2\over8m}\bigg(1+{1\over\sin^2\theta}\bigg)\bigg]dt\right\}
  \qquad
  \\   &
  ={1\over4\pi}\sum_{l=0}^\infty(2l+1)
    P_l\big(\cos\psi_\szwei(\vec q'',\vec q')\big)
    \e^{-\i\hbar Tl(l+1)/2m}
  \tag\NUM.\num\\   \global\plus
       &
  =\sum_{l=0}^\infty\sum_{m=-l}^l
   Y_l^{m\,*}(\theta',\phi')Y_l^m(\theta'',\phi'')
    \e^{-\i\hbar Tl(l+1)/2m}\enspace.
   \tag\NUM.\num\endalign$$
The $Y_l^m(\theta,phi)$ are the usual spherical harmonics on the
$\szwei$-sphere.

\subsubsection{Sphero-Conical Coordinates}
We consider the general elliptic coordinate system
\plus$$\gathered
   s_0^2+s_1^2+s_2^2=1\enspace,
  \\
  {s_0^2\over\rho_i-a}+{s_1^2\over\rho_i-b}
  +{s_2^2\over\rho_i-c}=0\enspace,
  \endgathered
  \tag\NUM.\num$$
where $c\leq\rho_2\leq b\leq\rho_1\leq a$. The
connection to the variables $\vec s$ on the sphere $\szwei$ is given by
\plus$$
  s_0^2={(\rho_1-a)(\rho_2-a)\over(a-c)(a-b)},\
  s_1^2={(\rho_1-b)(\rho_2-b)\over(b-a)(b-c)},\
  s_2^2={(\rho_1-c)(\rho_2-c)\over(c-a)(c-b)}\enspace.
  \tag\NUM.\num$$
One now can make the identification
$\rho_1=k^2\cn^2(\mu,k)$, $\rho_2=-{k'}^2\cn^2(\nu,k')$,
$b=-{k'}^2$, $a=k^2$ and $c=0$. This yields
\plus$$\left.\alignedat 3
  s_0&=\sn(\mu,k)\dn(\nu,k')\enspace,
     &\quad      &k^2+{k'}^2=1\enspace,
  \\
  s_1&=\dn(\mu,k)\sn(\nu,k')\enspace,
     &\quad      &\quad
  \\
  s_2&=\cn(\mu,k)\cn(\nu,k')\enspace.
     &\quad      &\quad
  \endalignedat\qquad\qquad\right\}
  \tag\NUM.\num$$
The functions $\sn(\mu,k)$, $\cn(\mu,k)$ and $\dn(\mu,k)$
denote Jacobi elliptic functions, where e.g.\ $\cn(\mu,k)$ has the
periods $4\i K,\,2K+2\i K',\,4K$, respectively ($K,K'$ the elliptic
integrals) [\ABS]. Note the relations $\cn^2\mu+\sn^2\mu=1$ and
$\dn^2\mu=1-k^2\sn^2\mu$. The metric tensor $g_{ab}$ in these
coordinates is therefore given by $(g_{ab})=(k^2\cn^2\mu+{k'}^2\cn^2\nu)
\bbbone_2$. The momentum operators are
\plus$$
  p_\mu=\hi\bigg({\partial\over\partial\mu}
         -{k^2\sn\mu\cn\mu\dn\mu\over
  k^2\cn^2\mu+{k'}^2\cn^2\nu}\bigg)\enspace,\quad
  p_\nu=\hi\bigg({\partial\over\partial\nu}
         -{{k'}^2\sn\nu\cn\nu\dn\nu\over
  k^2\cn^2\mu+{k'}^2\cn^2\nu}\bigg)\enspace,
  \tag\NUM.\num$$
\edef\numea{\NUM.\num}%
and the Hamiltonian has the form
\plus$$\myalign
  -\hbarm\Delta_\szwei
  &=-\hbarm{1\over k^2\cn^2\mu+{k'}^2\cn^2\nu}\bigg(
     {\partial^2\over\partial\mu^2}+{\partial^2\over\partial\nu^2}\bigg)
  \\   &
  ={1\over2m}{1\over\sqrt{k^2\cn^2\mu+{k'}^2\cn^2\nu}}
   (p_\mu^2+p_\nu^2){1\over\sqrt{k^2\cn^2\mu+{k'}^2\cn^2\nu}}\enspace.
  \tag\NUM.\num\endalign$$
The path integral can therefore be written down yielding [\GROm]
$$\myalign
  &\int\limits_{\mu(t')=\mu'}^{\mu(t'')=\mu''}
  \CD\mu(t)
  \int\limits_{\nu(t')=\nu'}^{\nu(t'')=\nu''}\CD\nu(t)
  (k^2\cn^2\mu+{k'}^2\cn^2\nu)
  \\   &\qquad\qquad\qquad\qquad\times
  \exp\Bigg[{\i m\over2\hbar}\int^{t''}_{t'}(k^2\cn^2\mu+{k'}^2\cn^2\nu)
  (\dot\mu^2+\dot\nu^2)dt\Bigg]
  \\   &
  =\sum_\kappa\sum_{l=0}^\infty
   A_{l,\kappa}^*(\mu')B_{l,\kappa}^*(\nu')
   A_{l,\kappa}(\mu'')B_{l,\kappa}(\nu'')
   \e^{-\i\hbar Tl(l+1)/2m}
  \tag\NUM.\num\\   \global\plus
       &
  ={1\over4\pi}\sum_{l=0}^\infty(2l+1)
    P_l\big(\cos\psi_\szwei(\vec q'',\vec q')\big)
    \e^{-\i\hbar Tl(l+1)/2m}\enspace.
  \tag\NUM.\num\endalign$$
The eigen-functions $\Psi_\lambda(\mu,\nu)=A_{l,\kappa}(\mu)
B_{l,\kappa}(\nu)$, are Lam\'e polynomials so that the functions
$A_{l,\kappa}$ and $B_{l,\kappa}$ are solutions of the Lam\'e
differential equations [\EMOTa]
\plus$$\left.\aligned
  {\d^2A_{l,\kappa}(\mu)\over\d\mu^2}&+
   \bigg[-{\kappa\over4}+l(l+1)-l(l+1)k^2\sn^2\mu\bigg]
                                   A_{l,\kappa }(\mu)=0\enspace,  \\
  {\d^2B_{l,\kappa}(\nu)\over\d\nu^2}&+
   \bigg[{\kappa\over4}-l(l+1){k'}^2\sn^2\nu\bigg]
                                   B_{l,\kappa}(\nu)=0\enspace.
  \endaligned\qquad\qquad\right\}
  \tag\NUM.\num$$
These functions are highly transcendental ones and cannot be expressed
in terms of standard power series expansions like (confluent)
hypergeometric functions.

\eject\noindent
\subsection{The Pseudosphere $\lzwei$}
\subsubsection{General Form of the Propagator and the Green Function}
The evaluation of the Green function on the Pseudosphere $\lzwei$ was
subject of a couple of papers. It was first evaluated by means of path
integration in [\GRSa], followed by [\GROc, \GRSc] (compare also
[\GUTc, \KUB]). The propagator and the Green function have the form
$$\myalign
  K^\lzwei\Big(d_\lzwei(\vec q'',\vec q');T\Big)
  &={1\over2\pi}\int_0^\infty pdp\,\tanh\pi p
  \\   \times
  \CP_{\i p-\half}&\big(\cosh d_\lzwei(\vec q'',\vec q')\big)
  \Energylzwei\enspace,
  \tag\NUM.\num\\   \global\plus
  G^\lzwei\Big(d_\lzwei(\vec q'',\vec q');E\Big)
  &={m\over\pi\hbar^2}
   \CQ_{-1/2-\i{\sqrt{2mE/\hbar^2-\viert}}}
    \big(\cosh d_\lzwei(\vec q'',\vec q')\big)\enspace.\qquad
  \tag\NUM.\num\endalign$$
The invariant hyperbolic distance in $\lzwei$ is given by
(for the definition of the coordinates see below)
\plus$$\myalign
       &\hbox{Pseudo-Polar Coordinates:}
  \\   &
  \cosh d_\lzwei(\vec q'',\vec q')
  =\cosh\tau''\cosh\tau'-\sinh\tau''\sinh\tau'\cos(\phi''-\phi')
  \enspace,
  \tag\NUM.\num a\\
       &\hbox{Horicyclic Coordinates:}
  \\   &\phantom{\cosh d_\lzwei}
   =\dfrac{(x''-x')^2+{y''}^2+{y'}^2}{2y'y''}
  \enspace,
  \tag\NUM.\num b\\
       &\hbox{Equidistant Coordinates:}
  \\   &\phantom{\cosh d_\lzwei}
   =\cosh(\tau_2''-\tau_2')\cosh\tau_1''\cosh\tau_1'
    -\sinh\tau_1''\sinh\tau_1'\enspace,
  \tag\NUM.\num c\\
       &\hbox{Pseudo-Ellipsoidal Coordinates:}
  \\   &\phantom{\cosh d_\lzwei}
  =\nc\mu''\nc\mu'\nc\nu''\nc\nu'
  -\dc\mu''\dc\mu'\sc\nu''\sc\nu'
  -\sc\mu''\sc\mu'\dc\nu''\dc\nu'\enspace.\qquad
  \tag\NUM.\num d\endalign$$
Because the discussion of the path integral solutions on the
pseudosphere $\lzwei$ has been done in some extend in Refs.~[\GROc,
\GRSa], we just cite the results, with the exception of the case of the
ellipsoidal coordinates. In the nomenclature we follow [\KAL].

\subsubsection{Horicyclic Coordinates (Poincar\'e Upper Half-Plane)}
We consider the coordinate system
\plus$$\left.\alignedat 3
   u_0&={1\over2y}(y^2+1-x)\enspace,\quad
   u_1={x\over y}\enspace,
      &\qquad     &y>0\enspace,
   \\
   u_2&={1\over2y}(y^2-1-x)\enspace,
      &\qquad
      &x\in\bbbr\enspace,
   \endalignedat\qquad\qquad\right\}
   \tag\NUM.\num$$
The metric has the form $(g_{ab})=\bbbone_2/y^2$, and the momentum
operators are
\plus$$
  p_x=\hi{\partial\over\partial x}\enspace,\qquad
  p_y=\hi\bigg({\partial\over\partial y}-{1\over y}\bigg)
  \enspace.
  \tag\NUM.\num$$
\edef\numee{\NUM.\num}%
\baselineskip=11.0pt\noindent
For the Hamiltonian we have
\plus$$
  -\hbarm\Delta_\lzwei
  =\hbarm y^2\bigg({\partial^2\over\partial x^2}
                  +{\partial^2\over\partial y^2}\bigg)
  ={1\over2m}(y^2p_x^2+yp_y^2y)\enspace.
  \tag\NUM.\num$$
We therefore obtain the path integral identity
[\GROc, \GRSa, \GUTc]
\plus$$\myalign
  &\int\limits_{x(t')=x'}^{x(t'')=x''}\CD x(t)
  \int\limits_{y(t')=y'}^{y(t'')=y''}{\CD y(t)\over y^2}
  \exp\left({\i m\over2\hbar}\int_{t'}^{t''}
  {\dot x^2+\dot y^2\over y^2}dt\right)
         \\   &
  ={\sqrt{y'y''}\over\pi^3}\int_{\bbbr} dk\int_0^\infty
  pdp\,\sinh\pi p\energylzwei
  \e^{\i k(x''-x')}K_{\i p}(\vert k\vert y')\,K_{\i p}(\vert k\vert y'')
  \enspace.
  \tag\NUM.\num\endalign$$

\subsubsection{Equidistant Coordinates (Hyperbolic Strip)}
We consider the coordinate system
\plus$$\alignedat 3
   u_0&=\cosh\tau_1\cosh\tau_2\enspace,\qquad
   u_1=\sinh\tau_1\enspace,\qquad
      &\qquad
      &\tau_1,\tau_2\in\bbbr\enspace,
   \\
   u_2&=\cosh\tau_1\sinh\tau_2\enspace.
      &\qquad
      &\qquad
   \endalignedat
   \tag\NUM.\num$$
The metric has the form $(g_{ab})=\diag(1,\cosh^2\tau_1)$, and the
momentum operators are
\plus$$
  p_{\tau_1}=\hi\bigg({\partial\over\partial\tau_1}
         +\half\tanh\tau_1\bigg)\enspace,\qquad
  p_{\tau_2}=\hi{\partial\over\partial\tau_2}\enspace.
  \tag\NUM.\num$$
\edef\numde{\NUM.\num}%
For the Hamiltonian we have
\plus$$\myalign
  -\hbarm\Delta_\lzwei
  &=\hbarm\bigg({\partial^2\over\partial\tau_1}
         +\tanh\tau_1{\partial\over\partial\tau_1}
  +{1\over\cosh^2\tau_1}{\partial^2\over\partial\tau_2}\bigg)
  \\   &
  ={1\over2m}\bigg(p_{\tau_1}^2+{1\over\cosh^2\tau_1}p_{\tau_2}^2\bigg)
  +{\hbar^2\over8m}\bigg(1+{1\over\cosh^2\tau_1}\bigg)\enspace.
  \tag\NUM.\num\endalign$$
Therefore we obtain the path integral identity [\GROc]
\plus$$\myalign
       &
  \int\limits_{\tau_1(t')=\tau_1'}^{\tau_1(t'')=\tau_1''}
  \cosh\tau_1\CD\tau_1(t)
  \int\limits_{\tau_2(t')=\tau_2'}^{\tau_2(t'')=\tau_2''}\CD\tau_2(t)
  \\   &\qquad\times
  \exp\Bigg\{\ih\int_{t'}^{t''}\bigg[{m\over2}
     \Big(\dot\tau_1^2+\cosh^2\tau_1\dot\tau_2^2\Big)
    -{\hbar^2\over8m}\bigg(1+{1\over\cosh^2\tau_1}\bigg)\bigg]dt\Bigg\}
  \\   &
  =\Big(\cosh\tau_1'\cosh\tau_1''\Big)^{-1/2}
  \e^{-\i\hbar T/8m}
  \int_{\bbbr}{dk\over2\pi}\e^{\i k(\tau_2''-\tau_2')}
  \\   &\qquad\times
  \int\limits_{\tau_1(t')=\tau_1'}^{\tau_1(t'')=\tau_1''}\CD\tau_1(t)
  \exp\Bigg[\ih\int_{t'}^{t''}\bigg({m\over2}\dot\tau_1^2
   -\hbarm{k^2+\viert\over\cosh^2\tau_1}\bigg)dt\Bigg]
  \\   &
  =\Big(\cosh\tau_1'\cosh\tau_1''\Big)^{-1/2}
  \int_{\bbbr}{dk\over2\pi}\e^{\i k(\tau_2''-\tau_2')}
  \\   &\qquad\times
  \half\int_{\bbbr}{p\sinh\pi p\,dp\over
  \cosh^2\pi k+\sinh^2\pi p}\energylzwei
  P_{\i k-\half}^{-\i p}(\tanh\tau_1')
  P_{\i k-\half}^{\i p}(\tanh\tau_1'')\enspace.
  \tag\NUM.\num\endalign$$
In the path integral evaluation one successively uses the path integral
solution of the special case of the modified P\"oschl-Teller potential
(the simple Rosen-Morse potential [\GROe]).

\subsubsection{Pseudospherical Coordinates (Upper Sheet of the
Two-Sheeted Hyperboloid, Poincar\'e Disc)}
We consider the coordinate system
\plus$$\left.\alignedat 3
   u_0&=\cosh\tau\enspace,
      &\qquad
      &\tau>0\enspace,
   \\
   u_1&=\sinh\tau\sin\phi\enspace,
      &\qquad
      &0\leq\phi<2\pi\enspace,
   \\
   u_2&=\sinh\tau\cos\phi\enspace.
      &\qquad
      &\qquad
   \endalignedat\qquad\qquad\right\}
   \tag\NUM.\num$$
These are the well-known two-dimensional (pseudo-) spherical
coordinates on $\lzwei$. The metric has the form
$(g_{ab})=\diag(1,\sinh^2\tau_1)$, and the momentum operators are
\plus$$
  p_\tau=\hi\bigg({\partial\over\partial\tau}
         +\half\coth\tau\bigg)\enspace,\qquad
  p_\phi=\hi{\partial\over\partial\phi}\enspace.
  \tag\NUM.\num$$
\edef\numdd{\NUM.\num}%
For the Hamiltonian we have
\plus$$\myalign
  -\hbarm\Delta_\lzwei
  &=\hbarm\bigg({\partial^2\over\partial\tau}
         +\coth\tau{\partial\over\partial\tau}
    +{1\over\cosh^2\tau}{\partial^2\over\partial\phi}\bigg)
  \\   &
  ={1\over2m}\bigg(p_\tau^2+{1\over\sinh^2\tau}p_\phi^2\bigg)
  +{\hbar^2\over8m}\bigg(1-{1\over\sinh^2\tau_1}\bigg)\enspace.
  \tag\NUM.\num\endalign$$
Therefore we obtain the path integral identity [\GROc, \GRSc]
$$\myalign
 &\int\limits_{\tau(t')=\tau'}^{\tau(t'')=\tau''}\sinh\tau\CD\tau(t)
  \int\limits_{\phi(t')=\phi'}^{\phi(t'')=\phi''}\CD\phi(t)
         \\   &\qquad\qquad\qquad\times
  \exp\left\{\ih\int_{t'}^{t''}
   \left[{m\over2}(\dot\tau^2+\sinh^2\tau\dot\phi^2)
    -{\hbar^2\over8m}\bigg(1-{1\over\sinh^2\tau}\bigg)
  \right]dt\right\}
         \\   &
  ={1\over2\pi^2}\int_0^\infty dp\sum_{l=-\infty}^\infty
  p\sinh\pi p\Energylzwei
  \bigg\vert \Gamma\bigg(\half+\i p+l\bigg)\bigg\vert ^2
         \\   &\qquad\qquad\qquad\times
  \e^{\i l(\phi''-\phi')}
  \CP^{-l}_{\i p-\half}(\cosh \tau')
  \CP^{-l}_{\i p-\half}(\cosh\tau'')
  \tag\NUM.\num\\   \global\plus
  &\hbox{(Pseudospherical polar coordinate system on the disc,
        $0\leq r<1,\,\psi\in[0,2\pi)$:)}
         \\   &
  =\ih \int_0^\infty  dT\,\e^{\i TE/\hbar}
   \int\limits_{r(t')=r'}^{r(t'')=r''}{4r\CD r(t)\over(1-r^2)^2}
  \int\limits_{\psi(t')=\psi'}^{\psi(t'')=\psi''}\CD\psi(t)
         \\   &\qquad\qquad\qquad\times
  \exp\left\{\ih\int_{t'}^{t''}\bigg[2m
  {\dot r^2+r^2\dot\psi^2\over (1-r^2)^2}
   +\hbar^2{(1-r^2)^2\over32mr^2}\bigg]dt\right\}
         \\   &
  ={1\over2\pi^2}\int_0^\infty dp\sum_{l=-\infty}^\infty
  p\sinh\pi p\Energylzwei
  \bigg\vert \Gamma\bigg(\half+\i p+l\bigg)\bigg\vert ^2
         \\   &\qquad\qquad\qquad\times
  \vphantom{\bigg)}
  \e^{\i l(\psi''-\psi')}
  \CP^{-l}_{\i p-\half}\left({1+{r'}^2\over1-{r'}^2}\right)
  \CP^{-l}_{\i p-\half}\left({1+{r''}^2\over1-{r''}^2}\right)
  \tag\NUM.\num\endalign$$
Here we have also stated the (equivalent) representation in terms
of the Poincar\'e disc via $z=x_1+\i x_2=r\e^{\i\psi}=\tanh{\tau\over2}
(\sin\phi+\i\cos\phi)$ which is important is various discussions
in the theory of quantum chaos [\AMSS--\BV, \GUTd].

\subsubsection{General Pseudo-Ellipsoidal Coordinates}
We consider the general elliptic coordinate system
\plus$$\gathered
   u_0^2-u_1^2-u_2^2=1\enspace,
  \\
  {u_0^2\over\rho_i-a}-{u_1^2\over\rho_i-b}
  -{u_2^2\over\rho_i-c}=0\enspace,\qquad(i=1,2)\enspace.
  \endgathered
  \tag\NUM.\num$$
Explicitly in terms of the variables $\vec u$ on the pseudosphere
$\lzwei$ they are given by
\plus$$
  u_0^2={(\rho_1-c)(\rho_2-c)\over(a-c)(b-c)}\enspace,\quad
  u_1^2=-{(\rho_1-b)(\rho_2-b)\over(a-b)(c-b)}\enspace,\quad
  u_2^2=-{(\rho_1-a)(\rho_2-a)\over(b-a)(c-a)}\enspace.
  \tag\NUM.\num$$
\edef\numdaa{\NUM.\num}%
One now can make the identification $\rho_1=-{k'}^2\nc^2(\mu,k)$,
$\rho_2=k^2\nc^2(\nu,k')$, $a=-{k'}^2$, $b=k^2$ and $c=0$. This yields
\plus$$\left.\alignedat 3
  u_0&=\nc(\mu,k)\nc(\nu,k')\enspace,
     &\quad      &k^2+{k'}^2=1\enspace,
  \\
  u_1&=\dc(\mu,k)\sc(\nu,k')\enspace,
     &\quad      &\quad
  \\
  u_2&=\sc(\mu,k)\dc(\nu,k')\enspace,
     &\quad      &\quad
  \endalignedat\qquad\qquad\right\}
  \tag\NUM.\num$$
\edef\numda{\NUM.\num}%
together with $0<k^2<k^2\nc^2\alpha<-{k'}^2<-{k'}^2\nc^2\alpha$, say.
The functions $\sc(\mu,k)$, $\dc(\mu,k)$ and $\nc(\mu,k)$ denote Jacobi
elliptic functions which correspond to the functions $\sn,\, \dn$ and
$\cn$ by taking the argument imaginary, where e.g.\ $\nc(\mu,k)$ has the
periods $4\i K,\,2K+2\i K',\,4K$ ($K,K'$ the elliptic integrals) [\ABS].
Note the relations $\nc^2\mu+\sc^2\mu=1$ and $\dc^2\mu=1+k^2\sc^2\mu$.
The metric tensor $g_{ab}$ in these coordinates is given by
\plus$$
  (g_{ab})=({k'}^2\nc^2\mu+k^2\nc^2\nu)\bbbone_2\enspace,
  \tag\NUM.\num$$
and the momentum operators are
\plus$$
  p_\mu=\hi\bigg({\partial\over\partial\mu}
         -{k'}^2{\sc\mu\nc\mu\dc\mu\over
  {k'}^2\nc^2\mu+k^2\nc^2\nu}\bigg)\enspace,\quad
  p_\nu=\hi\bigg({\partial\over\partial\nu}
         -k^2{\sc\nu\nc\nu\dc\nu\over
  {k'}^2\nc^2\mu+k^2\nc^2\nu}\bigg)\enspace.
  \tag\NUM.\num$$
The Hamiltonian has the form
\plus$$\myalign
  -\hbarm\Delta_\lzwei
  &=-\hbarm{1\over {k'}^2\nc^2\mu+k^2\nc^2\nu}
    \bigg({\partial^2\over\partial\mu^2}
         +{\partial^2\over\partial\nu^2}\bigg)
  \\   &
  ={1\over2m}{1\over\sqrt{{k'}^2\nc^2\mu+k^2\nc^2\nu}}
   (p_\mu^2+p_\nu^2){1\over\sqrt{{k'}^2\nc^2\mu+k^2\nc^2\nu}}
  \enspace.
  \tag\NUM.\num\endalign$$
The path integral can therefore be formulated yielding
$$\myalign
  &\ih\int_0^\infty dE\,\e^{\i ET/\hbar}
   \int\limits_{\mu(t')=\mu'}^{\mu(t'')=\mu''}\CD\mu(t)
  \int\limits_{\nu(t')=\nu'}^{\nu(t'')=\nu''}\CD\nu(t)
  ({k'}^2\nc^2\mu+k^2\nc^2\nu)
  \\   &\qquad\qquad\times
  \exp\Bigg[{\i m\over2\hbar}\int^{t''}_{t'}
  ({k'}^2\nc^2\mu+k^2\nc^2\nu)(\dot\mu^2+\dot\nu^2)dt\Bigg]
  \\   &
  =\int_0^\infty ds''
  \int\limits_{\mu(0)=\mu'}^{\mu(s'')=\mu''}\CD\mu(s)
  \int\limits_{\nu(0)=\nu'}^{\nu(s'')=\nu''}\CD\nu(s)
  \\   &\qquad\qquad\times
  \exp\Bigg\{\ih\int_0^{s''}\bigg[{m\over2}
  (\dot\mu^2+\dot\nu^2)+E({k'}^2\nc^2\mu+k^2\nc^2\nu)\bigg]ds\Bigg\}
  \\   &
  =\sum_\kappa\int_0^\infty{dp\over\hbar^2(p^2+\viert)/2m-E}
   \CA_{p,\kappa}^*(\mu')\CB_{p,\kappa}^*(\nu')
   \CA_{p,\kappa}(\mu'')\CB_{p,\kappa}(\nu'')
  \tag\NUM.\num\\   \global\plus
       &
   ={m\over\pi\hbar^2}
   \CQ_{-1/2-\i{\sqrt{2mE/\hbar^2-\viert}}}
    \big(\cosh d_\lzwei(\vec q'',\vec q')\big)\enspace,
  \tag\NUM.\num\endalign$$
where $d_\lzwei(\vec q'',\vec q')$ must be expressed in ellipsoidal
coordinates. The eigen-functions $\Psi_{p,\kappa}(\mu,\nu):=
\CA_{p,\kappa}(\mu)\CB_{p,\kappa}(\nu)$, are generalized hyperbolic
Lam\'e polynomials in $\mu$ and $\nu$ and are the normalized solution
of the Hamiltonian $-\hbarm\Delta_\lzwei$:
\plus$$\left.\aligned
  {\d^2\CA_{p,\kappa}(\mu)\over\d\mu^2}&+
   \bigg[-{\kappa\over4}+(p^2+\bviert){k'}^2\nc^2\mu\bigg]
   \CA_{l,\kappa }(\mu)=0\enspace,  \\
  {\d^2\CB_{p,\kappa}(\nu)\over\d\nu^2}&+
   \bigg[{\kappa\over4}+(p^2+\bviert)k^2\nc^2\nu\bigg]
   \CB_{l,\kappa}(\nu)=0\enspace.
  \endaligned\qquad\qquad\right\}
  \tag\NUM.\num$$
Depending on the domain, the values $\mu$ and $\nu$ can take on,
one can discriminate six different coordinate system which are encoded
in (\numdaa, \numda). For this purpose one writes the kinetic
energy-term in terms of $\rho_{1,2}$ [\KAL, \OLE]
\plus$$
   \dot u_0^2-\dot u_1^2-\dot u_2^2
   ={1\over4}(\rho_1-\rho_2)
   \bigg[{\dot\rho_1^2\over P_3(\rho_1)}
        -{\dot\rho_2^2\over P_3(\rho_2)}\bigg]\enspace,
  \tag\NUM.\num$$
where the polynomial $P_3(\rho)$ is defined by $P_3(\rho)=(\rho-a)(\rho-
b)(\rho-c)$. Therefore we obtain the alternative path integral
formulation
$$\myalign
  &\ih\int_0^\infty\e^{\i ET/\hbar}
  \int\limits_{\rho_1(t')=\rho_1'}^{\rho_1(t'')=\rho_1''}\CD\rho_1(t)
  \int\limits_{\rho_2(t')=\rho_2'}^{\rho_2(t'')=\rho_2''}\CD\rho_2(t)
  {(-\i)(\rho_1-\rho_2)\over4\sqrt{P_3(\rho_1)P_3(\rho_2)}}
  \\   &\qquad\times
  \exp\Bigg\{\ih\int^{t''}_{t'}\bigg[{m\over8}
   (\rho_1-\rho_2)
   \bigg({\dot\rho_1^2\over P_3(\rho_1)}
        -{\dot\rho_2^2\over P_3(\rho_2)}\bigg)-\Delta V_{PF}\bigg]
   dt\Bigg\}
  \\   &
  =\int_0^\infty ds''
  \int\limits_{\rho_1(0)=\rho_1'}^{\rho_1(s'')=\rho_1''}
  \sqrt{-P_3(\rho_1)}\,\CD\rho_1(s)
  \int\limits_{\rho_2(0)=\rho_2'}^{\rho_2(s'')=\rho_2''}
  \sqrt{P_3(\rho_2)}\,\CD\rho_2(s)
  \\   &\qquad\times
  \exp\left\{\ih\int_{0}^{s''}\left[{m\over2}
  \Big(P_3(\rho_2)\dot\rho_1^2-P_3(\rho_1)\dot\rho_2^2\Big)+ES
    -{\hbar^2\over8m}\sum_{i=1}^DM_i\Big(\Gamma_i^2+2\Gamma_i'\Big)
    \right]ds\right\}
  \\   &
  =\sum_\kappa\int_0^\infty{dp\over\hbar^2(p^2+\viert)/2m-E}
   \CA_{p,\kappa}^*(\rho_1')\CB_{p,\kappa}^*(\rho_2')
   \CA_{p,\kappa}(\rho_1'')\CB_{p,\kappa}(\rho_2'')
  \tag\NUM.\num\\   \global\plus
       &
   ={m\over\pi\hbar^2}
   \CQ_{-1/2-\i{\sqrt{2mE/\hbar^2-\viert}}}
    \big(\cosh d_\lzwei(\vec q'',\vec q')\big)\enspace,
  \tag\NUM.\num\endalign$$
with the functions $\CA_{p,\kappa}$ and $\CB_{p,\kappa}$ rewritten
in $\rho_{1,2}$ and $\cosh d_\lzwei$ expressed in $\rho_{1,2}$.
Furthermore denote $M_1=1/P_3(\rho_2), M_2=P_3(\rho_1)$, $S
=(\rho_1-\rho_2)/4P_3(\rho_1)P_3(\rho_2)$ and $\Gamma_i=P_3'(\rho_i)/
2P_3(\rho_i)$ ($i=1,2$). The six different coordinate systems are now
defined by by the roots of the polynomial $P_3(\rho)$ (simple real,
complex, double or triple real) with the requirement of the positive
definiteness of the two quantities
\plus$$
  {\rho_1-\rho_2\over P_3(\rho_1)}>0\enspace,\qquad
  -{\rho_1-\rho_2\over P_3(\rho_2)}>0\enspace.
  \tag\NUM.\num$$
They are given by [\KAL, \OLE]

\medskip
\medskip\noindent{\sl \the\secno.\the\subno.\the\subsubno.1.}
     {\sl Elliptic Coordinates.}
\plus$$\gathered
  c<b<\rho_2<a<\rho_1\enspace,
  \\
  {u_1^2\over\rho_i-b}+{u_2^2\over\rho_i-a}-{u_0^2\over\rho_i-c}=0
  \enspace,\qquad(i=1,2)\enspace.
  \endgathered
  \tag\NUM.\num$$

\medskip\noindent{\sl \the\secno.\the\subno.\the\subsubno.2.}
     {\sl Hyperbolic Coordinates.}
\plus$$\gathered
  \rho_2<c<b<a<\rho_1\enspace,
  \\
  {u_1^2\over\rho_i-c}+{u_2^2\over\rho_i-a}-{u_0^2\over\rho_i-b}=0
  \enspace,\qquad(i=1,2)\enspace.
  \endgathered
  \tag\NUM.\num$$

\medskip\noindent{\sl \the\secno.\the\subno.\the\subsubno.3.}
     {\sl Semi-Hyperbolic Coordinates.}
\plus$$\gathered
  \rho_2<a<\rho_1\enspace,\qquad b=\gamma+\i\delta
                  \enspace,\qquad c=\gamma-\i\delta
  \\
  {u_2^2\over\rho_i-a}-
   {2\delta u_1u_0+(\rho_i-\gamma)(u_0^2-u_1^2)\over
     [(\rho_i-\gamma)^2+\delta^2]}=0
  \enspace,\qquad(i=1,2)\enspace.
  \endgathered
  \tag\NUM.\num$$

\medskip\noindent{\sl \the\secno.\the\subno.\the\subsubno.4.}
     {\sl Elliptic-Parabolic Coordinates.}
\plus$$\gathered
  c=b<\rho_2<a<\rho_1\enspace,
  \\
  -{u_2^2\over\rho_i-a}+{u_0^2-u_1^2\over\rho_i-b}
   -{(u_0-u_1)^2\over(\rho_i-b)^2}=0
  \enspace,\qquad(i=1,2)\enspace.
  \endgathered
  \tag\NUM.\num$$

\medskip\noindent{\sl \the\secno.\the\subno.\the\subsubno.5.}
     {\sl Hyperbolic-Parabolic Coordinates.}
\plus$$\gathered
  \rho_2<c=b<a<\rho_1\enspace,
  \\
  -{u_2^2\over\rho_i-a}+{u_0^2-u_1^2\over\rho_i-b}
   -{(u_0-u_1)^2\over(\rho_i-b)^2}=0
  \enspace,\qquad(i=1,2)\enspace.
  \endgathered
  \tag\NUM.\num$$

\medskip\noindent{\sl \the\secno.\the\subno.\the\subsubno.6.}
     {\sl Semi-Circular-Parabolic Coordinates.}
\plus$$\gathered
  \rho_2<c=b=a<\rho_1\enspace,
  \\
  \bigg({u_0-u_1\over\rho_i-a}+u_2\bigg)^2=u_0^2-u_1^2
  \enspace,\qquad(i=1,2)\enspace.
  \endgathered
  \tag\NUM.\num$$
This concludes the discussion of the separable coordinate systems on
two-dimensional spaces of constant curvature.


\eject
\PLUS\glno=0                      
\section{Separation of Variables in Three Dimensions}
In this Section we give the path integral representations in the
three-dimensional spaces of constant curvature $\edrei$, $\sdrei$ and
$\ldrei$. The notation and presentation will be as in te previous
Section, with the only difference that the stating of the invariant
distances will be given separately in each case.

Some new path integral evaluations will be presented, however, our
emphasize again is more concerned with the connection of the various
path integral identities with each other. The wave-function expansions
then represent again the various possible group matrix element
representations yielding identities connecting the coordinate
realizations. We also note the general form a potential must have to be
separable in a coordinate system. The two-dimensional analogue then
follows from omitting the $z$-coordinate.

\subsection{The Flat Space $\edrei$}
\ssf
\subsubsection{General Form of the Propagator and the Green Function}
Similarly as in the two-dimensional case, the propagator and the Green
function on $\edrei$ are well-known, and are given by
$$\myalign
  K^\edrei\Big(d_\edrei(\vec q'',\vec q');\Big)
  &=\bigg({m\over2\pi\i\hbar T}\bigg)^{3/2}
    \exp\bigg[-{m\over2\i\hbar T}d^2_\edrei(\vec q'',\vec q')\bigg]
  \enspace,
  \tag\NUM.\num\\   \global\plus
  G^\edrei\Big(d_\edrei(\vec q'',\vec q');E\Big)
  &={m\over4\pi\hbar^2 d_\edrei(\vec q'',\vec q')}
    \exp\bigg(-{d_\edrei(\vec q'',\vec q')\over\hbar}\sqrt{-2mE}\,\bigg)
  \enspace.
  \tag\NUM.\num\endalign$$
$\vec q$ denotes any of the eleven coordinate systems which allows a
separation of variables of the Laplacian in three dimensions. Due to the
rather many coordinate systems, the explicit expression for
$d_\edrei(\vec q'',\vec q')$ will be given together with the definition
of the coordinates. In the nomenclature we follow [\KAL, \MF].

\subsubsection{Cartesian Coordinates}
Again, we start with the simplest case, cartesian coordinates
$(x,y,z)=\vec x\in\bbbr^3$. Then $(g_{ab})=\bbbone_3$,
\plus$$
  d_\edrei^2(\vec q'',\vec q')=\vert\vec x''-\vec x'\vert^2\enspace,
  \tag\NUM.\num$$
and $\vec p=-\i\hbar\nabla$. This gives for the Hamiltonian
\plus$$
  -\hbarm\Delta_\edrei
  =-\hbarm\nabla^2={1\over2m}\vec p^2\enspace,
  \tag\NUM.\num$$
and for the path integral we have [\FEY, \FH]
$$\myalign
  \int\limits_{\vec x(t')=\vec x'}^{\vec x(t'')=\vec x''}\CD\vec x(t)
  \exp\left({\i m\over2\hbar}\int_{t'}^{t''}{\dot{\vec x}}^2dt\right)
       &
  =\bigg({m\over2\pi\i\hbar T}\bigg)^{3/2}\exp\bigg({\i m\over2\hbar T}
         \vert\vec x''-\vec x'\vert^2\bigg)
  \tag\NUM.\num\\   \global\plus
      &
  =\int_{\bbbr^3}{d\vec p\over(2\pi)^3}
    \exp\bigg[-{\i\hbar T\over2m}\vec p^2
        +\i\vec p\cdot(\vec x''-\vec x')\bigg]\enspace.
  \qquad
  \tag\NUM.\num\endalign$$
Note the formulation via a path integration over the Euclidean group
[\BJc]. The most general potential separable in cartesian coordinates
has the form
\plus$$
  V=u(x)+v(y)+w(z)\enspace.
  \tag\NUM.\num$$

\subsubsection{Circular Cylinder Coordinates}
Next we consider circular cylinder coordinates which are very similar
to the two-dimensional polar coordinates
\plus$$\left.\alignedat 3
  x&=r\cos\phi\enspace,
  &\qquad
  &r>0\enspace,
  \\
  y&=r\sin\phi\enspace,
  &\qquad
  &0\leq\phi<2\pi\enspace,
  \\
  z&=z\enspace,
  &\qquad
  &z\in\bbbr\enspace.
  \endalignedat\qquad\qquad\right\}
  \tag\NUM.\num$$
Here we have for $d_\edrei(\vec q'',\vec q')$
\plus$$
  d^2_\edrei(\vec q'',\vec q')=
  \vert z''-z'\vert^2+{r'}^2+{r''}^2-2r'r''\cos(\phi''-\phi')
  \enspace.
  \tag\NUM.\num$$
The metric reads $(g_{ab})=\diag(1,r^2,1)$, and the momentum
operators are given by (\numdf) and $p_z=-\i\hbar\partial_z$.
This gives for the Hamiltonian
\plus$$
  -\hbarm\Delta_\edrei
  =-\hbarm\bigg(
   {\partial^2\over\partial r^2}+{1\over r}{\partial\over\partial r}
    +{1\over r^2}{\partial^2\over\partial\phi^2}
    +{\partial^2\over\partial z^2}\bigg)
  ={1\over2m}\bigg(p_r^2+{1\over r^2}p_\phi^2+p_z^2\bigg)
   -{\hbar^2\over8mr^2}\enspace.
  \tag\NUM.\num$$
We therefore obtain the path integral identity
[\ART, \BJb, \GRSb, \PI, \STEc]
$$\myalign
  &\int\limits_{r(t')=r'}^{r(t'')=r''}r\CD r(t)
  \int\limits_{\phi(t')=\phi'}^{\phi(t'')=\phi''}\CD\phi(t)
  \int\limits_{z(t')=z'}^{z(t'')=z''}\CD z(t)
  \\   &\qquad\qquad\times
  \exp\left\{\ih\int_{t'}^{t''}\bigg[{m\over2}\Big(\dot r^2
      +r^2\dot\phi^2+\dot z^2\Big)+{\hbar^2\over8mr^2}\bigg]dt\right\}
  \\   &
  =\bigg({m\over2\pi\i\hbar T}\bigg)^{3/2}
   \exp\bigg[{\i m\over2\hbar T}\Big(\vert z''-z'\vert^2
     +{r'}^2+{r''}^2\Big)\bigg]
   \sum_{l\in\bbbz}\e^{\i l(\phi''-\phi')}
   I_l\bigg({mr'r''\over\i\hbar T}\bigg)\qquad
  \tag\NUM.\num\\   \global\plus
       &
  =\int_{\bbbr}{dp_z\over2\pi}\e^{\i p_z(z''-z')-\i\hbar p_z^2T/2m}
  \sum_{l\in\bbbz}{\e^{\i l(\phi''-\phi')}\over2\pi}
   \int_0^\infty pdp\,J_l(pr')J_l(pr'')\e^{-\i\hbar p^2T/2m}
   \enspace.
  \tag\NUM.\num\endalign$$
A separable potential has te form
\plus$$
  V=u(r)+{1\over r^2}v(\phi)+w(z)\enspace.
  \tag\NUM.\num$$

\subsubsection{Elliptic Cylinder Coordinates}
We consider the coordinate system
\plus$$\left.\alignedat 3
  x&=d\cosh\mu\cos\nu\enspace,
   &\qquad
   &\mu>0\enspace,
  \\
  y&=d\sinh\mu\sin\nu\enspace,
   &\qquad
   &-\pi<\nu\leq\pi
  \\
  z&=z\enspace,
  &\qquad
  &z\in\bbbr\enspace.
  \endalignedat\qquad\qquad\right\}
  \tag\NUM.\num$$
(alternatively $\mu\in\bbbr$, $0<\nu<\pi$ [\MESCH]).
Here the invariant distance is
\plus$$
  d_\edrei^2(\vec q'',\vec q')
  =\vert z''-z'\vert^2
   +d^2(\cosh\mu'\cosh\mu''\cos\nu'\cos\nu''
   +\sinh\mu'\sinh\mu''\sin\nu'\sin\nu'')\enspace.
  \tag\NUM.\num$$
The metric is $(g_{ab})=\diag[d^2(\sinh^2\mu+\sin^2\nu),
d^2(\sinh^2\mu+\sin^2\nu),1]$, and we obtain for the momentum operators
(\numec), and $p_z=-\i\hbar\partial_z$. Consequently for the Hamiltonian
\plus$$\myalign
  -\hbarm\Delta_\edrei
  &=-\hbarm\bigg[{1\over d^2(\sinh^2\mu+\sin^2\nu)}
    \bigg({\partial^2\over\partial\mu^2}
    +{\partial^2\over\partial\nu^2}\bigg)
    +{\partial^2\over\partial z^2}\bigg]
  \\   &
  ={1\over2m}\bigg[{1\over d^2}(\sinh^2\mu+\sin^2\nu)^{-1/2}
  (p_\mu^2+p_\nu^2)(\sinh^2\mu+\sin^2\nu)^{-1/2}
  +p_z^2\bigg]\enspace.\qquad
  \tag\NUM.\num\endalign$$
The path integral {\it construction\/} is straightforward, however,
again no explicit path {\it integration\/} is possible for the
coordinates $\mu$ and $\nu$. The corresponding wave-functions yield
again as eigen-functions Mathieu functions. Because we know the kernel
in $\edrei$ in terms of the invariant distance $d_\edrei$ we can state
the following path integral identity
$$\myalign
  &\int\limits_{\mu(t')=\mu'}^{\mu(t'')=\mu''}\CD\mu(t)
  \int\limits_{\nu(t')=\nu'}^{\nu(t'')=\nu''}\CD\nu(t)
  d^2(\sinh^2\mu+\sin^2\nu)
  \int\limits_{z(t')=z'}^{z(t'')=z''}\CD z(t)
  \\   &\qquad\qquad\times
  \exp\left\{{\i m\over2\hbar}\int_{t'}^{t''}\Big[d^2
   (\sinh^2\mu+\sin^2\nu)(\dot\mu^2+\dot\nu^2)+\dot z^2\Big]dt\right\}
  \\   &
  =\int_{\bbbr}{dp_z\over2\pi}\e^{\i p_z(z''-z')-\i\hbar p_z^2T/2m}
  {1\over2\pi}\sum_{\nu\in\Lambda}\int_0^\infty pdp\,
  \e^{-\i\hbar p^2T/2m}
  \\   &\qquad\qquad\times
  \me_\nu^*(\eta',\hbox{${d^2p^2\over4}$})
  \me_\nu(\eta'',\hbox{${d^2p^2\over4}$})
  \Me_\nu^{(1)\,*}(\xi',\hbox{${d^2p^2\over4}$})
  \Me_\nu^{(1)}(\xi'',\hbox{${d^2p^2\over4}$})\qquad
  \tag\NUM.\num\\   \global\plus
       &
  =\bigg({m\over2\pi\i\hbar T}\bigg)^{3/2}\exp\bigg[{\i m\over2\hbar T}
             d^2_\edrei(\vec q'',\vec q')\bigg]\enspace,
  \tag\NUM.\num\endalign$$
and $d_\edrei$ must be taken in elliptic cylinder coordinates, and we
have used the same notation as for the two-dimensional elliptic
coordinates. A potential separable in these coordinates reads
\plus$$
  V={u(\cosh\mu)+v(\cos\nu)\over\sinh^2\mu+\sin^2\nu}+v(z)\enspace.
  \tag\NUM.\num$$

\subsubsection{Parabolic Cylinder Coordinates}
The last example for cylinder coordinates in three dimensions are the
parabolic cylinder coordinates
\plus$$\alignedat 3
  x&=\xi\eta\enspace,\qquad
  y=\bhalf(\eta^2-\xi^2)\enspace,
   &\qquad
   &\xi\in\bbbr,\eta>0\enspace,
  \\
  z&=z\enspace,
   &\qquad
   &z\in\bbbr\enspace,
  \endalignedat
  \tag\NUM.\num$$
which is the obvious generalization of the two-dimensional case.
Therefore
\plus$$\multline
  d_\edrei^2(\vec q'',\vec q')
  =\vert z''-z'\vert^2
  \\
   +\viert\Big[({\eta''}^2+{\xi''}^2)^2+({\eta'}^2+{\xi'}^2)^2
   -2({\eta'}^2-{\xi'}^2)({\eta''}^2-{\xi''}^2)
   -8\eta'\eta''\xi'\xi''\Big]\enspace.
  \endmultline
  \tag\NUM.\num$$
We have $(g_{ab})=\diag(\xi^2+\eta^2,\xi^2+\eta^2,1)$, hence for the
momentum operators (\numeb) and $p_z=-\i\hbar\partial_z$. This gives
for the Hamiltonian
\plus$$\myalign
  -\hbarm\Delta_\edrei
  &=-\hbarm\bigg[{1\over\xi^2+\eta^2}
   \bigg({\partial^2\over\partial\xi^2}
        +{\partial^2\over\partial\eta^2}\bigg)
    +{\partial^2\over\partial z^2}\bigg]
  \\   &
  ={1\over2m}\Big[(\xi^2+\eta^2)^{-1/2}
             (p_\xi^2+p_\eta^2)(\xi^2+\eta^2)^{-1/2}
             +p_z^2\Big]\enspace.
  \tag\NUM.\num\endalign$$
The arising path integral in these coordinates must be treated
similarly as in the two-dimensional case, the only difference being
the $z$-path integration which is trivial. We obtain
\plus$$\myalign
  &\int\limits_{\xi(t')=\xi'}^{\xi(t'')=\xi''}\CD\xi(t)
  \int\limits_{\eta(t')=\eta'}^{\eta(t'')=\eta''}\CD\eta(t)
  (\xi^2+\eta^2)
  \int\limits_{z(t')=z'}^{z(t'')=z''}\CD z(t)
  \\   &\qquad\qquad\times
  \exp\left\{{\i m\over2\hbar}\int_{t'}^{t''}\Big[
   (\xi^2+\eta^2)(\dot\xi^2+\dot\eta^2)+\dot z^2\Big]dt\right\}
  \\   &
  =\int_{\bbbr}{dp_z\over2\pi}\e^{\i p_z(z''-z')-\i\hbar p_z^2T/2m}
  \\   &\qquad\qquad\times
  \sum_{e,o}\int_{\bbbr} d\zeta\int_{\bbbr}dp\,
  \e^{-\i\hbar p^2T/2m}\Psi_{p,\zeta}^{(e,o)\,*}(\xi',\eta')
  \Psi_{p,\zeta}^{(e,o)}(\xi'',\eta'')\enspace,
  \tag\NUM.\num\endalign$$
in the notation of 4.1.5.
A separable potential has the form
\plus$$
  V={u(\xi)+v(\eta)\over\xi^2+\eta^2}+v(z)\enspace.
  \tag\NUM.\num$$

\subsubsection{Sphero-Conical Coordinates}
We consider the coordinate system
\plus$$\left.\alignedat 3
  x&=r\sn(\mu,k)\dn(\nu,k')\enspace,
     &\quad      &r>0\enspace,
  \\
  y&=r\cn(\mu,k)\cn(\nu,k')\enspace,
     &\quad      &k^2+{k'}^2=1\enspace,
  \\
  z&=r\dn(\mu,k)\sn(\nu,k')\enspace,
     &\quad      &\quad
  \endalignedat\qquad\qquad\right\}
  \tag\NUM.\num$$
in the notation of 4.2.3. The invariant distance is given by
\plus$$\myalign
  d_\edrei^2(\vec q'',\vec q')
  &={r'}^2+{r''}^2-2r'r''\Big(
  \sn\mu''\sn\mu'\dn\nu''\dn\nu'
  \\   &\qquad
  +\cn\mu''\cn\mu'\cn\nu''\cn\nu'
  +\dn\mu''\dn\mu'\sn\nu''\sn\nu'
  \Big)\enspace.
  \tag\NUM.\num\endalign$$
The metric tensor $g_{ab}$ in these coordinates has the form
\plus$$
  (g_{ab})=\diag\big[1,r^2(k^2\cn^2\mu+{k'}^2\cn^2\nu),
  r^2(k^2\cn^2\mu+{k'}^2\cn^2\nu)]
  \enspace,
  \tag\NUM.\num$$
and the momentum operators are $p_r=-\i\hbar(\partial/\partial r+1/r)$,
and $p_\mu$, $p_\nu$ as in (\numea). The path integral can now be
formulated yielding
\minus
$$\myalign
  &\int\limits_{r(t')=r'}^{r(t'')=r''}r^2\CD r(t)
  \int\limits_{\mu(t')=\mu'}^{\mu(t'')=\mu''}
  \CD\mu(t)
  \int\limits_{\nu(t')=\nu'}^{\nu(t'')=\nu''}\CD\nu(t)
  (k^2\cn^2\mu+{k'}^2\cn^2\nu)
  \\   &\qquad\qquad\times
  \exp\Bigg[{\i m\over2\hbar}\int^{t''}_{t'}\Big(\dot
  r^2+r^2(k^2\cn^2\mu+{k'}^2\cn^2\nu)(\dot\mu^2+\dot\nu^2)\Big)dt\Bigg]
  \\   &
  =\sum_{l=0}^\infty\sum_\kappa
  A_{l,\kappa}^*(\mu')B_{l,\kappa}^*(\nu')
  A_{l,\kappa}(\mu'')B_{l,\kappa}(\nu'')
  \\   &\qquad\qquad\times
  {1\over r'r''}\int\limits_{r(t')=r'}^{r(t'')=r''}\CD r(t)
  \exp\left[\ih\int^{t''}_{t'}\bigg({m\over2}\dot r^2
  -\hbarm{(l+\half)^2-\viert\over r^2}\bigg)dt\right]
  \\   &
  ={m\over\i\hbar T\sqrt{r'r''}}
   \exp\bigg[-{m\over2\i\hbar T}({r'}^2+{r''}^2)\bigg]
  \\   &\qquad\qquad\times
   \sum_\kappa\sum_{l=0}^\infty
   A_{l,\kappa}^*(\mu')B_{l,\kappa}^*(\nu')
   A_{l,\kappa}(\mu'')B_{l,\kappa}(\nu'')
   I_{l+1/2}\bigg({mr'r''\over\i\hbar T}\bigg)
  \tag\NUM.\num\\   \global\plus
      &
  =\sum_\kappa\sum_{l=0}^\infty
   A_{l,\kappa}^*(\mu')B_{l,\kappa}^*(\nu')
   A_{l,\kappa}(\mu'')B_{l,\kappa}(\nu'')
   \int_0^\infty{pdp\over\sqrt{r'r''}}
   J_{l+\half}(pr')J_{l+\half}(pr'')\e^{-\i\hbar p^2T/2m}
  \tag\NUM.\num\\   \global\plus
      &
  =\bigg({m\over2\pi\i\hbar T}\bigg)^{3/2}
    \exp\bigg[-{m\over2\i\hbar T}d^2_\edrei(\vec q'',\vec q')\bigg]
  \enspace,
  \tag\NUM.\num\endalign$$
with $d_\edrei(\vec q'',\vec q')$ in sphero-conical coordinates.
Here a separable potential must have the form
\plus$$
  V=u(r)+{1\over r^2}{v(\cn\alpha)+w(\cn\beta)\over
         k^2\cn^2\alpha+{k'}^2\cn^2\beta}
  \enspace.
  \tag\NUM.\num$$

\subsubsection{Spherical Coordinates}
We consider the spherical coordinates
\plus$$\left.\alignedat 3
   x&=r\sin\theta\cos\phi\enspace,
    &\qquad
    &r>0\enspace,
   \\
   y&=r\sin\theta\sin\phi\enspace,
    &\qquad
    &0<\theta<\pi\enspace,
   \\
   z&=r\cos\theta\enspace,
    &\qquad
    &0\leq\phi<2\pi\enspace.
   \endalignedat\qquad\qquad\right\}
   \tag\NUM.\num$$
These are the usual three-dimensional polar coordinates. Here
\plus$$
  d_\edrei^2(\vec q'',\vec q')=
  {r'}^2+{r''}^2-2r'r''\Big(\cos\theta'\cos\theta''
       -\sin\theta'\sin\theta''\cos(\phi''-\phi')\Big)\enspace.
  \tag\NUM.\num$$
The metric tensor is $(g_{ab})=\diag(1,r^2,r^2\sin^2\theta)$, and the
momentum operators have the form $p_r=-\i\hbar(\partial/\partial
r+1/r)$ together with (\numed). For the Hamiltonian we obtain
\plus$$\myalign
  -\hbarm\Delta_\edrei
  &=-\hbarm\bigg[{\partial^2\over\partial r^2}
                 +{2\over r}{\partial\over\partial r}
    +{1\over r^2}\bigg({\partial^2\over\partial\theta^2}
                 +\cot\theta{\partial\over\partial\theta}\bigg)
    +{1\over r^2\sin^2\theta}{\partial^2\over\partial\phi^2}\bigg]
  \\   &
  ={1\over2m}\bigg(p_r^2+{1\over r^2}p_\theta^2
          +{1\over r^2\sin^2\theta}p_\phi^2\bigg)
          -{\hbar^2\over8mr^2}\bigg(1+{1\over\sin^2\theta}\bigg)
  \enspace.
  \tag\NUM.\num\endalign$$
The corresponding path integral is well-known [\DURd, \GOOb, \PI] and we
have the identity
$$\myalign
  &\int\limits_{r(t')=r'}^{r(t'')=r''}r^2\CD r(t)
   \int\limits_{\theta(t')=\theta'}^{\theta(t'')=\theta''}
  \sin\theta\CD\theta(t)
  \int\limits_{\phi(t')=\phi'}^{\phi(t'')=\phi''}\CD\phi(t)
  \\   &\qquad\times
  \exp\left\{\ih\int_{t'}^{t''}\bigg[{m\over2}\Big(\dot r^2
      +r^2\dot\theta^2+r^2\sin^2\theta\dot\phi^2\Big)
  +{\hbar^2\over8mr^2}\bigg(1+{1\over\sin^2\theta}\bigg)\bigg]dt\right\}
  \\   &
  ={m\over4\pi\i\hbar T\sqrt{r'r''}}
   \exp\bigg[-{m\over2\i\hbar T}({r'}^2+{r''}^2)\bigg]
  \\   &\qquad\times
   \sum_{l=0}^\infty(2l+1)P_l\big(\cos\psi_\szwei(\vec q'',\vec q')\big)
   I_{l+1/2}\bigg({mr'r''\over\i\hbar T}\bigg)
  \tag\NUM.\num\\   \global\plus
       &
  =\sum_{l=0}^\infty\sum_{m=-l}^l
   Y_l^{m\,*}(\theta',\phi')Y_l^m(\theta'',\phi'')
   {1\over\sqrt{r'r''}}\int_0^\infty pdp\,
   J_{l+\half}(pr')J_{l+\half}(pr'')\e^{-\i\hbar p^2T/2m}\enspace.
  \tag\NUM.\num\endalign$$
A potential which is separable in spherical coordinates reads
\plus$$
  V=u(r)+{1\over r^2}v(\theta)+{1\over r^2\sin^2\theta}w(\phi)\enspace.
  \tag\NUM.\num$$

\subsubsection{Parabolic Coordinates}
We consider the coordinate system
\plus$$\left.\alignedat 3
  x&=\xi\eta\cos\phi\enspace,\qquad
  y=\xi\eta\sin\phi\enspace,
   &\quad        &\xi,\eta>0\enspace,
  \\
  z&=\half(\xi^2-\eta^2)\enspace,
   &\quad        &0\leq\phi<2\pi\enspace.
  \endalignedat\qquad\qquad\right\}
  \tag\NUM.\num$$
This gives for the metric tensor $(g_{ab})=\diag(\xi^2+\eta^2,\xi^2+
\eta^2,\xi^2\eta^2)$, and
\plus$$\multline
  d_\edrei^2(\vec q'',\vec q')=
  \viert\Big[({\eta''}^2+{\xi''}^2)^2+({\eta'}^2+{\xi'}^2)^2
   \\
   -2({\eta'}^2-{\xi'}^2)({\eta''}^2-{\xi''}^2)
   -8\eta'\eta''\xi'\xi''\cos(\phi''-\phi')\Big]\enspace.
  \endmultline
  \tag\NUM.\num$$
For the momentum operators we get
\plus$$
  p_\xi=\hi\bigg({\partial\over\partial\xi}+{\xi\over\xi^2+\eta^2}
          +{1\over2\xi}\bigg)\enspace,\qquad
  p_\eta=\hi\bigg({\partial\over\partial\eta}+{\eta\over\xi^2+\eta^2}
          +{1\over2\eta}\bigg)\enspace,
  \tag\NUM.\num$$
together with $p_\phi=-\i\hbar\partial_\phi$.
For the Hamiltonian we obtain
\plus$$\myalign
  -\hbarm\Delta_\edrei
  &=-\hbarm\bigg[{1\over\xi^2+\eta^2}\bigg(
   {\partial^2\over\partial\xi^2}+{\partial\over\partial\xi}
   +{\partial^2\over\partial\eta^2}+{\partial\over\partial\eta}\bigg)
   +{1\over\xi^2\eta^2}{\partial^2\over\partial\phi^2}\bigg]
  \\   &
  ={1\over2m}\bigg[{1\over\sqrt{\xi^2+\eta^2}}(p_\xi^2+p_\eta^2)
    {1\over\sqrt{\xi^2+\eta^2}}+{1\over\xi^2\eta^2}p_\phi^2\bigg]
    -{\hbar^2\over8m\xi^2\eta^2}\enspace.
  \tag\NUM.\num\endalign$$
We obtain the path integral identity [\CHe, \GROm]
\plus$$\myalign
  &\int\limits_{\xi(t')=\xi'}^{\xi(t'')=\xi''}\CD\xi(t)
  \int\limits_{\eta(t')=\eta'}^{\eta(t'')=\eta''}\CD\eta(t)
  (\xi^2+\eta^2)\xi\eta
  \int\limits_{\phi(t')=\phi'}^{\phi(t'')=\phi''}\CD\phi(t)
  \\   &\ \times
  \exp\left\{\ih\int_{t'}^{t''}\bigg[{m\over2}\Big(
   (\xi^2+\eta^2)(\dot\xi^2+\dot\eta^2)+\xi^2\eta^2\dot\phi^2\Big)
  +{\hbar^2\over8m\xi^2\eta^2}\bigg]dt\right\}
  \\   &
  =(\xi'\xi''\eta'\eta'')^{-1/2}
  \sum_{l\in\bbbz}{\e^{-\i l(\phi''-\phi')}\over2\pi}
   \int\limits_{\xi(t')=\xi'}^{\xi(t'')=\xi''}\CD\xi(t)
  \int\limits_{\eta(t')=\eta'}^{\eta(t'')=\eta''}\CD\eta(t)
  (\xi^2+\eta^2)
  \\   &\ \times
  \exp\left\{\ih\int_{t'}^{t''}\bigg[{m\over2}
   (\xi^2+\eta^2)(\dot\xi^2+\dot\eta^2)
  -{\hbar^2\over2m}{l^2-\viert\over\xi^2\eta^2}\bigg]dt\right\}
  \\   &
  =\sum_{l\in\bbbz}{\e^{-\i l(\phi''-\phi')}\over2\pi}
  \int_{\bbbr} d\zeta\int_0^\infty{dp\over p}
  {\vert\Gamma\big({1+\vert l\vert\over2}+{\i\zeta\over2p})\big\vert^4
   \e^{\pi/ap}\over4\pi^2\xi'\xi''\eta'\eta''\Gamma^4(1+\vert l\vert)}
  \e^{-\i\hbar p^2T/2m}
  \\   &\ \times
  M_{-\i\zeta/2p,\vert l\vert/2}(-\i p{\xi''}^2)
  M_{\i\zeta/2p,\vert l\vert/2}(\i p{\xi''}^2)
  M_{-\i\zeta/2p,\vert l\vert/2}(-\i p{\eta''}^2)
  M_{\i\zeta/2p,\vert l\vert/2}(\i p{\eta''}^2)\enspace,
  \\   &
  \tag\NUM.\num\endalign$$
where the last line was obtained in a similar way as the
corresponding result in 4.1.5, and use has been made of (A.2.2).
In parabolic coordinates a potential is separable if it has the form
\plus$$
  V={u(\xi)+v(\eta)\over\sqrt{x^2+y^2+z^2}}+{w(\phi)\over x^2+y^2}
  \enspace.
  \tag\NUM.\num$$

\subsubsection{Prolate Spheroidal Coordinates}
We consider the coordinate system
\plus$$\left.\alignedat 3
  x&=d\sinh\mu\sin\nu\cos\phi\enspace,
   &\quad        &\mu>0\enspace,
  \\
  y&=d\sinh\mu\sin\nu\sin\phi\enspace,
   &\quad        &0<\nu<\pi\enspace,
  \\
  z&=d\cosh\mu\cos\nu\enspace,
   &\quad        &0\leq\phi<2\pi\enspace.
  \endalignedat\qquad\qquad\right\}
  \tag\NUM.\num$$
This yields $(g_{ab})=d^2\diag(\sinh^2\mu+\sin^2\nu, \sinh^2\mu+
\sin^2\nu,\sinh^2\mu\sin^2\nu)$, and for the momentum operators
we obtain
\plus$$
  p_\mu=\hi\Bigg({\partial\over\partial\mu}
          +{\sinh\mu\cosh\mu\over\sinh^2\mu+\sin^2\nu}
          +\half\coth\mu\Bigg)\,,\
  p_\nu=\hi\Bigg({\partial\over\partial\nu}
          +{\sin\nu\cos\nu\over\sinh^2\mu+\sin^2\nu}
          +\half\cot\nu\Bigg)\,,
  \tag\NUM.\num$$
\edef\numef{\NUM.\num}%
and $p_\phi=-\i\hbar\partial_\phi$. The invariant distance is given by
\plus$$
  d_\edrei^2(\vec q'',\vec q')=
  d^2\Big[\cosh\mu'\cosh\mu''\cos\nu'\cos\nu''
        +\sinh\mu'\sinh\mu''\sin\nu'\sin\nu''\cos(\phi''-\phi')\Big]
  \enspace.
  \tag\NUM.\num$$
The Hamiltonian has the form
\plus$$\align
  &-\hbarm\Delta_\edrei
  \\   &
  =-{\hbar^2\over2md^2}\bigg[{1\over\sinh^2\mu+\sin^2\nu}\bigg(
  {\partial^2\over\partial\mu^2}+\coth\mu{\partial\over\partial\mu}+
  {\partial^2\over\partial\nu^2}+\cot\nu{\partial\over\partial\nu}\bigg)
  +{1\over\sinh^2\mu\sin^2\nu}{\partial^2\over\partial^2\phi}\bigg]
  \\   &
  ={1\over2md^2}\bigg[{1\over\sqrt{\sinh^2\mu+\sin^2\nu}}
   (p_\mu^2+p_\nu^2){1\over\sqrt{\sinh^2\mu+\sin^2\nu}}
   +{1\over\sinh^2\mu+\sin^2\nu}p_\phi^2\bigg]
  \\   &\qquad\qquad\qquad\qquad\qquad\qquad\qquad\qquad\qquad\qquad
   -{\hbar^2\over8md^2\sinh^2\mu\sin^2\nu}\enspace.
  \tag\NUM.\num\endalign$$
The path integral {\it construction\/} is straightforward, however no
explicit path {\it integration\/} is possible. Actually, an expansion
into the corresponding wave-functions in the coordinates $\mu$ and
$\nu$ yields the spheroidal functions $\ps_n^m(\cos\nu,\gamma)$ and
$S_n^{m\,(1)}(\cosh\mu,\gamma)$ ($\gamma^2=2mEd^2/\hbar^2=p^2d^2$),
respectively, as the eigen-function of the Hamiltonian, a specific
class of higher transcendental functions [\MESCH], similar to the
Mathieu functions. However, because we know on the one side the
eigen-functions of the Hamiltonian in terms of these functions
[\MESCH], and on the other the kernel in $\edrei$ in terms of the
invariant distance $d_\ezwei$ we can state the following path integral
identity (note the implemented time-transformation)
\minus
$$\myalign
  &\int\limits_{\mu(t')=\mu'}^{\mu(t'')=\mu''}\CD\mu(t)
  \int\limits_{\nu(t')=\nu'}^{\nu(t'')=\nu''}\CD\nu(t)
  d^3(\sinh^2\mu+\sin^2\nu)\sinh\mu\sin\nu
  \int\limits_{\phi(t')=\phi'}^{\phi(t'')=\phi''}\CD\phi(t)
  \\   &\quad\times
  \exp\Bigg\{\ih\int_{t'}^{t''}\bigg[{m\over2}d^2\Big(
   (\sinh^2\mu+\sin^2\nu)(\dot\mu^2+\dot\nu^2)
   +\sinh^2\mu\sin^2\nu\dot\phi^2\Big)
  \\   &\qquad\qquad\qquad\qquad\qquad\qquad
        \qquad\qquad\qquad\qquad\qquad\qquad
   +{\hbar^2\over8md^2\sinh^2\mu\sin^2\nu}\bigg]dt\Bigg\}
  \\   &
  =(\sinh\mu'\sinh\mu''\sin\nu'\sin\nu'')^{-1/2}
   \sum_{l\in\bbbz}{\e^{\i l(\phi''-\phi')}\over2\pi}
  \\   &\quad\times
   \int_{\bbbr}{dE\over2\pi\hbar}\e^{-\i ET/\hbar}\int_0^\infty ds''
   \int\limits_{\mu(0)=\mu'}^{\mu(s'')=\mu''}\CD\mu(s)
  \int\limits_{\nu(0)=\nu'}^{\nu(s'')=\nu''}\CD\nu(s)
  \\   &\quad\times
  \exp\left\{\ih\int_{0}^{s''}
   \bigg[{m\over2}(\dot\mu^2+\dot\nu^2)+Ed^2(\sinh^2\mu+\sin^2\nu)
   -\hbarm\bigg({l^2-\viert\over\sinh^2\mu}
        +{l^2-\viert\over\sin^2\nu}\bigg)\bigg]ds\right\}
  \\   &
  \tag\NUM.\num\\   \global\plus
       &
  =\sum_{l\in\bbbz}{\e^{\i l(\phi''-\phi')}\over2\pi}
   \sum_{n\in\Lambda}{2\over\pi}\int_0^\infty p^2dp\,
  {2n+1\over2}{(n-l)!\over(n+l)!} \e^{-\i\hbar p^2T/2m}
  \\   &\quad\times
  \ps_n^l(\cos\nu',pd) \ps_n^l(\cos\nu'',pd)
  S_n^{l\,(1)\,*}(\cosh\mu',pd)S_n^{l\,(1)\,*}(\cosh\mu'',pd)
  \tag\NUM.\num\\   \global\plus
       &
  =\bigg({m\over2\pi\i\hbar T}\bigg)^{3/2}\exp\bigg[{\i m\over2\hbar T}
             d^2_\edrei(\vec q'',\vec q')\bigg]\enspace,
  \tag\NUM.\num\endalign$$
and $d_\edrei(\vec q'',\vec q')$ must be taken in prolate spheroidal
coordinates. The functions $\ps_n^l(\cos\nu,\gamma)$ and $S_n^{l\,(1)}
(\cosh\mu,\gamma)$ are mutually determined by the separation parameter
$\lambda=\lambda_n^l(\gamma)$ giving an infinite countable set
$\{\lambda_n\},\,(n\in\Lambda)$. The functions $\ps_n^l(\cos\nu,\gamma)$
yield in the limit $\gamma\to0$ the associated Legendre-polynomials
$P_n^l(\cos\nu)$, and the functions $S_n^{l\,(1)}(\cosh\mu,\gamma)$ the
spherical Bessel functions, i.e.\ $S_n^{l\,(1)}(z/\gamma,\gamma)\propto
\sqrt{\pi/2z}J_{l+\half}(z)$ $(\gamma\to0$), therefore obeying the
correct boundary-conditions. A separable potential must have the form
\plus$$
  V={u(\cosh\mu)+v(\cos\nu)\over\sinh^2\mu+\sin^2\nu}
   +{w(\phi)\over\sinh^2\mu\sin^2\nu}\enspace.
  \tag\NUM.\num$$

Note that by the replacement $z\to d(\cosh\mu\cos\nu+1)$ we obtain the
spheroidal coordinate system in which the Coulomb-problem in $\bbbr^3$
is separable [\LL], e.g.\ c.f.\ [\GROm] and references therein.

\subsubsection{Oblate Spheroidal Coordinates}
We consider the coordinate system
\plus$$\left.\alignedat 3
  x&=d\cosh\xi\sin\nu\cos\phi\enspace,
   &\quad        &\xi>0\enspace,
  \\
  y&=d\cosh\xi\sin\nu\sin\phi\enspace,
   &\quad        &0<\nu<\pi\enspace,
  \\
  z&=d\sinh\xi\cos\nu\enspace,
   &\quad        &0\leq\phi<2\pi
  \endalignedat\qquad\qquad\right\}
  \tag\NUM.\num$$
(alternatively $\mu\in\bbbr$, $0<\nu<\pi/2$ [\MESCH]).
This yields $(g_{ab})=d^2\diag(\cosh^2\xi-\sin^2\nu, \cosh^2\xi-
\sin^2\nu,\cosh^2\xi\sin^2\nu)$, and for the momentum operators
we obtain
\plus$$
  p_\xi=\hi\Bigg({\partial\over\partial\mu}
          +{\sinh\xi\cosh\xi\over\cosh^2\xi-\sin^2\nu}
          +\half\tanh\xi\Bigg)\,,\
  p_\nu=\hi\Bigg({\partial\over\partial\nu}
          +{\sin\nu\cos\nu\over\cosh^2\xi-\sin^2\nu}
          +\half\cot\nu\Bigg)\,,
  \tag\NUM.\num$$
and $p_\phi=-\i\hbar\partial_\phi$. The invariant distance is given by
\plus$$
  d_\edrei^2(\vec q'',\vec q')=
  d^2\Big[\sinh\xi'\sinh\xi''\cos\nu'\cos\nu''
        +\cosh\xi'\cosh\xi''\sin\nu'\sin\nu''\cos(\phi''-\phi')\Big]
  \enspace.
  \tag\NUM.\num$$
The Hamiltonian has the form
\plus$$\myalign
  &-\hbarm\Delta_\edrei
  \\   &
  =-{\hbar^2\over2md^2}\bigg[{1\over\cosh^2\xi-\sin^2\nu}\bigg(
  {\partial^2\over\partial\xi^2}+\tanh\xi{\partial\over\partial\xi}+
  {\partial^2\over\partial\nu^2}+\cot\nu{\partial\over\partial\nu}\bigg)
  +{1\over\cosh^2\xi\sin^2\nu}{\partial^2\over\partial^2\phi}\bigg]
  \\   &
  ={1\over2md^2}\bigg[{1\over\sqrt{\cosh^2\xi-\sin^2\nu}}
   (p_\xi^2+p_\nu^2){1\over\sqrt{\cosh^2\xi-\sin^2\nu}}
   +{1\over\cosh^2\xi-\sin^2\nu}p_\phi^2\bigg]
  \\   &\qquad\qquad\qquad\qquad\qquad\qquad\qquad\qquad\qquad\qquad
   -{\hbar^2\over8md^2\cosh^2\xi\sin^2\nu}\enspace.
  \tag\NUM.\num\endalign$$
The path integral construction is {\it straightforward\/}, however
again no explicit path {\it integration\/} is possible. An expansion
into the corresponding wave-functions in the coordinates $\xi$ and $\nu$
yields eigen-function of the Hamiltonian in terms of spheroidal
functions $\ps_n^m(\cos\nu,\i\gamma)=\sphpsi_n^m(\cos\nu,\gamma)$ and
$S_n^{m\,(1)}(-\i\cosh\xi,\i\gamma)=\Si_n^{m\,(1)}(\cosh\xi,\gamma)$
($\gamma^2=2mEd^2/\hbar^2=pd$), respectively. Because we know on the
one side the eigen-functions of the Hamiltonian in terms of these
functions [\MESCH], and on the other the kernel in $\edrei$ in terms of
the invariant distance $d_\edrei$ we can state the following path
integral identity
\minus
$$\align
  &\int\limits_{\xi(t')=\xi'}^{\xi(t'')=\xi''}\CD\xi(t)
  \int\limits_{\nu(t')=\nu'}^{\nu(t'')=\nu''}\CD\nu(t)
  d^3(\cosh^2\xi-\sin^2\nu)\cosh\xi\sin\nu
  \int\limits_{\phi(t')=\phi'}^{\phi(t'')=\phi''}\CD\phi(t)
  \\   &\quad\times
  \exp\Bigg\{\ih\int_{t'}^{t''}\bigg[{m\over2}d^2\Big(
   (\cosh^2\xi-\sin^2\nu)(\dot\xi^2+\dot\nu^2)
   +\cosh^2\xi\sin^2\nu\dot\phi^2\Big)
  \\   &\qquad\qquad\qquad\qquad\qquad\qquad
        \qquad\qquad\qquad\qquad\qquad\qquad
   +{\hbar^2\over8md^2\cosh^2\xi\sin^2\nu}\bigg]dt\Bigg\}
  \allowdisplaybreak
  \\   &
  =(\cosh\xi'\cosh\xi''\sin\nu'\sin\nu'')^{-1/2}
   \sum_{l\in\bbbz}{\e^{\i l(\phi''-\phi')}\over2\pi}
  \\   &\quad\times
  \int_{\bbbr}{dE\over2\pi\hbar}\e^{-\i ET/\hbar}\int_0^\infty ds''
   \int\limits_{\xi(0)=\xi'}^{\xi(s'')=\xi''}\CD\xi(s)
  \int\limits_{\nu(0)=\nu'}^{\nu(s'')=\nu''}\CD\nu(s)
  \\   &\quad\times
  \exp\left\{\ih\int_{0}^{s''}
   \bigg[{m\over2}(\dot\xi^2+\dot\nu^2)+Ed^2(\cosh^2\xi-\sin^2\nu)
   -\hbarm\bigg({l^2-\viert\over\cosh^2\xi}
        -{l^2-\viert\over\sin^2\nu}\bigg)\bigg]ds\right\}
  \\   &
  \tag\NUM.\num\\   \global\plus
       &
  =\sum_{l\in\bbbz}{\e^{\i l(\phi''-\phi')}\over2\pi}
   \sum_{n\in\Lambda}{2\over\pi}\int_0^\infty p^2dp\,
  {2n+1\over2}{(n-l)!\over(n+l)!}\e^{-\i\hbar p^2T/2m}
  \\   &\quad\times
  \sphpsi_n^l(\cos\nu',pd)\sphpsi_n^l(\cos\nu'',pd)
  \Si_n^{l\,(1)\,*}(\cosh\xi',pd)\Si_n^{l\,(1)\,*}(\cosh\xi'',pd)
  \tag\NUM.\num\\   \global\plus
       &
  =\bigg({m\over2\pi\i\hbar T}\bigg)^{3/2}\exp\bigg[{\i m\over2\hbar T}
             d^2_\edrei(\vec q'',\vec q')\bigg]\enspace,
  \tag\NUM.\num\endalign$$
and $d_\edrei(\vec q'',\vec q')$ must be taken in oblate spheroidal
coordinates. Due to the construction of the spectral expansion the
functions $\sphpsi_n^l(\cos\nu,\gamma)$ and $\Si_n^{l\,(1)}(\cosh\mu,
\gamma)$ are obeying the correct boundary-conditions.
Here a separable potential must have the form
\plus$$
  V={u(\sinh\mu)+v(\cos\nu)\over\cosh^2\mu-\sin^2\nu}
   +{w(\phi)\over\cosh^2\mu\sin^2\nu}\enspace.
  \tag\NUM.\num$$

\subsubsection{Ellipsoidal Coordinates}
The last two coordinate system are the most complicated ones and
are similar in some of their features, c.f.\ Section~3.1. First we
consider the coordinate system
\plus$$\left.\aligned
  x&=\sqrt{{(\xi_1^2-a^2)(\xi_2^2-a^2)(\xi_3^2-a^2)\over
            a^2(a^2-b^2)}}\enspace,\quad
  y=\sqrt{{(\xi_1^2-b^2)(\xi_2^2-b^2)(\xi_3^2-b^2)\over
            b^2(b^2-a^2)}}\enspace,
  \\
  z&={\xi_1\xi_2\xi_3\over ab}\enspace,
  \quad
  \xi_1^2\geq a^2\geq\xi_2^2\geq b^2\geq\xi_3^2\geq c^2=0\enspace.
  \endaligned\qquad\right\}
  \tag\NUM.\num$$
The metric tensor is given by
\plus$$
  (g_{ab})=\diag\left(
  {(\xi_1^2-\xi_2^2)(\xi_1^2-\xi_3^2)\over P_2(\xi_1^2)},
  {(\xi_2^2-\xi_1^2)(\xi_2^2-\xi_3^2)\over P_2(\xi_2^2)},
  {(\xi_3^2-\xi_1^2)(\xi_3^2-\xi_2^2)\over P_2(\xi_3^2)}\right)\enspace,
  \tag\NUM.\num$$
with $P_2(\xi^2)=(\xi^2-a^2)(\xi^2-b^2)$. With the identification
$\xi_1=a\dn(\lambda,k)/\cn(\lambda,k)$, $\xi_2=a\dn(\mu,k')$ and
$\xi_3=b\sn(\nu,k)$ ($b=ka,\,\sqrt{a^2-b^2}=k'a=d$) this can be
rewritten into
\plus$$\left.\aligned
  x&=d{\sn(\lambda,k)\sn(\mu,k')\dn(\nu,k)\over\cn(\lambda,k)}\enspace,
  \quad
  y=d{\cn(\mu,k')\cn(\nu,k)\over\cn(\lambda,k)}\enspace,
  \\
  z&=a{\dn(\lambda,k)\dn(\mu,k')\sn(\nu,k)\over\cn(\lambda,k)}\enspace.
  \endaligned\qquad\qquad\right\}
  \tag\NUM.\num$$
Due to the very complicated structure it is of now use to write down
all subsequent necessary quantities for the path integral.
Instead we exploit the results of Section~3.2, in particular the
separation formula (\numca). We identify $(g_{ab})\equiv\diag(h_1^2,
h_2^2,h_3^2)$, furthermore
\minus$$\align
  &\Gamma_i={\xi_i P_2'(\xi_i^2)\over P_2(\xi^2)}\enspace,
  \qquad(i=1,2,3)
  \tag\NUM.\num\\   \global\plus
  &S=\left\vert\matrix
     1  &1/(\xi_1^2-a^2)  &1/(\xi_1^2-b^2)(a^2-b^2)  \\
     1  &1/(\xi_2^2-a^2)  &1/(\xi_2^2-b^2)(a^2-b^2)  \\
     1  &1/(\xi_3^2-a^2)  &1/(\xi_3^2-b^2)(a^2-b^2)
   \endmatrix\right\vert\enspace,
  \tag\NUM.\num\\   \global\plus
  &\left.\aligned
   &M_1={1\over a^2-b^2}\left({1\over(\xi_2^2-a^2)(\xi_3^2-b^2)}
    -{1\over(\xi_2^2-b^2)(\xi_3^2-a^2)}\right)\enspace,
   \\
   &M_2={1\over a^2-b^2}\left({1\over(\xi_1^2-b^2)(\xi_3^2-a^2)}
    -{1\over(\xi_1^2-a^2)(\xi_3^2-b^2)}\right)\enspace,
   \\
   &M_3={1\over a^2-b^2}\left({1\over(\xi_1^2-a^2)(\xi_2^2-b^2)}
    -{1\over(\xi_1^2-b^2)(\xi_2^2-a^2)}\right)\enspace.
   \endaligned\qquad\qquad\right\}
  \tag\NUM.\num\endalign$$
and obtain the following path integral identity
\minus$$\myalign
  &\prod_{i=1}^3\int\limits_{\xi_i(t')=\xi_i'}^{\xi_i(t'')=\xi_i''}
   h_i\CD\xi_i(t)
   \exp\left\{\ih\int_{t'}^{t''}\left[{m\over2}\sum_{i=1}^3
      h_i^2\dot\xi_i^2-\Delta V_{PF}(\{\xi\})\right]dt\right\}
  \\   &
  =\prod_{i=1}^3\int\limits_{\xi_i(t')=\xi_i'}^{\xi_i(t'')=\xi_i''}
   \sqrt{S\over M_i}\,\CD\xi_i(t)
   \exp\left\{\ih\int_{t'}^{t''}\left[{m\over2}S\sum_{i=1}^3
   {\dot\xi_i^2\over M_i}-\Delta V_{PF}(\{\xi\})\right]dt\right\}
  \\   &
  =(S'S'')^{-1/4}\int_{\bbbr}{dE\over2\pi\hbar}
  \e^{-\i ET/\hbar}\int_0^\infty ds''
   \prod_{i=1}^3\int\limits_{\xi_i(0)=\xi_i'}^{\xi_i(s'')=\xi_i''}
   M_i^{-1/2}\CD\xi_i(s)
  \\   &\qquad\qquad\times
   \exp\left\{\ih\int_{0}^{s''}\left[{m\over2}\sum_{i=1}^3
   {\dot\xi_i^2\over M_i}+ES
    -{\hbar^2\over8m}\sum_{i=1}^3M_i\Big(\Gamma_i^2+2\Gamma_i'\Big)
    \right]ds\right\}\qquad\qquad
  \tag\NUM.\num\\   \global\plus
       &
  =\int_0^\infty dp\sum_{\kappa,\lambda}
   \Psi_{\kappa,\lambda,p}^*(\xi_1',\xi_2',\xi_3')
   \Psi_{\kappa,\lambda,p}(\xi_1'',\xi_2'',\xi_3'')
   \e^{-\i\hbar p^2T/2m}
  \tag\NUM.\num\\   \global\plus
       &
  =\bigg({m\over2\pi\i\hbar T}\bigg)^{3/2}
    \exp\bigg[-{m\over2\i\hbar T}d^2_\edrei(\vec q'',\vec q')\bigg]
  \enspace,
  \tag\NUM.\num\endalign$$
and $d_\edrei(\vec q'',\vec q')$ must be expressed in ellipsoidal
coordinates via $d_\edrei=\vert \vec x(\{\xi''\}-\vec x(\{\xi'\})\vert$.
The functions $\Psi_{\kappa,\lambda,p}(\xi_1,\xi_2,\xi_3)=
A_{\kappa,\lambda,p}(\xi_1)B_{\kappa,\lambda,p}(\xi_2)
C_{\kappa,\lambda,p}(\xi_3)$ are solutions of a three-fold Lam\'e
equation with separation parameters $\lambda,\kappa,p$, respectively
[\MF, p.1305]. A potential separable in ellipsoidal coordinates must
have the form
\plus$$
  V={(\xi_2^2-\xi_3^2)u(\xi_1)+(\xi_1^2-\xi_3^2)u(\xi_2)
    +(\xi_1^2-\xi_2^2)u(\xi_3)\over
     (\xi_1^2-\xi_2^2)(\xi_1^2-\xi_3^2)(\xi_2^2-\xi_3^2)}\enspace.
  \tag\NUM.\num$$
\edef\numek{\NUM.\num}%

\eject\noindent
\baselineskip=12pt
\subsubsection{Paraboloidal Coordinates}
As the last coordinate system in $\edrei$ we now consider
\plus$$\left.\aligned
  x&=\sqrt{{(\xi_1^2-a^2)(\xi_2^2-a^2)(\xi_3^2-a^2)\over
            (a^2-b^2)}}\enspace,
  \qquad
  y=\sqrt{{(\xi_1^2-b^2)(\xi_2^2-b^2)(\xi_3^2-b^2)\over
            (b^2-a^2)}}\enspace,
  \\
  z&=\half\big(\xi_1^2+\xi_2^2+\xi_3^2-a^2-b^2\big)\enspace,
  \quad
  \xi_1^2\geq a^2\geq\xi_2^2\geq b^2\geq\xi_3^2\geq c^2=0\enspace.
  \endaligned\qquad\right\}
  \tag\NUM.\num$$
\edef\numel{\NUM.\num}%
The metric tensor is given by
\plus$$
  (g_{ab})=\diag\left(
   \xi_1^2{(\xi_1^2-\xi_2^2)(\xi_1^2-\xi_3^2)\over P_2(\xi_1^2)},
   \xi_2^2{(\xi_2^2-\xi_1^2)(\xi_2^2-\xi_3^2)\over P_2(\xi_2^2)},
   \xi_3^2{(\xi_3^2-\xi_1^2)(\xi_3^2-\xi_2^2)\over P_2(\xi_3^2)}
   \right)\enspace.
  \tag\NUM.\num$$
Similarly as before we can make the identification $\xi_1=a\dn(\lambda,
k)/\cn(\lambda,k)$, $\xi_2=a\dn(\nu,k')$ and $\xi_3= a\sqrt{1-{k'}^2/
\cn^2(\mu,k)}$ ($b=ka,\,\sqrt{a^2-b^2}=k'a=\sqrt{d}\,$) and (\numel)
can be rewritten into
\plus$$\left.\aligned
  x&=d{\sn(\lambda,k)\sn(\nu,k')\over\cn(\lambda,k)\cn(\mu,k)}\enspace,
  \qquad
  y=d{\sn(\mu,k)\cn(\nu,k')\over\cn(\lambda,k)\cn(\mu,k)}\enspace,
  \\
  z&={d\over2}\bigg[
     {\sn^2(\lambda,k)\over\cn^2(\lambda,k)}
     -{\sn^2(\mu,k)\over\cn^2(\mu,k)}
     +{\dn^2(\nu,k')\over {k'}^2}\bigg]\enspace.
  \endaligned\qquad\qquad\right\}
  \tag\NUM.\num$$
We proceed similarly as before and use (\numca) for the path integral
formulation. We identify $(g_{ab})\equiv\diag(h_1^2,h_2^2,h_3^2)$,
furthermore
\minus$$\align
  &\Gamma_i={\xi_i P_2'(\xi_i^2)\over P_2(\xi_i^2)}-{1\over\xi_i}
   \enspace,\qquad(i=1,2,3)
  \tag\NUM.\num\\   \global\plus
  &S=\left\vert\matrix
  \xi_1^2  &\xi_1^2/(\xi_1^2-a^2)  &\xi_1^2/(\xi_1^2-b^2)(a^2-b^2)  \\
  \xi_2^2  &\xi_2^2/(\xi_2^2-a^2)  &\xi_2^2/(\xi_2^2-b^2)(a^2-b^2)  \\
  \xi_3^2  &\xi_3^2/(\xi_3^2-a^2)  &\xi_3^2/(\xi_3^2-b^2)(a^2-b^2)
   \endmatrix\right\vert\enspace,
  \tag\NUM.\num\\   \global\plus
  &\left.\aligned
   &M_1={\xi_2^2\xi_3^2\over
             a^2-b^2}\left({1\over(\xi_2^2-a^2)(\xi_3^2-b^2)}
    -{1\over(\xi_2^2-b^2)(\xi_3^2-a^2)}\right)\enspace,
   \\
   &M_2={\xi_1^2\xi_3^2\over
         a^2-b^2}\left({1\over(\xi_1^2-b^2)(\xi_3^2-a^2)}
    -{1\over(\xi_1^2-a^2)(\xi_3^2-b^2)}\right)\enspace,
   \\
   &M_3={\xi_1^2\xi_2^2\over
         a^2-b^2}\left({1\over(\xi_1^2-a^2)(\xi_2^2-b^2)}
    -{1\over(\xi_1^2-b^2)(\xi_2^2-a^2)}\right)\enspace.
   \endaligned\qquad\qquad\right\}
  \tag\NUM.\num\endalign$$
and obtain the following path integral identity
\minus$$\myalign
  &\prod_{i=1}^3\int\limits_{\xi_i(t')=\xi_i'}^{\xi_i(t'')=\xi_i''}
   h_i\CD\xi_i(t)
   \exp\left\{\ih\int_{t'}^{t''}\left[{m\over2}\sum_{i=1}^3
      h_i^2\dot\xi_i^2-\Delta V_{PF}(\{\xi\})\right]dt\right\}
  \\   &
  =\prod_{i=1}^3\int\limits_{\xi_i(t')=\xi_i'}^{\xi_i(t'')=\xi_i''}
   M_i^{-1/2}\CD\xi_i(s)
   \exp\left\{\ih\int_{t'}^{t''}\left[{m\over2}S\sum_{i=1}^3
   {\dot\xi_i^2\over M_i}-\Delta V_{PF}(\{\xi\})\right]dt\right\}
  \\   &
  =(S'S'')^{-1/4}\int_{\bbbr}{dE\over2\pi\hbar}
  \e^{-\i ET/\hbar}\int_0^\infty ds''
   \prod_{i=1}^3\int\limits_{\xi_i(0)=\xi_i'}^{\xi_i(s'')=\xi_i''}
   M_i^{-1/2}\CD\xi_i(s)
  \\   &\qquad\qquad\times
   \exp\left\{\ih\int_{0}^{s''}\left[{m\over2}\sum_{i=1}^3
   {\dot\xi_i^2\over M_i}+ES
    -{\hbar^2\over8m}\sum_{i=1}^3M_i\Big(\Gamma_i^2+2\Gamma_i'\Big)
    \right]ds\right\}\qquad\qquad
  \tag\NUM.\num\\   \global\plus
       &
  =\int_0^\infty dp\sum_{\kappa,\lambda}
   \Psi_{\kappa,\lambda,p}^*(\xi_1',\xi_2',\xi_3')
   \Psi_{\kappa,\lambda,p}(\xi_1'',\xi_2'',\xi_3'')
   \e^{-\i\hbar p^2T/2m}
  \tag\NUM.\num\\   \global\plus
       &
  =\bigg({m\over2\pi\i\hbar T}\bigg)^{3/2}
    \exp\bigg[-{m\over2\i\hbar T}d^2_\edrei(\vec q'',\vec q')\bigg]
  \enspace,
  \tag\NUM.\num\endalign$$
and $d_\edrei(\vec q'',\vec q')$ must be expressed in paraboloidal
coordinates via $d_\edrei=\vert \vec x(\{\xi''\}-\vec x(\{\xi'\})\vert$.
The functions $\Psi_{\kappa,\lambda,p}(\xi_1,\xi_2,\xi_3)=
A_{\kappa,\lambda,p}(\xi_1)B_{\kappa,\lambda,p}(\xi_2)C_{\kappa,\lambda,
p}(\xi_3)$ are solutions of a three-fold Lam\'e equation with
separation parameters $\lambda,\kappa,p$ [\MF, p.1305]. A potential
separable in ellipsoidal coordinates has the form of (\numek).
\goodbreak

\subsection{The Sphere $\sdrei$}
\ssf
\subsubsection{General Form of the Propagator and the Green Function}
We consider the sphere $\sdrei$ as the other space of constant positive
curvature. There exist six coordinate systems which admit separation of
variables on $\sdrei$ which will be discussed in the following. The
propagator and the Green function on $\szwei$ are best calculated in
terms of polar coordinates. One obtains [\BJd, \GRSh, \SCHUa]
$$\myalign
  K^\sdrei\Big(\psi_\sdrei(\vec q'',\vec q');T\Big)
  &={\e^{\i\hbar T/2m}\over4\pi^2}
  {\d\over\d\cos\psi_\sdrei}
  \theta_3\left({\psi_\sdrei(\vec q'',\vec q')\over2}\bigg\vert
  -{\hbar T\over2\pi m}\right)\enspace,
  \tag\NUM.\num\\   \global\plus
  G^\sdrei\Big(\psi_\sdrei(\vec q'',\vec q');E\Big)
  &={m\over2\pi\hbar^2}
  {\sin\big[\big(\pi-\psi_\sdrei(\vec q'',\vec q')\big)(a+\half)\big]
  \over\sin[\pi(a+\half)]\sin\psi_\sdrei(\vec q'',\vec q')}
  \enspace.
  \tag\NUM.\num\endalign$$
where $a=-\half+\sqrt{2mE/\hbar^2+1}$, and $\theta_3(z\vert\tau)$ is
a Jacobi theta-function [\GRA, p.931]
\plus$$\theta_3(u,q)=\theta_3(u\vert\tau)
  =1+2\sum_{n=1}^\infty q^{n^2}\cos(2\pi nu)\enspace,
  \qquad(q=\e^{\i\pi\tau})\enspace.
  \tag\NUM.\num$$

\subsubsection{Spherical Cylinder Coordinates}
We first consider the coordinate system
\plus$$\left.\alignedat 3
  s_0&=\cos\theta\cos\phi_1\enspace,
     &\qquad
     &0<\theta<\pi/2\enspace,
  \\
  s_1&=\cos\theta\sin\phi_1\enspace,
     &\qquad
     &\qquad
  \\
  s_2&=\sin\theta\cos\phi_2\enspace,
     &\qquad
     &0\leq\phi_{1,2}<2\pi\enspace,
  \\
  s_3&=\sin\theta\sin\phi_2\enspace.
     &\qquad
     &\qquad
  \endalignedat\qquad\qquad\right\}
  \tag\NUM.\num$$
The metric reads $(g_{ab})=\diag(1,\cos^2\theta,\sin^2\theta)$,
and the invariant distance is given by
\plus$$
  \cos\psi_\sdrei(\vec q'',\vec q')
  =\cos\theta'\cos\theta''\cos(\phi_1''-\phi_1')
  +\sin\theta'\sin\theta''\cos(\phi_2''-\phi_2')\enspace.
  \tag\NUM.\num$$
The momentum operators are
\plus$$
  p_\theta=\hi\bigg({\partial\over\partial\theta}
             +\cot\theta-\tan\theta\bigg)\enspace,
  \tag\NUM.\num$$
and $p_{\phi_{1,2}}=-\i\hbar\partial_{\phi_{1,2}}$.
Therefore for the Hamiltonian
\plus$$\myalign
  -\hbarm\Delta_\sdrei
  &=-\hbarm\bigg[{\partial^2\over\partial\theta^2}
     +2(\cot\theta-\tan\theta){\partial\over\partial\theta}
    +{1\over\cos^2\theta}{\partial^2\over\partial\phi_1^2}
    +{1\over\sin^2\theta}{\partial^2\over\partial\phi_2^2}\bigg]
  \\   &
  ={1\over2m}\bigg(p_\theta^2+{p_{\phi_1}^2\over\cos^2\theta}
                +{p_{\phi_2}^2\over\sin^2\theta}\bigg)
    -{\hbar^2\over8m}\bigg(4+{1\over\cos^2\theta}+{1\over\sin^2\theta}
   \bigg)\enspace.
   \tag\NUM.\num\endalign$$
We obtain for the path integral
\plus$$\myalign
  &\int\limits_{\theta(t')=\theta'}^{\theta(t'')=\theta''}
  \sin\theta\cos\theta\CD\theta(t)
  \int\limits_{\phi_1(t')=\phi_1'}^{\phi_1(t'')=\phi_1''}\CD\phi_1(t)
  \int\limits_{\phi_2(t')=\phi_2'}^{\phi_2(t'')=\phi_2''}\CD\phi_2(t)
  \\   &\qquad\times
  \exp\left\{\ih\int_{t'}^{t''}\bigg[{m\over2}\Big(
   \dot\theta^2+\cos^2\theta\dot\phi_1^2+\sin^2\theta\dot\phi_2^2\Big)
  +{\hbar^2\over8m}\bigg(4+{1\over\cos^2\theta}
                          +{1\over\sin^2\theta}\bigg)\bigg]dt\right\}
  \\   &
  =(\sin\theta'\sin\theta''\cos\theta'\cos\theta'')^{-1/2}
  \sum_{l,k\in\bbbz}{\e^{\i[l(\phi_1''-\phi_1')+k(\phi_2''-\phi_2')]}
          \over4\pi^2}
  \\   &\qquad\times
   \int\limits_{\theta(t')=\theta'}^{\theta(t'')=\theta''}\CD\theta(t)
  \exp\left\{\ih\int_{t'}^{t''}\bigg[{m\over2}\dot\theta^2
    -{\hbar^2\over2m}\bigg({l^2-\viert\over\cos^2\theta}
          +{k^2-\viert\over\sin^2\theta}\bigg)\bigg]dt
   +{\i\hbar T\over2m}\right\}
  \\   &
  =\sum_{l,k\in\bbbz}{\e^{\i[l(\phi_1''-\phi_1')+k(\phi_2''-\phi_2')]}
          \over4\pi^2}
  \sum_{n=0}^\infty2(\vert l\vert!+\vert k\vert!+2n+1)
     {n!\Gamma(\vert l\vert +\vert k\vert +n+1)\over
      \Gamma(\vert l\vert +n+1)\Gamma(\vert k\vert +n+1)}
  \\   &\qquad\times
  (\sin\theta'\sin\theta'')^{\vert k\vert}
  (\cos\theta'\cos\theta'')^{\vert l\vert}
  P_n^{(\vert k\vert ,\vert l\vert )}(1-2\sin^2\theta')
  P_n^{(\vert k\vert ,\vert l\vert )}(1-2\sin^2\theta'')
  \\   &\qquad\times
  \exp\bigg\{-{\i\hbar T\over2m}\Big[(2n+\vert l\vert+\vert k\vert+1)^2
          -1\Big]\bigg\}\enspace.
  \tag\NUM.\num\endalign$$

\subsubsection{Sphero-Conical Coordinates}
We consider the  coordinate system
\plus$$\left.\alignedat 3
  s_0&=\cos\theta\enspace,
     &\quad      &-\pi/2<\theta<\pi/2\enspace,
  \\
  s_1&=\sin\theta\sn(\mu,k)\dn(\nu,k')\enspace,
     &\quad      &k^2+{k'}^2=1\enspace,
  \\
  s_2&=\sin\theta\cn(\mu,k)\cn(\nu,k')\enspace,
     &\quad      &\quad
  \\
  s_3&=\sin\theta\dn(\mu,k)\sn(\nu,k')\enspace,
     &\quad      &\quad
  \endalignedat\qquad\qquad\right\}
  \tag\NUM.\num$$
in the notation of 4.2.3. The metric has the form $(g_{ab})=\diag[1,\sin
^2\theta(k^2\cn^2\mu+{k'}^2\cn^2\nu),\sin^2\theta(k^2\cn^2\mu+{k'}^2\cn^
2\nu)]$ and
\plus$$\multline
  \cos\psi_\sdrei(\vec q'',\vec q')
  =\cos\theta'\cos\theta''
  -\sin\theta'\sin\theta''
  \\   \times
   \Big[
   \sn\mu''\sn\mu'\dn\nu''\dn\nu'
   +\cn\mu''\cn\mu'\cn\nu''\cn\nu'
   +\dn\mu''\dn\mu'\sn\nu''\sn\nu'\Big]\enspace.
  \endmultline
  \tag\NUM.\num$$
For the momentum operators we have
\plus$$
  p_\theta=\hi\bigg({\partial\over\partial\theta}+\cot\theta\bigg)
  \enspace,
  \tag\NUM.\num$$
\edef\numej{\NUM.\num}%
together with $p_\mu$ and $p_\nu$ as in (\numea). We have for the
Hamiltonian
\plus$$\myalign
  &-\hbarm\Delta_\sdrei
  =-\hbarm\bigg[{\partial^2\over\partial\theta^2}
  -2\cot\theta{\partial\over\partial\theta}
  +{1\over\sin^2\theta(k^2\cn^2\mu+{k'}^2\cn^2\nu)}
  \bigg({\partial^2\over\partial\mu^2}
       +{\partial^2\over\partial\nu^2}\bigg)\bigg]
  \\   &
  ={1\over2m}\bigg(p_\theta^2+
   {1\over\sqrt{k^2\cn^2\mu+{k'}^2\cn^2\nu}}
   {p_\mu^2+p_\nu^2\over\sin^2\theta}
   {1\over\sqrt{k^2\cn^2\mu+{k'}^2\cn^2\nu}}\bigg)
  -{\hbar^2\over2m}\enspace.
  \tag\NUM.\num\endalign$$
The path integral can therefore be written down yielding
\plus$$\myalign
  &\int\limits_{\theta(t')=\theta'}^{\theta(t'')=\theta''}
  \sin^2\theta\CD\theta(t)
  \int\limits_{\mu(t')=\mu'}^{\mu(t'')=\mu''}\CD\mu(t)
  \int\limits_{\nu(t')=\nu'}^{\nu(t'')=\nu''}\CD\nu(t)
  (k^2\cn^2\mu+{k'}^2\cn^2\nu)
  \\   &\quad\times
  \exp\Bigg\{\ih\int^{t''}_{t'}\bigg[{m\over2}\Big(\dot\theta^2
  +\sin^2\theta(k^2\cn^2\mu+{k'}^2\cn^2\nu)(\dot\mu^2+\dot\nu^2)\Big)
  +{\hbar^2\over2m}\bigg]dt\Bigg\}
  \\   &
  =(\sin\theta'\sin\theta'')^{-1/2}\e^{-\i\hbar T/2m}
   \sum_\kappa\sum_{l=0}^\infty
   A_{l,\kappa}^*(\mu')B_{l,\kappa}^*(\nu')
   A_{l,\kappa}(\mu'')B_{l,\kappa}(\nu'')
  \\   &\quad\times
   \int\limits_{\theta(t')=\theta'}^{\theta(t'')=\theta''}\CD\theta(t)
  \exp\Bigg[\ih\int^{t''}_{t'}\bigg({m\over2}\dot\theta^2
  -{\hbar^2\over2m}{(l+\half)^2-\viert\over\cos^2\theta}\bigg)dt\Bigg]
  \\   &
  =\sum_\kappa\sum_{l=0}^\infty
   A_{l,\kappa}^*(\mu')B_{l,\kappa}^*(\nu')
   A_{l,\kappa}(\mu'')B_{l,\kappa}(\nu'')
  \\   &\quad\times
  \sum_{N=0}^\infty(N+1){\Gamma(N+l+2)\over(N-l)!}
  P_{N+1/2}^{-l-1/2}(\sin\theta'')P_{N+1/2}^{-l-1/2}(\sin\theta')
  \exp\bigg[-{\i\hbar T\over2m}N(N+2)\bigg]\enspace,
  \\   &
  \tag\NUM.\num\endalign$$
in the notation of 4.2.3, and we have set $N=n+l$, $n\in\bbbn_0$.

\subsubsection{Spherical Coordinates}
We consider the usual three-dimensional polar coordinates
\plus$$\left.\alignedat 3
  s_0&=\cos\theta_1\enspace,
     &\qquad
     &0<\theta_{1,2}<\pi\enspace,
  \\
  s_1&=\sin\theta_1\sin\theta_2\sin\phi\enspace,
     &\qquad
     &0\leq\phi<2\pi\enspace,
  \\
  s_2&=\sin\theta_1\sin\theta_2\cos\phi\enspace,
     &\qquad
     &\qquad
  \\
  s_3&=\sin\theta_1\cos\theta_2\enspace.
     &\qquad
     &\qquad
  \endalignedat\qquad\qquad\right\}
  \tag\NUM.\num$$
Here we have $(g_{ab})=\diag(1,\sin^2\theta_1,\sin^2\theta_1\sin^2\theta
_2)$ and
\plus$$
  \cos\psi_\sdrei(\vec q'',\vec q')=\cos\theta_1'\cos\theta_1''
  +\sin\theta_1'\sin\theta_1''\Big(\cos\theta_2'\cos\theta_2''
  +\sin\theta_2'\sin\theta_2''\cos(\phi''-\phi')\Big)\enspace.
  \tag\NUM.\num$$
For the momentum operators we have (\numej) for $p_{\theta_1}$, and for
$p_{\theta_2}$ and $p_\phi$ as in (\numed). For the Hamiltonian we
obtain
\plus$$\myalign
  &-\hbarm\Delta_\sdrei
  =\hbarm
   \bigg[{\partial^2\over\partial\theta_1}
         +2\cot\theta_1{\partial\over\partial\theta_1}
  +{1\over\sin^2\theta_1}
   \bigg({\partial^2\over\partial\theta_2}
         +\cot\theta_2{\partial\over\partial\theta_2}
    +{1\over\sin^2\theta_2}{\partial^2\over\partial\phi}\bigg)\bigg]
  \\   &
  ={1\over2m}\bigg(p_{\theta_1}^2+{p_\theta^2\over\sin^2\theta_1}
   +{p_\phi^2\over\sin^2\theta_1\sin^2\theta_2}\bigg)
    -{\hbar^2\over8m}\bigg(4+{1\over\sin^2\theta_1}
                          +{1\over\sin^2\theta_1\sin^2\theta_2}\bigg)
  \enspace.\qquad
  \tag\NUM.\num\endalign$$
This gives the path integral formulation
$$\myalign
  &\int\limits_{\theta_1(t')=\theta_1'}^{\theta_1(t'')=\theta_1''}
  \sin^2\theta_1\CD\theta_1(t)
   \int\limits_{\theta_2(t')=\theta_2'}^{\theta_2(t'')=\theta_2''}
  \sin\theta_2\CD\theta_2(t)
  \int\limits_{\phi(t')=\phi'}^{\phi(t'')=\phi''}\CD\phi(t)
  \\   &\qquad\qquad\times
  \exp\Bigg\{\ih\int_{t'}^{t''}\bigg[{m\over2}\bigg(
   \dot\theta_1^2+\sin^2\theta_1\dot\theta_2^2
    +\sin^2\theta_1\sin^2\theta_2\dot\phi^2\bigg)
  \\   &\qquad\qquad\qquad\qquad\qquad\qquad\qquad\qquad\qquad
  +{\hbar^2\over8m}\bigg(4+{1\over\sin^2\theta}
            +{1\over\sin^2\theta_1\sin^2\theta_2}\bigg)\bigg]dt\Bigg\}
  \\   &
  ={1\over2\pi^2}\sum_{l=0}^\infty(l+1)
   C_l^1\big(\cos\psi_\sdrei(\vec q'',\vec q')\big)
  \Energysdrei
  \tag\NUM.\num\\   \global\plus
       &
  =\sum_{m_1,m_2}\sum_{l=0}^\infty
  \Psi^{\sdrei\,*}_{l,m_1,m_2}(\theta_1',\theta_2',\phi')
  \Psi^\sdrei_{l,m_1,m_2}(\theta_1'',\theta_2'',\phi'')
  \Energysdrei\enspace,
  \tag\NUM.\num\endalign$$
where the wave-functions are given by [\EMOTa, p.240]
\plus$$\myalign
  &\Psi^\sdrei_{l,m_1,m_2}(\theta_1,\theta_2,\phi)
  =N_\sdrei^{-1/2}
    \e^{\i m_2\phi}(\sin\theta_1)^{m_1}(\sin\theta_2)^{m_2}
    C_{l-m_1}^{m_1+2}(\cos\theta_1)
    C_{m_1-m_2}^{m_2+{3\over2}}(\cos\theta_2),
  \\   &
  \tag\NUM.\num a\\
  &N_\sdrei={2\pi^32^{-1-2m_1-2m_2}\over
            (l+1)(m_1+{3\over2}(l-m_1)!(m_1-m_2)!}
            {\Gamma(l+m_1+2)\Gamma(m_1+m_2+1)\over
            \Gamma^2(m_1+1)\Gamma^2(m_2+{3\over2})}\enspace.
  \\   &
  \tag\NUM.\num b\endalign$$

\subsubsection{Twofold Confocal Ellipsoidal Coordinates 1)}
We consider the two-fold confocal ellipsoidal coordinates defined by
\plus$$
  {s_1^2+s_3^2\over\rho_i-b}
  +{s_2^2\over\rho_i-a}+{s_0^3\over\rho_i-c}=0\enspace,\qquad
  (i=1,2)\enspace,
  \tag\NUM.\num$$
where $c<\rho_2<b<\rho_1<a$. The corresponding metric reads
\plus$$
  {ds^2\over dt^2}=-\viert(\rho_1-\rho_2)
  \bigg({\dot\rho_1^2\over P_3(\rho_1)}
  -{\dot\rho_2^2\over P_3(\rho_2)}\bigg)
  -(\rho_1-b)(\rho_2-b)\dot\rho_3^2\enspace,
  \tag\NUM.\num$$
where $P_3(\rho)=(\rho-a)(\rho-b)(\rho-c)$. Using the notations of the
sphero-conical coordinates we can identify
\plus$$\left.\alignedat 3
  s_0&=\sn(\mu,k)\dn(\nu,k')\enspace,
     &\quad      &k^2+{k'}^2=1\enspace,
  \\
  s_1&=\dn(\mu,k)\sn(\nu,k')\cos\phi\enspace,
     &\quad      &0\leq\phi<2\pi\enspace,
  \\
  s_2&=\dn(\mu,k)\sn(\nu,k')\sin\phi\enspace,
     &\quad      &\quad
  \\
  s_3&=\cn(\mu,k)\cn(\nu,k')\enspace.
     &\quad      &\quad
  \endalignedat\qquad\qquad\right\}
  \tag\NUM.\num$$
We omit details in the following. Obviously, these coordinates
correspond to sphero-conical coordinates with an additional
circular coordinate. This gives the path integral formulation
\plus$$\myalign
  &\int\limits_{\mu(t')=\mu'}^{\mu(t'')=\mu''}\CD\mu(t)
  \int\limits_{\nu(t')=\nu'}^{\nu(t'')=\nu''}\CD\nu(t)
  (k^2\cn^2\mu+{k'}^2\cn^2\nu)\sqrt{-{k'}^2}\,\dn\mu\sn\mu
  \int\limits_{\phi(t')=\phi'}^{\phi(t'')=\phi''}\CD\phi(t)
  \\   &\ \times
  \exp\Bigg\{\ih\int^{t''}_{t'}\bigg[{m\over2}\Big(
  (k^2\cn^2\mu+{k'}^2\cn^2\nu)(\dot\mu^2+\dot\nu^2)
  -{k'}^2\dn^2\mu\sn^2\nu\dot\phi^2\Big)
  -\Delta V_{PF}\bigg]dt\Bigg\}
  \\   &
  ={\e^{\i\hbar T/2m}\over4\pi^2}{\d\over\d\cos\psi_\sdrei}
  \theta_3\left({\psi_\sdrei(\vec q'',\vec q')\over2}\bigg\vert
  -{\hbar T\over2\pi m}\right)\enspace,
  \tag\NUM.\num\endalign$$
and $\psi_\sdrei(\vec q'',\vec q')$ has to be taken in these
coordinates. Note that the $\phi$-path integration can be explicitly
done, leading to a path integral in $\mu,\nu$ with similar to the
sphero-conical one with additional $1/\dn^2\mu$ and $1/\sn^2\nu$-terms,
respectively.

\subsubsection{Twofold Confocal Ellipsoidal Coordinates 2)}
We consider the two-fold confocal ellipsoidal coordinates defined by
\plus$$
  {s_1^2\over\rho_i-b}
  +{s_2^2+s_3^2\over\rho_i-a}+{s_0^2\over\rho_i-c}=0\enspace,\qquad
  (i=1,2)\enspace,
  \tag\NUM.\num$$
where $c<\rho_2<b<\rho_1<a$. The corresponding metric reads
\plus$$
  {ds^2\over dt^2}=-\viert(\rho_1-\rho_2)
  \bigg({\dot\rho_1^2\over P_3(\rho_1)}
  -{\dot\rho_2^2\over P_3(\rho_2)}\bigg)
  +(\rho_1-a)(\rho_2-a)\dot\rho_3^2\enspace.
  \tag\NUM.\num$$
There is only little difference to the previous case. Proceeding
similarly as before we have
\plus$$\left.\alignedat 3
  s_0&=\sn(\mu,k)\dn(\nu,k')\cos\phi\enspace,
     &\quad      &k^2+{k'}^2=1\enspace,
  \\
  s_1&=\sn(\mu,k)\dn(\nu,k')\sin\phi\enspace,
     &\quad      &0\leq\phi<2\pi\enspace,
  \\
  s_2&=\dn(\mu,k)\sn(\nu,k')\enspace,
     &\quad      &\quad
  \\
  s_3&=\cn(\mu,k)\cn(\nu,k')\enspace.
     &\quad      &\quad
  \endalignedat\qquad\qquad\right\}
  \tag\NUM.\num$$
We omit details in the following. This gives the path integral
formulation
\plus$$\myalign
  &\int\limits_{\mu(t')=\mu'}^{\mu(t'')=\mu''}\CD\mu(t)
  \int\limits_{\nu(t')=\nu'}^{\nu(t'')=\nu''}\CD\nu(t)
  (k^2\cn^2\mu+{k'}^2\cn^2\nu)\sqrt{k^2}\,\sn\mu\dn\mu
  \int\limits_{\phi(t')=\phi'}^{\phi(t'')=\phi''}\CD\phi(t)
  \\   &\ \times
  \exp\Bigg\{\ih\int^{t''}_{t'}\bigg[{m\over2}\Big(
  (k^2\cn^2\mu+{k'}^2\cn^2\nu)(\dot\mu^2+\dot\nu^2)
  +k^2\sn^2\mu\dn^2\nu\dot\phi^2\Big)
  -\Delta V_{PF}\bigg]dt\Bigg\}
  \\   &
  ={\e^{\i\hbar T/2m}\over4\pi^2}{\d\over\d\cos\psi_\sdrei}
  \theta_3\left({\psi_\sdrei(\vec q'',\vec q')\over2}\bigg\vert
  -{\hbar T\over2\pi m}\right)\enspace,
  \tag\NUM.\num\endalign$$
and $\psi_\sdrei(\vec q'',\vec q')$ has to be taken in these
coordinates. The difference of the last two coordinate system lies in
the orientation with respect to an axis.

\subsubsection{Confocal Ellipsoidal Coordinates}
Let us consider the coordinate system defined by
\plus$$
  {s_0^2\over\rho_i-d}+{s_1^2\over\rho_i-c}
  +{s_2^2\over\rho_i-a}+{s_3^2\over\rho_i-b}
  =0\enspace,\qquad(i=1,2,3)\enspace,
  \tag\NUM.\num$$
where $d<\rho_3<c<\rho_2<b<\rho_1<a$ (c.f.~Section 3.1). Explicitly
\plus$$\left.\aligned
  s_0^2&={(\rho_1-d)(\rho_2-d)(\rho_3-d)\over(a-d)(b-d)(c-d)}\enspace,
  \quad
  s_1^2={(\rho_1-c)(\rho_2-c)(\rho_3-c)\over(a-c)(b-c)(d-c)}\enspace,
  \\
  s_2^2&={(\rho_1-a)(\rho_2-a)(\rho_3-a)\over(d-a)(c-a)(b-a)}\enspace,
  \quad
  s_3^2={(\rho_1-b)(\rho_2-b)(\rho_3-b)\over(d-b)(c-b)(a-b)}\enspace.
  \endaligned\qquad\qquad\right\}
  \tag\NUM.\num$$
The corresponding line element is given by
\plus$$
  {ds^2\over dt^2}=-\viert\left[
   {(\rho_1-\rho_2)(\rho_1-\rho_3)\over P_4(\rho_1)}\dot\rho_1^2
  +{(\rho_2-\rho_3)(\rho_2-\rho_1)\over P_4(\rho_2)}\dot\rho_2^2
  +{(\rho_3-\rho_1)(\rho_3-\rho_2)\over P_4(\rho_3)}\dot\rho_3^2
  \right]\enspace,
  \tag\NUM.\num$$
where $P_4(\rho)=(\rho-a)(\rho-b)(\rho-c)(\rho-d)$.
We apply the separation formula (\numca). We identify
$(g_{ab})\equiv\diag(h_1^2,h_2^2,h_3^2)$, furthermore
\minus$$\align
  &\Gamma_i={\rho_i P_4'(\rho_i)\over P_4(\rho)}\enspace,
  \qquad(i=1,2,3)
  \tag\NUM.\num\\   \global\plus
  &S={(\rho_1-\rho_2)(\rho_1-\rho_3)(\rho_2-\rho_3)\over
    4P_4(\rho_1)P_4(\rho_2)P_4(\rho_2)}\enspace,
  \tag\NUM.\num\\   \global\plus
  &M_1={\rho_2-\rho_3\over P_4(\rho_2)P_4(\rho_3)}\enspace,
  \quad
   M_2={\rho_3-\rho_1\over P_4(\rho_1)P_4(\rho_3)}\enspace,
  \quad
   M_3={\rho_1-\rho_2\over P_4(\rho_1)P_4(\rho_2)}\enspace,
  \tag\NUM.\num\endalign$$
and obtain the following path integral identity
$$\myalign
  &\prod_{i=1}^3\int\limits_{\rho_i(t')=\rho_i'}^{\rho_i(t'')=\rho_i''}
  h_i\CD\rho_i(t)\exp\left\{\ih\int_{t'}^{t''}\left[{m\over2}
  \sum_{i=1}^3h_i^2\dot\rho_i^2-\Delta V_{PF}(\{\rho\})\right]dt\right\}
  \\   &
  =(S'S'')^{-1/4}\int_{-\infty}^\infty{dE\over2\pi\hbar}
   \e^{-\i ET/\hbar}\int_0^\infty ds''
  \prod_{i=1}^3\int\limits_{\rho_i(0)=\rho_i'}^{\rho_i(s'')=\rho_i''}
   M_i^{-1/2}\CD\rho_i(s)
  \\   &\qquad\qquad\times
   \exp\left\{\ih\int_{0}^{s''}\left[{m\over2}\sum_{i=1}^3
   {\dot\rho_i^2\over M_i}+ES
    -{\hbar^2\over8m}\sum_{i=1}^3M_i\Big(\Gamma_i^2+2\Gamma_i'\Big)
    \right]ds\right\}\qquad\qquad
  \tag\NUM.\num\\   \global\plus
       &
  ={\e^{\i\hbar T/2m}\over4\pi^2}{\d\over\d\cos\psi_\sdrei}
  \theta_3\left({\psi_\sdrei(\vec q'',\vec q')\over2}\bigg\vert
  -{\hbar T\over2\pi m}\right)\enspace,
  \tag\NUM.\num\endalign$$
and $\psi_\sdrei(\vec q'',\vec q')$ must be expressed in confocal
ellipsoidal coordinates. Actually, ``The linear element of \the\secno.
\the\subno.\the\subsubno.5 and \the\secno.\the\subno.\the\subsubno.6
can be considered as degenerations of \the\secno.\the\subno.\the
\subsubno.7 when $b=c$ and $a=b$, respectively\dots'' [\OLE].

\subsection{The Pseudosphere $\ldrei$}
\ssf
\subsubsection{General Form of the Propagator and the Green Function}
The propagator, respectively the Green function, on three-dimensional
hyperbolic space, i.e.\ the three-dimensional pseudosphere
$\ldrei$ are given by [\GROn, \GRSc, \GUTc]
$$\myalign
  K^\ldrei\Big(d_\ldrei(\vec q'',\vec q');T\Big)
   &=\bigg({m\over2\pi\i\hbar T}\bigg)^{3\over2}
   {d_\ldrei(\vec q'',\vec q')\over\sinh d_\ldrei(\vec q'',\vec q')}
   \exp\bigg[{\i m\over2\hbar T}d_\ldrei^2(\vec q'',\vec q')
                                           -{\i\hbar T\over2m}\bigg]
  \\   &
  \tag\NUM.\num\\   \global\plus
  G^\ldrei\Big(d_\ldrei(\vec q'',\vec q');E\Big)
  &={-m\over\pi^2\hbar^2\sinh d_\ldrei(\vec q'',\vec q')}
  \\   &\qquad\times
  \CQ_{-\i\sqrt{2mE/\hbar^2-1}-1/2}^{1/2}
        \big(\cosh d_\ldrei(\vec q'',\vec q')\big)\enspace.
  \tag\NUM.\num\endalign$$
where $\vec q$ can denote any of the 34 coordinate systems which
separate on $\ldrei$.

\subsubsection{Spherical Coordinates}
We consider the coordinate system
\plus$$\left.\alignedat 3
  u_0&=\cosh\tau_1\cosh\tau_2\enspace,
     &\quad      &\tau_1,\tau_2\in\bbbr\enspace,
  \\
  u_1&=\cosh\tau_1\sinh\tau_2\enspace,
     &\quad      &\quad
  \\
  u_2&=\sinh\tau_1\sin\phi\enspace,
     &\quad      &0\leq\phi<2\pi\enspace,
  \\
  u_3&=\sinh\tau_1\cos\phi\enspace.
     &\quad      &\quad
  \endalignedat\qquad\qquad\right\}
  \tag\NUM.\num$$
The metric tensor is $(g_{ab})=\diag(1,\cosh^2\tau_1,\sinh^2\tau_1)$.
For the momentum operators we have
\plus$$
  p_{\tau_1}=\hi\bigg({\partial\over\partial\tau_1}
       +\half\coth\tau_1+\half\tanh\tau_1\bigg)\enspace,\qquad
  p_{\tau_2}=\hi{\partial\over\partial\tau_2}\enspace,\qquad
  p_\phi=\hi{\partial\over\partial\phi}\enspace.
  \tag\NUM.\num$$
The hyperbolic distance is given by
\plus$$\cosh d_\ldrei\vec q'',\vec q')
  =\cosh\tau_1'\cosh\tau_1''\cosh(\tau_2''-\tau_2')
   -\sinh\tau_1'\sinh\tau_1'\cos(\phi''-\phi')\enspace.
  \tag\NUM.\num$$
For the Hamiltonian we have
\plus$$\myalign
  -\hbarm\Delta_\ldrei&=
  -\hbarm\bigg[{\partial^2\over\partial\tau_1^2}
   +\Big(\tanh\tau_1+\coth\tau_1\Big){\partial\over\partial\tau_1}
   +{1\over\cosh^2\tau_1}{\partial^2\over\partial\tau_2^2}
    +{1\over\sinh^2\tau_1}{\partial^2\over\partial\phi^2}\bigg]
  \\   &
  ={1\over2m}\bigg(p_{\tau_1}^2+{1\over\cosh^2\tau_1}p_{\tau_2}^2
     +{1\over\sinh^2\tau_1}p_\phi^2\bigg)
  +{\hbar^2\over8m}\bigg(4+{1\over\cosh^2\tau_1}
        -{1\over\sinh^2\tau_1}\bigg)\enspace.
  \\   &
  \tag\NUM.\num\endalign$$
We obtain the path integral formulation
\plus$$\myalign
       &
  \int\limits_{\tau_1(t')=\tau_1'}^{\tau_1(t'')=\tau_1''}
  \cosh\tau_1\sinh\tau_1\CD\tau_1(t)
  \int\limits_{\tau_2(t')=\tau_2'}^{\tau_2(t'')=\tau_2''}\CD\tau_2(t)
  \int\limits_{\phi(t')=\phi'}^{\phi(t'')=\phi''}\CD\phi(t)
  \\   &\qquad\times
  \exp\Bigg\{\ih\int_{t'}^{t''}\bigg[{m\over2}
     \Big(\dot\tau_1^2+\cosh^2\tau_1\dot\tau_2^2
          +\sinh^2\tau_1\dot\phi^2\Big)
  \\   &\qquad\qquad\qquad\qquad\qquad\qquad\qquad\qquad\qquad
    -{\hbar^2\over8m}\bigg(4+{1\over\cosh^2\tau_1}
     -{1\over\sinh^2\tau_1}\bigg)\bigg]dt\Bigg\}
  \\   &
  =\Big(\sinh\tau_1'\sinh\tau_1''
        \cosh\tau_1'\cosh\tau_1''\Big)^{-1/2}\e^{-\i\hbar T/2m}
  \sum_{l\in\bbbz}{\e^{\i l(\phi''-\phi')}\over2\pi}
  \int_{\bbbr}{dp_1\over2\pi}\e^{\i p_1(\tau_2''-\tau_2')}
  \\   &\qquad\times
  \int\limits_{\tau_1(t')=\tau_1'}^{\tau_1(t'')=\tau_1''}\CD\tau_1(t)
  \exp\Bigg\{\ih\int_{t'}^{t''}\bigg[{m\over2}
     \dot\tau_1^2
     -\hbarm\bigg({p_1^2+\viert\over\cosh^2\tau_1}
     +{l^2-\viert\over\sinh^2\tau_1}\bigg)\bigg]dt\Bigg\}
  \\   &
  =\Big(\sinh\tau_1'\sinh\tau_1''
        \cosh\tau_1'\cosh\tau_1''\Big)^{-1/2}
  \sum_{l\in\bbbz}{\e^{\i l(\phi''-\phi')}\over2\pi}
  \int_{\bbbr}{dp_1\over2\pi}\e^{\i p_1(\tau_2''-\tau_2')}
  \\   &\qquad\times \int_0^\infty dp\,
  \Psi_p^{(p_1,l)\,*}(\tau_1')\Psi_p^{(p_1,l)}(\tau_1'')
  \Energyldrei\enspace.
  \tag\NUM.\num\endalign$$
In the path integration, the cylindrical $\phi$- and free particle
$\tau_2$-path integration have been separated straightforwardly. In the
last path integral the path integral solutions for modified
P\"oschl-Teller potential with $\nu=p_1$, $\eta=\vert l\vert$ have
been used. The wave-functions $\Psi_p^{(p_1,l)}(\tau_1)$ are
given by
\plus$$\myalign
  \Psi_p^{(p_1,l)}(\tau_1)
  &=N_p^{(p_1,l)}
    (\cosh\tau_1)^{\i p_1-1/2}
    (\sinh\tau_1)^{\vert l\vert-1/2}
  \\   &\qquad\times
    {_2}F_1\bigg[\half(\vert l\vert+\i p_1-\i p+1),
                 \half(\vert l\vert+\i p_1+\i p+1);\vert l\vert+1;
    -\sinh^2\tau_1\bigg]\enspace,
  \\   &
  \tag\NUM.\num a\\
  N_p^{(p_1,l)}
  &={1\over \vert l\vert !}\sqrt{p\sinh\pi p\over2\pi^2}\,
    \Gamma\bigg({1+\vert l\vert+\i p_1+\i p\over2}\bigg)
    \Gamma\bigg({1+\vert l\vert+\i p_1-\i p\over2}\bigg)\enspace.
  \tag\NUM.\num b\endalign$$

\subsubsection{Horicyclic Coordinates}
We consider the coordinate system
\plus$$\left.\alignedat 3
  u_0&=\half\left({1\over y}+y+{x_1^2+x_2^2\over y}\right)\enspace,
     &\quad      &y>0\enspace,
  \\
  u_1&={x_1\over y}\enspace,\quad
  u_2={x_2\over y}\enspace,
     &\quad      &(x_1,x_2)=\vec x\in\bbbr^2\enspace,
  \\
  u_3&=\half\left({1\over y}-y-{x_1^2+x_2^2\over y}\right)\enspace,
     &\quad      &\quad
  \endalignedat\qquad\qquad\right\}
  \tag\NUM.\num$$
The metric tensor is $(g_{ab})=\bbbone/y^2$. For the momentum operators
we have $p_{x_{1,2}}=-\i\hbar\partial_{x_{1,2}}$ and
\plus$$
  p_y=\hi\bigg({\partial\over\partial y}-{3\over2y}\bigg)
 \enspace.
  \tag\NUM.\num$$
\edef\numeg{\NUM.\num}%
The hyperbolic distance is given by
\plus$$
  \cosh d_\ldrei(\vec q'',\vec q')
  ={{y'}^2+{y''}^2+(x_1''-x_1')^2+(x_2''-x_2')^2\over2y'y''}
  \enspace.
  \tag\NUM.\num$$
For the Hamiltonian we have
\plus$$\myalign
  -\hbarm\Delta_\ldrei&=
  -\hbarm y^2\bigg({\partial^2\over\partial y^2}
   -{1\over y}{\partial\over\partial y}
   +{\partial^2\over\partial x_1^2}
   +{\partial^2\over\partial x_2^2}\bigg)
  \\   &
  ={1\over2m}\Big(yp_y^2y+y^2p_{x_1}^2+y^2p_{x_2}^2\Big)
   +{3\hbar^2\over8m}\enspace.
   \tag\NUM.\num\endalign$$
The path integral discussion for this coordinate system and its
$D$-dimensional generalization has been discussed in [\GROn]. The path
integral is solved by exploiting the path integral solution of the
radial harmonic oscillator; we cite the result. We obtain the identity
[\GROn]
\plus$$\myalign
       &
  \int\limits_{y(t')=y'}^{y(t'')=y''}{\CD y(t)\over y^3}
  \int\limits_{x_1(t')=x_1'}^{x_1(t'')=x_1''}\CD x_1(t)
  \int\limits_{x_2(t')=x_2'}^{x_2(t'')=x_2''}\CD x_2(t)
  \exp\Bigg[\ih\int_{t'}^{t''}\bigg({m\over2}
     {{\dot{\vec x}}^2+\dot y^2\over y^2}
     -{3\hbar^2\over 8m}\bigg)dt\Bigg]
  \\   &
  =y'y''\int_{\bbbr^2}{d\vec k\over (2\pi)^2}
  \e^{\i\vec k\cdot(\vec x''-\vec x')}
  \\   &\qquad\times
  {2\over\pi^2}\int_0^\infty dp\,p\sinh\pi p\Energyldrei
  K_{\i p}(\vert\vec k\vert y')K_{\i p}(\vert\vec k\vert y'')
  \enspace.
  \tag\NUM.\num\endalign$$

\subsubsection{Pseudospherical-Conical Coordinates}
We consider the coordinate system
\plus$$\left.\alignedat 3
  u_0&=\cosh\tau
     \enspace,
     &\quad      &\tau>0\enspace,
  \\
  u_1&=\sinh\tau\sn(\mu,k)\dn(\nu,k')\enspace,
     &\quad      &k^2+{k'}^2=1\enspace,
  \\
  u_2&=\sinh\tau\cn(\mu,k)\cn(\nu,k')\enspace,
     &\quad      &\quad
  \\
  u_3&=\sinh\tau\dn(\mu,k)\sn(\nu,k')\enspace,
     &\quad      &\quad
  \endalignedat\qquad\qquad\right\}
  \tag\NUM.\num$$
in the notation of 4.2.3. The hyperbolic distance reads
\plus$$\myalign
  \cosh d_\ldrei(\vec q'',\vec q')
  &=\cosh\tau''\cosh\tau'-\sinh\tau''\sinh\tau'
  \Big(\sn\mu''\sn\mu'\dn\nu''\dn\nu'
  \\   &\qquad
  +\cn\mu''\cn\mu'\cn\nu''\cn\nu'+\dn\mu''\dn\mu'\sn\nu''\sn\nu'
  \Big)\enspace.
  \tag\NUM.\num\endalign$$
The metric tensor $g_{ab}$ is given by
\plus$$
  (g_{ab})=\diag\big[1,\sinh^2\tau(k^2\cn^2\mu+{k'}^2\cn^2\nu),
  \sinh^2\tau(k^2\cn^2\mu+{k'}^2\cn^2\nu)] \enspace,
  \tag\NUM.\num$$
and the momentum operators are
\plus$$
  p_\tau=\hi\bigg({\partial\over\partial\tau}+\coth\tau\bigg)\enspace,
  \tag\NUM.\num$$
\edef\numeh{\NUM.\num}%
and with $p_\mu$, $p_\nu$ as in (\numea). The Hamiltonian has the form
\plus$$\myalign
  &-\hbarm\Delta_\ldrei
  =-\hbarm\bigg[{\partial^2\over\partial\tau^2}
     +2\coth\tau{\partial\over\partial\tau}
     +{1\over\sinh^2\tau}{1\over k^2\cn^2\mu+{k'}^2\cn^2\nu}
  \bigg({\partial^2\over\partial\mu^2}
   +{\partial^2\over\partial\nu^2}\bigg)\bigg]
  \\   &
  ={1\over2m}\bigg[p_\tau^2+{1\over\sinh^2\tau}
   {1\over\sqrt{k^2\cn^2\mu+{k'}^2\cn^2\nu}}
   (p_\mu^2+p_\nu^2){1\over\sqrt{k^2\cn^2\mu+{k'}^2\cn^2\nu}}
   \bigg]+\hbarm\enspace.
  \\   &
  \tag\NUM.\num\endalign$$
The path integral can be written down yielding
\plus$$\myalign
  &
  \int\limits_{\tau(t')=\tau'}^{\tau(t'')=\tau''}\sinh^2\tau\CD\tau(t)
  \int\limits_{\mu(t')=\mu'}^{\mu(t'')=\mu''}\CD\mu(t)
  \int\limits_{\nu(t')=\nu'}^{\nu(t'')=\nu''}\CD\nu(t)
  (k^2\cn^2\mu+{k'}^2\cn^2\nu)
  \\   &\qquad\times
  \exp\Bigg\{{\i m\over2\hbar}\int^{t''}_{t'}\Big[
  \dot\tau^2+\sinh^2\tau(k^2\cn^2\mu+{k'}^2\cn^2\nu)
  (\dot\mu^2+\dot\nu^2)\Big]dt-{\i\hbar T\over2m}\Bigg\}
  \\   &
  =(\sinh\tau'\sinh\tau'')^{-1}\e^{-\i\hbar T/2m}
  \sum_{l=0}^\infty\sum_\kappa
  A_{l,\kappa}^*(\mu')B_{l,\kappa}^*(\nu')
  A_{l,\kappa}(\mu'')B_{l,\kappa}(\nu'')
  \\   &\qquad\times
  \int\limits_{\tau(t')=\tau'}^{\tau(t'')=\tau''}\CD\tau(t)
  \exp\left[\ih\int^{t''}_{t'}\bigg({m\over2}\dot\tau^2
  -\hbarm{(l+\half)^2-\viert\over\sinh^2\tau}\bigg)dt\right]
  \\   &
  =(\sinh\tau'\sinh\tau'')^{-1/2}\sum_{l=0}^\infty\sum_\kappa
  A_{l,\kappa}^*(\mu')B_{l,\kappa}^*(\nu')
  A_{l,\kappa}(\mu'')B_{l,\kappa}(\nu'')
  \\   &\qquad\times
  {1\over\pi}\int_0^\infty dp\,p\sinh\pi p
  \big\vert\Gamma(\i p+l+1)\vert^2 \Energyldrei
  \\   &\qquad\times
  \CP_{\i p-1/2}^{-\half-l}(\cosh\tau')
  \CP_{\i p-1/2}^{-\half-l}(\cosh\tau'')      \enspace.
  \tag\NUM.\num\endalign$$

\subsubsection{Hyperbolic-Pseudo-Conical Coordinates}
\plus$$\left.\alignedat 3
  u_0&=\cosh\tau\nc(\mu,k)\nc(\nu,k')\enspace,
     &\quad      &k^2+{k'}^2=1\enspace,
  \\
  u_1&=\sinh\tau\enspace,
     &\quad      &\tau\in\bbbr\enspace,
  \\
  u_2&=\cosh\tau\dc(\mu,k)\sc(\nu,k')\enspace,
     &\quad      &\quad
  \\
  u_3&=\cosh\tau\sc(\mu,k)\dc(\nu,k')\enspace.
     &\quad      &\quad
  \endalignedat\qquad\qquad\right\}
  \tag\NUM.\num$$
Actually, these coordinates encode six different coordinate systems in a
similar way as discussed in 4.3.5. We omit the details and write down
the corresponding path integral formulation which reads as follows
$$\myalign
 &\int\limits_{\tau(t')=\tau'}^{\tau(t'')=\tau''}\cosh^2\tau\CD\tau(t)
   \int\limits_{\mu(t')=\mu'}^{\mu(t'')=\mu''}\CD\mu(t)
  \int\limits_{\nu(t')=\nu'}^{\nu(t'')=\nu''}\CD\nu(t)
  ({k'}^2\nc^2\mu+k^2\nc^2\nu)
  \\   &\qquad\times
  \exp\Bigg\{{\i m\over2\hbar}\int^{t''}_{t'}\Big[
  \dot\tau^2+\cosh^2\tau({k'}^2\nc^2\mu+k^2\nc^2\nu)
   (\dot\mu^2+\dot\nu^2)\Big]dt-{\i\hbar T\over2m}\Bigg\}
  \\   &
  =(\cosh\tau'\cosh\tau'')^{-1}\e^{-\i\hbar T/2m}
  \sum_\kappa\int_0^\infty dk
  \CA_{k,\kappa}^*(\mu')\CB_{k,\kappa}^*(\nu')
  \CA_{k,\kappa}(\mu'')\CB_{k,\kappa}(\nu'')
  \\   &\qquad\times
  \int\limits_{\tau(t')=\tau'}^{\tau(t'')=\tau''}\CD\tau(t)
  \exp\Bigg[\ih\int_{t'}^{t''}\bigg({m\over2}\dot\tau^2
   -\hbarm{k^2+\viert\over\cosh^2\tau}\bigg)dt\Bigg]
  \\   &
  =(\cosh\tau'\cosh\tau'')^{-1}\sum_\kappa\int_0^\infty dk
  \CA_{k,\kappa}^*(\mu')\CB_{k,\kappa}^*(\nu')
  \CA_{k,\kappa}(\mu'')\CB_{k,\kappa}(\nu'')
  \\   &\qquad\times
  \half\int_{\bbbr}{p\sinh\pi p\,dp\over\cosh^2\pi k+\sinh^2\pi p}
  \energyldrei
  P_{\i k-\half}^{-\i p}(\tanh\tau')P_{\i k-\half}^{\i p}(\tanh\tau'')
  \\   &
  \tag\NUM.\num\\   \global\plus
       &
  =\bigg({m\over2\pi\i\hbar T}\bigg)^{3\over2}
  {d_\ldrei(\vec q'',\vec q')\over\sinh d_\ldrei(\vec q'',\vec q')}
  \exp\bigg[{\i m\over2\hbar T}d_\ldrei^2(\vec q'',\vec q')
                                           -{\i\hbar T\over2m}\bigg]
  \enspace,
  \tag\NUM.\num\endalign$$
where $d_\ldrei(\vec q'',\vec q')$ must be expressed in hyperbolic
pseudo-conical coordinates.

\subsubsection{Pseudospherical Coordinates}
We consider the coordinate system
\plus$$\left.\alignedat 3
  u_0&=\cosh\tau\enspace,
     &\quad      &\tau>0\enspace
     \\
  u_1&=\sinh\tau\sin\theta\sin\phi\enspace,
     &\quad      &0<\theta<\pi\enspace,
     \\
  u_2&=\sinh\tau\sin\theta\cos\phi\enspace,
     &\quad      &0\leq\phi<2\pi\enspace,
     \\
  u_3&=\sinh\tau\cos\theta\enspace.
     &\quad      &\quad
  \endalignedat\qquad\qquad\right\}
  \tag\NUM.\num$$
This is the usual three-dimensional (pseudo-) spherical polar
coordinates system on $\ldrei$. We have $(g_{ab})=\diag(1,\sinh^2\tau,
\sinh^2\tau\sin^2\theta)$ and the invariant distance is given by
\plus$$\multline
  \cosh d_\ldrei(\vec q'',\vec q')
  \\
  =\cosh\tau''\cosh\tau'-\sinh\tau''\sinh\tau'\big[
   \sin\theta''\sin\theta'\cos(\phi''-\phi')+\cos\theta''\cos\theta'
   \big]\enspace.
  \endmultline
  \tag\NUM.\num$$
For the momentum operators we have the operators (\numed,\numeh).
This gives for the Hamiltonian
\hfuzz=5pt
\plus$$\myalign
  -\hbarm\Delta_\ldrei
  &=-\hbarm\bigg[{\partial^2\over\partial\tau^2}
  +2\coth\tau{\partial\over\partial\tau}
  +{1\over\sinh^2\tau}\bigg({\partial^2\over\partial\theta^2}
   +\cot\theta{\partial\over\partial\theta}
    +{1\over\sin^2\theta}{\partial^2\over\partial\phi^2}\bigg)\bigg]
  \\   &
  ={1\over2m}\bigg(p_\tau^2+{p_\theta^2\over\sinh^2\tau}
     +{p_\phi^2\over\sinh^2\tau\sin^2\theta}\bigg)
  +{\hbar^2\over8m}\bigg(4-{1\over\sinh^2\tau}
                -{1\over\sinh^2\tau\sin^2\theta}\bigg)\enspace.
  \\   &
  \tag\NUM.\num\endalign$$
\hfuzz=3pt
In the path integral evaluation one successively uses the path integral
solution related special case of the modified P\"oschl-Teller potential.
Therefore we obtain the path integral formulation (and c.f.\ [\BJb,
\GRSc] for its $D$-dimensional generalization)
\plus$$\myalign
       &
  \int\limits_{\tau(t')=\tau'}^{\tau(t'')=\tau''}\sinh^2\tau\CD\tau(t)
   \int\limits_{\theta(t')=\theta'}^{\theta(t'')=\theta''}
      \sin\theta\CD\theta(t)
  \int\limits_{\phi(t')=\phi'}^{\phi(t'')=\phi''}\CD\phi(t)
  \\   &\qquad\times
  \exp\Bigg\{\ih\int_{t'}^{t''}\bigg[{m\over2}\Big(\dot\tau^2
     +\sinh^2\tau\dot\theta^2+\sinh^2\tau\sin^2\theta\dot\phi^2\Big)
  \\   &\qquad\qquad\qquad\qquad\qquad\qquad\qquad\qquad\qquad
    -{\hbar^2\over8m}\bigg(4-{1\over\sinh^2\tau}
                -{1\over\sinh^2\tau\sin^2\theta}\bigg)\bigg]dt\Bigg\}
  \\   &
  =(\sinh\tau'\sinh\tau'')^{-1}\e^{-\i\hbar T/2m}
  \sum_{l=0}^\infty\sum_{m=-l}^l
  Y_l^{m\,*}(\theta',\phi')Y_l^m(\theta'',\phi'')
  \\   &\qquad\times
  \int\limits_{\tau(t')=\tau'}^{\tau(t'')=\tau''}\CD\tau(t)
  \exp\Bigg[\ih\int_{t'}^{t''}\bigg({m\over2}\dot\tau^2
  -\hbarm{(l+\half)^2-\viert\over\sinh^2\tau}\bigg)dt\Bigg]
  \\   &
  =(\sinh\tau'\sinh\tau'')^{-1/2}\sum_{l=0}^\infty\sum_{m=-l}^l
  Y_l^{m\,*}(\theta',\phi')Y_l^m(\theta'',\phi'')
  \\   &\qquad\times
  {1\over\pi}\int_0^\infty dp\,p\sinh\pi p
  \big\vert\Gamma(\i p+l+1)\vert^2 \Energyldrei
  \\   &\qquad\times
  \CP_{\i p-1/2}^{-\half-l}(\cosh\tau')
  \CP_{\i p-1/2}^{-\half-l}(\cosh\tau'')\enspace.
  \tag\NUM.\num\endalign$$

\subsubsection{Hyperbolic-Spherical Coordinates}
We consider the coordinate system
\plus$$\left.\alignedat 3
  u_0&=\cosh\tau_1\cosh\tau_2\enspace,
     &\quad      &\tau_1\in\bbbr\enspace,
  \\
  u_1&=\sinh\tau_1\enspace,
     &\quad      &\tau_2>0\enspace,
  \\
  u_2&=\cosh\tau_1\sinh\tau_2\sin\phi\enspace,
     &\quad      &0\leq\phi<2\pi\enspace,
  \\
  u_3&=\cosh\tau_1\sinh\tau_2\cos\phi\enspace.
     &\quad      &\quad
  \endalignedat\qquad\qquad\right\}
  \tag\NUM.\num$$
The hyperbolic distance is given by
\plus$$\multline
  \cosh d_\ldrei(\vec q'',\vec q')
  =\cosh\tau_1''\cosh\tau_1'\cosh\tau_2''\cosh\tau_2'
  \\
  -\sinh\tau_1''\sinh\tau_1
  -\cosh\tau_1''\cosh\tau_1'\sinh\tau_2''\sinh\tau_2'
  \cos(\phi''-\phi')\enspace.
  \endmultline
  \tag\NUM.\num$$
Here $(g_{ab})=\diag(1,\cosh^2\tau_1,\cosh^2\tau_1\sinh^2\tau_2)$, and
for the momentum operators we have
\plus$$
  p_{\tau_1}
    =\hi\bigg({\partial\over\partial\tau_1}
         +\tanh\tau_1\bigg)\enspace,
  \tag\NUM.\num$$
\edef\numei{\NUM.\num}%
and $p_{\tau_2},p_\phi$ as in (\numdd). For the Hamiltonian we have
\plus$$\myalign
  &-\hbarm\Delta_\ldrei
  =\hbarm
   \bigg[{\partial^2\over\partial\tau_1}
         +2\tanh\tau_1{\partial\over\partial\tau_1}
  +{1\over\cosh^2\tau_1}
   \bigg({\partial^2\over\partial\tau_2^2}
         +\coth\tau_2{\partial\over\partial\tau_2}
    +{1\over\sinh^2\tau_2}{\partial^2\over\partial\phi}\bigg)\bigg]
  \\   &
  ={1\over2m}\bigg(p_{\tau_1}^2+{p_\tau^2\over\cosh^2\tau_1}
   +{p_\phi^2\over\cosh^2\tau_1\sinh^2\tau_2}\bigg)
  +{\hbar^2\over8m}\bigg(4+{1\over\cosh^2\tau_1}
                          -{1\over\cosh^2\tau_1\sinh^2\tau_2}\bigg)
  \enspace.
  \\   &
  \tag\NUM.\num\endalign$$
Therefore we obtain the path integral formulation
\plus$$\myalign
       &
  \int\limits_{\tau_1(t')=\tau_1'}^{\tau_1(t'')=\tau_1''}
  \cosh^2\tau_1\CD\tau_1(t)
  \int\limits_{\tau_2(t')=\tau_2'}^{\tau_2(t'')=\tau_2''}
  \sinh\tau_2\CD\tau_2(t)
  \int\limits_{\phi(t')=\phi'}^{\phi(t'')=\phi''}\CD\phi(t)
  \\   &\qquad\times
  \exp\Bigg\{\ih\int_{t'}^{t''}\bigg[{m\over2}
     \Big(\dot\tau_1^2+\cosh^2\tau_1\dot\tau_2^2
          +\cosh^2\tau_1\sinh^2\tau_2\dot\phi^2)\Big)
  \\   &\qquad\qquad\qquad\qquad\qquad\qquad\qquad\qquad\qquad
    -{\hbar^2\over8m}\bigg(4+{1\over\cosh^2\tau_1}
     -{1\over\cosh^2\tau_1\sinh^2\tau_2}\bigg)\bigg]dt\Bigg\}
  \\   &
  =\Big(\cosh\tau_1'\cosh\tau_1'')^{-1}
  \sum_{l\in\bbbz}{\e^{\i l(\phi''-\phi')}\over2\pi}
  \\   &\ \times
  {1\over\pi}
  \int_{\bbbr} dp_1\,p_1\sinh\pi p_1
  \Big\vert\Gamma(\bhalf+\i p_1-l)\Big\vert^2
  \CP_{\i p_1-\half}^{-l}(\cosh\tau_2')
  \CP_{\i p_1-\half}^{-l}(\cosh\tau_2'')
  \\   &\ \times
  \half\int_{\bbbr}{p\sinh\pi p\,dp\over
  \cosh^2\pi p_1+\sinh^2\pi p}
  \exp\bigg[-{\i\hbar T\over2m}(p^2+1)\bigg]
  P_{\i p_1-\half}^{-\i p}(\tanh\tau_3')
  P_{\i p_1-\half}^{\i p}(\tanh\tau_3'')
  \enspace.
  \\   &
  \tag\NUM.\num\endalign$$
In the path integral evaluation first the path integral solution of the
simple Manning-Rosen potential, and second of the simple Rosen-Morse
potential was used.

\subsubsection{Equidistant Coordinates}
We consider the coordinate system
\plus$$\left.\alignedat 3
  u_0&=\cosh\tau_1\cosh\tau_2\cosh\tau_3\enspace,
     &\quad      &\tau_1,\tau_2,\tau_3\in\bbbr\enspace,
  \\
  u_1&=\sinh\tau_1\enspace,\quad
  u_2=\cosh\tau_1\sinh\tau_2\enspace,
     &\quad      &\quad
  \\
  u_3&=\cosh\tau_1\cosh\tau_2\sinh\tau_3\enspace.
     &\quad      &\quad
  \endalignedat\qquad\qquad\right\}
  \tag\NUM.\num$$
The hyperbolic distance is given by
\plus$$\myalign
  \cosh d_\ldrei(\vec q'',\vec q')
  &=\cosh\tau_1'\cosh\tau_1''\Big[
   \cosh\tau_2'\cosh\tau_2''
   \\   &\qquad
   +\sinh\tau_2''\sinh\tau_2'\cos(\tau_3''-\tau_3')
   \Big]-\sinh\tau_1'\sinh\tau_1''\enspace.
  \tag\NUM.\num\endalign$$
This coordinate system was called in [\VISM] the hyperbolic,
respectively the Lobachevskian one. Here $(g_{ab})=\diag(1,\cosh^2\tau_1
,\cosh^2\tau_1\cosh^2\tau_2)$, and for the momentum operators we have
for $p_{\tau_1}$ as in (\numei), and for $p_{\tau_2},p_{\tau_3}$ as in
(\numde). For the Hamiltonian we get
\plus$$\myalign
  &-\hbarm\Delta_\ldrei
  =\hbarm
   \bigg[{\partial^2\over\partial\tau_1}
         +2\tanh\tau_1{\partial\over\partial\tau_1}
  +{1\over\cosh^2\tau_1}
   \bigg({\partial^2\over\partial\tau_2}
         +\tanh\tau_2{\partial\over\partial\tau_2}
  +{1\over\cosh^2\tau_2}{\partial^2\over\partial\tau_3}\bigg)\bigg]
  \\   &
  ={1\over2m}\bigg(p_{\tau_1}^2+{p_{\tau_2}^2\over\cosh^2\tau_1}
   +{p_{\tau_3}^2\over\cosh^2\tau_1\cosh^2\tau_2}\bigg)
  +{\hbar^2\over8m}\bigg(4+{1\over\cosh^2\tau_1}
                          +{1\over\cosh^2\tau_1\cosh^2\tau_2}\bigg)
  \enspace.
  \\   &
  \tag\NUM.\num\endalign$$
Therefore we obtain the path integral formulation
\plus$$\myalign
       &
  \int\limits_{\tau_1(t')=\tau_1'}^{\tau_1(t'')=\tau_1''}
  \cosh^2\tau_1\CD\tau_1(t)
  \int\limits_{\tau_2(t')=\tau_2'}^{\tau_2(t'')=\tau_2''}
  \cosh\tau_2\CD\tau_2(t)
  \int\limits_{\tau_3(t')=\tau_3'}^{\tau_3(t'')=\tau_3''}\CD\tau_3(t)
  \\   &\qquad\times
  \exp\Bigg\{\ih\int_{t'}^{t''}\bigg[{m\over2}
     \Big(\dot\tau_1^2+\cosh^2\tau_1\dot\tau_2^2
          +\cosh^2\tau_1\cosh^2\tau_2\dot\tau_3^2\Big)
  \\   &\qquad\qquad\qquad\qquad\qquad\qquad
    -{\hbar^2\over8m}\bigg(4+{1\over\cosh^2\tau_1}
     +{1\over\cosh^2\tau_1\cosh^2\tau_2}\bigg)\bigg]dt\Bigg\}
  \\   &
  =\Big(\cosh^2\tau_1'\cosh^2\tau_1''
        \cosh\tau_2'\cosh\tau_2''\Big)^{-1/2}
  \int_{\bbbr}{dk\over2\pi}\e^{\i k(\tau_3''-\tau_3')}
  \\   &\ \times
  \half\int_{\bbbr}{p_1\sinh\pi p_1\,dp_1\over
  \cosh^2\pi k+\sinh^2\pi p_1}
  P_{\i k-\half}^{-\i p_1}(\tanh\tau_2')
  P_{\i k-\half}^{\i p_1}(\tanh\tau_2'')
  \\   &\ \times
  \half\int_{\bbbr}{p\sinh\pi p\,dp\over
  \cosh^2\pi p_1+\sinh^2\pi p}\energyldrei
  P_{\i p_1-\half}^{-\i p}(\tanh\tau_3')
  P_{\i p_1-\half}^{\i p}(\tanh\tau_3'')
  \enspace.
  \tag\NUM.\num\endalign$$
In the path integral evaluation one successively uses the path integral
solution related special case of the modified P\"oschl-Teller potential.
In Appendix~3 we give the path integral solution of the D-dimensional
generalization of this coordinate system.

\subsubsection{Hyperbolic-Horicyclic Coordinates}
We consider the coordinate system
\plus$$\left.\alignedat 3
  u_0&={\cosh\tau\over2}\left({1\over y}+y+{x^2\over y}\right)\enspace,
     &\quad      &\tau\in\bbbr\enspace,
  \\
  u_1&=\sinh\tau\enspace,\quad
  u_2={x\over y}\cosh\tau\enspace,
     &\quad      &x\in\bbbr\enspace,
  \\
  u_3&={\cosh\tau\over2}\left({1\over y}-y-{x^2\over y}\right)\enspace,
     &\quad      &y>0\enspace.
  \endalignedat\qquad\qquad\right\}
  \tag\NUM.\num$$
The hyperbolic distance is given by
\plus$$
  \cosh d_\ldrei(\vec q'',\vec q')
  ={{y'}^2+{y''}^2+(x''-x')^2\over2y'y''}\cosh\tau'\cosh\tau''
    -\sinh\tau'\sinh\tau''\enspace.
  \tag\NUM.\num$$
Here $(g_{ab})=\diag(1,\cosh^2\tau/y^2,\cosh^2\tau/y^2)$ and for the
momentum operators we have the operators (\numee,\numei), therefore for
the Hamiltonian
\plus$$\myalign
  -\hbarm\Delta_\ldrei
  &=-\hbarm\bigg[{\partial^2\over\partial\tau^2}
  +2\tanh\tau{\partial\over\partial\tau}
  +{y^2\over\cosh^2\tau}\bigg({\partial\over\partial x^2}+
   {\partial\over\partial y^2}\bigg)\bigg]
  \\   &
  ={1\over2m}\bigg[p_\tau^2+{1\over\cosh^2\tau}
   \big(yp_y^2y+y^2p_x^2\big)\bigg]+\hbarm\enspace.
  \tag\NUM.\num\endalign$$
We therefore obtain the path integral formulation
\plus$$\myalign
  &
  \int\limits_{\tau(t')=\tau'}^{\tau(t'')=\tau''}\cosh^2\tau\CD\tau(t)
  \int\limits_{y(t')=y'}^{y(t')=y''}{\CD y(t)\over y^2}
  \int\limits_{x(t')=x'}^{x(t')=x''}\CD x(t)
  \\   &\qquad\times
  \exp\Bigg[{\i m\over2\hbar}\int_{t'}^{t''}\bigg(\dot\tau^2
    +\cosh^2\tau{\dot x^2+\dot y^2\over y^2}\bigg)dt
    -{\i\hbar T\over2m}\Bigg]
  \\   &
 =(\cosh\tau'\cosh\tau'')^{-1}{\sqrt{y'y''}\over\pi^3}\int_0^\infty dp_0
  \e^{\i p_0(x''-x')}
  \\   &\qquad\times
  \int_0^\infty dp_1\,p_1\sinh\pi p_1
  K_{\i p_1}(\vert p_0\vert y')K_{\i p_1}(\vert p_0\vert y'')
  \\   &\qquad\times
  \int\limits_{\tau(t')=\tau'}^{\tau(t'')=\tau''}
  \exp\Bigg[\ih\int_{t'}^{t''}\bigg({m\over2}\dot\tau^2
  -\hbarm{p_1^2+\viert\over\cosh^2\tau}\bigg)dt-{\i\hbar T\over2m}\Bigg]
  \\   &
 =(\cosh\tau'\cosh\tau'')^{-1}{\sqrt{y'y''}\over\pi^3}\int_0^\infty dp_0
  \e^{\i p_0(x''-x')}
  \\   &\ \times
  \int_0^\infty dp_1\,p_1\sinh\pi p_1
  K_{\i p_1}(\vert p_0\vert y')K_{\i p_1}(\vert p_0\vert y'')
  \\   &\ \times
  \half\int_{\bbbr}{p\sinh\pi p\,dp\over\cosh^2\pi p_2+\sinh^2\pi p}
  P_{\i p_1-\half}^{-\i p}(\tanh\tau_3')
  P_{\i p_1-\half}^{\i p}(\tanh\tau_3'')
  \energyldrei\enspace.
  \\   &
  \tag\NUM.\num\endalign$$
In this path integral solution the path integral identities from the
two-dimensional Poincar\'e upper half-plane, and the simple Rosen-Morse
potential have been used.

\subsubsection{Horicyclic-Spherical Coordinates}
We consider the coordinate system
\plus$$\left.\alignedat 3
  u_0&=\half\left({1\over y}+y+{r^2\over y}\right)\enspace,
     &\quad      &y>0\enspace,
  \\
  u_1&={r\sin\phi\over y}\enspace,\quad
  u_2={r\cos\phi\over y}\enspace,
     &\quad      &r>0\enspace,
  \\
  u_3&=\half\left({1\over y}-y-{r^2\over y}\right)\enspace,
     &\quad      &0\leq\phi<2\pi\enspace.
  \endalignedat\qquad\qquad\right\}
  \tag\NUM.\num$$
The hyperbolic distance is given by
\plus$$
  \cosh d_\ldrei(\vec q'',\vec q')
  ={{y'}^2+{y''}^2+{r'}^2+{r''}^2-2r'r''\cos(\phi''-\phi')\over2y'y''}
  \enspace.
  \tag\NUM.\num$$
Here $(g_{ab})=\diag(1,1,r^2)/y^2$ and the momentum operators are given
by (\numdf,\numeg). For the Hamiltonian we obtain
\plus$$\myalign
  -\hbarm\Delta_\ldrei
  &=-\hbarm y^2\bigg[{\partial^2\over\partial y^2}
   -{1\over y}{\partial\over\partial y}
   +y^2\bigg(
   {\partial^2\over\partial r^2}+{1\over r}{\partial\over\partial r}
    +{1\over r^2}{\partial^2\over\partial\phi^2}\bigg)\bigg]
  \\   &
  ={1\over2m}\bigg(yp_y^2y+y^2p_r^2+{y^2\over r^2}p_\phi^2\bigg)
   -{\hbar^2 y^2\over8mr^2}+{3\hbar^2\over8m}\enspace.
  \tag\NUM.\num\endalign$$
This coordinate system is, of course very similar to the horicyclic
system and has been considered in its $D$-dimensional generalization in
[\GROn] as well. We just cite the result: We obtain the path integral
identity
\plus$$\myalign
       &
  \int\limits_{y(t')=y'}^{y(t'')=y''}{\CD y(t)\over y^3}
  \int\limits_{r(t')=r'}^{r(t'')=r''}r\CD r(t)
  \int\limits_{\phi(t')=\phi'}^{\phi(t'')=\phi''}\CD\phi(t)
  \\   &\qquad\times
  \exp\Bigg[\ih\int_{t'}^{t''}\bigg({m\over2}
     {\dot y^2+\dot r^2+r^2\dot\phi^2\over y^2}
     +{\hbar^2y^2\over 8mr^2}\bigg)dt-{3\i\hbar T\over8m}\Bigg]
  \\   &
  ={y'y''\over\pi^3}\sum_{l\in\bbbz}\e^{\i l(\phi''-\phi')}
  \int_0^\infty dk\,kJ_l(kr')J_l(kr'')
  \\   &\qquad\times
  \int_0^\infty dp\,p\sinh\pi p K_{\i p}(ky')K_{\i p}(ky'')
  \Energyldrei\enspace.
  \tag\NUM.\num\endalign$$
In the path integral evaluation after the trivial $\phi$-path
integration the path integral solution of the free radial
two-dimensional motion has been used.

\subsubsection{Horicyclic-Spheroidal Coordinates}
We consider the coordinate system
\plus$$\left.\alignedat 3
  u_0&=\half\left({1\over y}+y+{\cosh^2\mu-\sin^2\nu\over y}\right)
     \enspace,
     &\quad      &y>0\enspace,
  \\
  u_1&={\cosh\mu\cos\nu\over y}\enspace,\quad
  u_2={\sinh\mu\sin\nu\over y}\enspace,
     &\quad      &\mu>0\enspace,
  \\
  u_3&=\half\left({1\over y}-y-{\cosh^2\mu-\sin^2\nu\over y}\right)
     \enspace,
     &\quad      &-\pi\leq\nu<\pi\enspace.
  \endalignedat\qquad\qquad\right\}
  \tag\NUM.\num$$
The hyperbolic distance is given by
\plus$$\myalign
  \cosh d_\ldrei(\vec q'',\vec q')
 &={1\over2y'y''}\Big[
   {y'}^2+{y''}^2+(\cosh^2\mu''-\sin^2\nu'')+(\cosh^2\mu'-\sin^2\nu')
 \\   &\qquad
 -2\cosh\mu''\cosh\mu'\cos\nu''\cos\nu'
 -2\sinh\mu''\sinh\mu'\sin\nu''\sin\nu'\Big]
  \enspace.
  \\   &
  \tag\NUM.\num\endalign$$
Here we have $(g_{ab})=\diag(1,\sinh^2\mu+\sin^2\nu,\sinh^2\mu+\sin^2\nu
)$ together with the momentum operator $p_y$ (\numeg), and for $p_\mu$
and $p_\nu$ as in (\numef). The Hamiltonian has the form
\plus$$\myalign
  &-\hbarm\Delta_\ldrei
  -\hbarm y^2\bigg[{\partial^2\over\partial y^2}
   -{1\over y}{\partial\over\partial y}
  +{1\over\sinh^2\mu+\sin^2\nu}\bigg(
    {\partial^2\over\partial\mu^2}
    +{\partial^2\over\partial\nu^2}\bigg)\bigg]
  \\   &
  ={1\over2m}\bigg[yp_y^2y
  +y^2(\sinh^2\mu+\sin^2\nu)^{-1/2}(p_\mu^2+p_\nu^2)
      (\sinh^2\mu+\sin^2\nu)^{-1/2}\bigg]-{3\hbar^2\over8m}\enspace.
  \qquad
  \tag\NUM.\num\endalign$$
Using the result form the two-dimensional elliptic coordinate system
in $\ezwei$ we obtain the path integral identity
$$\myalign
       &
  \int\limits_{y(t')=y'}^{y(t'')=y''}{\CD y(t)\over y^3}
  \int\limits_{\nu(t')=\nu'}^{\nu(t'')=\nu''}\CD\nu(t)
  \int\limits_{\mu(t')=\mu'}^{\mu(t'')=\mu''}\CD\mu(t)
  (\sinh^2\mu+\sin^2\nu)
  \\   &\qquad\qquad\times
  \exp\Bigg[{\i m\over2\hbar}\int_{t'}^{t''}
     {\dot y^2+(\sinh^2\mu+\sin^2\nu)(\dot\mu^2+\dot\nu^2)\over y^2}dt
     -{3\i\hbar T\over 8m}\Bigg]
  \\   &
  ={y'y''\over\pi^3}\sum_{\nu\in\Lambda}\int_0^\infty kdk\,
  \me_\nu^*(\eta',\hbox{${k^2\over4}$})
  \me_\nu(\eta'',\hbox{${k^2\over4}$})
  \Me_\nu^{(1)\,*}(\xi',\hbox{${k^2\over4}$})
  \Me_\nu^{(1)}(\xi'',\hbox{${k^2\over4}$})
  \\   &\qquad\qquad\times
  \int_0^\infty dp\,p\sinh\pi p K_{\i p}(ky')K_{\i p}(ky'')\energyldrei
  \tag\NUM.\num\\   \global\plus
       &
  =\bigg({m\over2\pi\i\hbar T}\bigg)^{3/2}
   {d_\ldrei(\vec q'',\vec q')\over\sinh d_\ldrei(\vec q'',\vec q')}
   \exp\bigg[{\i m\over2\hbar T}d_\ldrei^2(\vec q'',\vec q')
                                           -{\i\hbar T\over2m}\bigg]
  \enspace,
  \tag\NUM.\num\endalign$$
where $d_\ldrei(\vec q''\,\vec q')$ must be expressed in
hyperbolic-spheroidal coordinates.

\subsubsection{Horicyclic-Parabolic Coordinates}
We consider the coordinate system
\plus$$\left.\alignedat 3
  u_0&=\half\left({1\over y}+y+{(\xi^2+\eta^2)^2\over4y}\right)
     \enspace,
     &\quad      &y>0\enspace,
  \\
  u_1&={\xi\eta\over y}\enspace,\quad
  u_2={\eta^2-\xi^2\over2y}\enspace,
     &\quad      &\xi\in\bbbr,\eta>0\enspace,
  \\
  u_3&=\half\left({1\over y}-y-{(\xi^2+\eta^2)^2\over4y}\right)
     \enspace.
     &\quad      &\quad
  \endalignedat\qquad\qquad\right\}
  \tag\NUM.\num$$
The hyperbolic distance is given by
\plus$$\multline
  \cosh d_\ldrei(\vec q'',\vec q')
  ={1\over2y'y''}\Big[{y'}^2+{y''}^2+\viert({\xi''}^2+{\eta''}^2)^2
   \\
   +\viert({\xi'}^2+{\eta'}^2)^2
   -\half({\xi''}^2-{\eta''}^2)({\xi'}^2-{\eta'}^2)
   -2\xi''\xi'\eta''\eta'\Big]\enspace.
  \endmultline
  \tag\NUM.\num$$
Here we have $(g_{ab})=\diag(1,\xi^2+\eta^2,\xi^2+\eta^2)$ together
with the momentum operators (\numdg,\numeg). The Hamiltonian has the
form
\plus$$\myalign
  -\hbarm\Delta_\ldrei
  &-\hbarm y^2\bigg[{\partial^2\over\partial y^2}
   -{1\over y}{\partial\over\partial y}
  +{1\over\xi^2+\eta^2}\bigg(
    {\partial^2\over\partial\xi^2}
    +{\partial^2\over\partial\eta^2}\bigg)\bigg]
  \\   &
  ={1\over2m}\bigg[yp_y^2y
  +y^2(\xi^2+\eta^2)^{-1/2}(p_\xi^2+p_\eta^2)
      (\xi^2+\eta^2)^{-1/2}\bigg]-{3\hbar^2\over8m}\enspace.
  \tag\NUM.\num\endalign$$
Using the path integral solution corresponding to the two-dimensional
parabolic coordinates in $\ezwei$ we obtain the identity
\plus$$\myalign
  &\int\limits_{y(t')=y''}^{y(t'')=y''}{\CD y(t)\over y^3}
  \int\limits_{\eta(t')=\eta'}^{\eta(t'')=\eta''}\CD\eta(t)
   \int\limits_{\xi(t')=\xi'}^{\xi(t'')=\xi''}\CD\xi(t)(\xi^2+\eta^2)
  \\   &\qquad\qquad\times
  \exp\Bigg[{\i m\over2\hbar}\int_{t'}^{t''}
     {\dot y^2+(\xi^2+\eta^2)(\dot\xi^2+\dot\eta^2)\over y^2}dt
     -{3\hbar^2\over8m}\Bigg]
  \\   &
  ={2y'y''\over\pi^2}\sum_{e,o}\int_{\bbbr} d\zeta\int_{\bbbr}dk\,
   \Psi_{k,\zeta}^{(e,o)\,*}(\xi',\eta')
   \Psi_{k,\zeta}^{(e,o)}(\xi'',\eta'')
  \\   &\qquad\qquad\times
  \int_0^\infty dp\,p\sinh\pi p K_{\i p}(\vert k\vert y')
  K_{\i p}(\vert k\vert y'')\Energyldrei \enspace,\qquad
  \tag\NUM.\num\endalign$$
in the notation of 4.1.5.

\subsubsection{Confocal Ellipsoidal Coordinates}
Let us consider the coordinate system defined by
\plus$$
  {u_0^2\over\rho_i-d}-{u_1^2\over\rho_i-c}
  -{u_2^2\over\rho_i-a}-{u_3^2\over\rho_i-b}
  =0\enspace,\qquad(i=1,2,3)\enspace,
  \tag\NUM.\num$$
where $d<c<\rho_3<b<\rho_2<a<\rho_1$ (c.f.~Section 3.1). Explicitly
\plus$$\left.\aligned
  u_0^2&={(\rho_1-d)(\rho_2-d)(\rho_3-d)\over(a-d)(b-d)(c-d)}\enspace,
  \quad
  u_1^2=-{(\rho_1-c)(\rho_2-c)(\rho_3-c)\over(a-c)(b-c)(d-c)}\enspace,
  \\
  u_2^2&=-{(\rho_1-a)(\rho_2-a)(\rho_3-a)\over(d-a)(c-a)(b-a)}\enspace,
  \quad
  u_3^2=-{(\rho_1-b)(\rho_2-b)(\rho_3-b)\over(d-b)(c-b)(a-b)}\enspace.
  \endaligned\qquad\right\}
  \tag\NUM.\num$$
The corresponding line element is given by
\plus$$
  {ds^2\over dt^2}=\viert\left[
   {(\rho_1-\rho_2)(\rho_1-\rho_3)\over P_4(\rho_1)}\dot\rho_1^2
  +{(\rho_2-\rho_3)(\rho_2-\rho_1)\over P_4(\rho_2)}\dot\rho_2^2
  +{(\rho_3-\rho_1)(\rho_3-\rho_2)\over P_4(\rho_3)}\dot\rho_3^2
  \right]\enspace,
  \tag\NUM.\num$$
where $P_4(\rho)=(\rho-a)(\rho-b)(\rho-c)(\rho-d)$. Note the
similarities as well the differences in comparison to the confocal
ellipsoidal coordinates on the sphere $\sdrei$. We apply the separation
formula (\numca). We identify
$(g_{ab})\equiv\diag(h_1^2,h_2^2,h_3^2)$, furthermore
\minus$$\align
  &\Gamma_i={\rho_i P_4'(\rho_i)\over P_4(\rho)}\enspace,
  \qquad(i=1,2,3)
  \tag\NUM.\num\\   \global\plus
  &S={(\rho_1-\rho_2)(\rho_1-\rho_3)(\rho_2-\rho_3)\over
    4P_4(\rho_1)P_4(\rho_2)P_4(\rho_2)}\enspace,
  \tag\NUM.\num\\   \global\plus
  &M_1={\rho_2-\rho_3\over P_4(\rho_2)P_4(\rho_3)}\enspace,
  \quad
   M_2={\rho_3-\rho_1\over P_4(\rho_1)P_4(\rho_3)}\enspace,
  \quad
   M_3={\rho_1-\rho_2\over P_4(\rho_1)P_4(\rho_2)}\enspace,
  \tag\NUM.\num\endalign$$
and obtain the following path integral identity
$$\myalign
  &\prod_{i=1}^3\int\limits_{\rho_i(t')=\rho_i'}^{\rho_i(t'')=\rho_i''}
  h_i\CD\rho_i(t)\exp\left\{\ih\int_{t'}^{t''}\left[{m\over2}
  \sum_{i=1}^3h_i^2\dot\rho_i^2-\Delta V_{PF}(\{\rho\})\right]dt\right\}
  \\   &
  =(S'S'')^{-1/4}\int_{-\infty}^\infty{dE\over2\pi\hbar}
   \e^{-\i ET/\hbar}\int_0^\infty ds''
  \prod_{i=1}^3\int\limits_{\rho_i(0)=\rho_i'}^{\rho_i(s'')=\rho_i''}
   M_i^{-1/2}\CD\rho_i(s)
  \\   &\qquad\qquad\times
   \exp\left\{\ih\int_{0}^{s''}\left[{m\over2}\sum_{i=1}^3
   {\dot\rho_i^2\over M_i}+ES
    -{\hbar^2\over8m}\sum_{i=1}^3M_i\Big(\Gamma_i^2+2\Gamma_i'\Big)
    \right]ds\right\}\qquad\qquad
  \tag\NUM.\num\\   \global\plus
       &
  =\bigg({m\over2\pi\i\hbar T}\bigg)^{3\over2}
  {d_\ldrei(\vec q'',\vec q')\over\sinh d_\ldrei(\vec q'',\vec q')}
  \exp\bigg[{\i m\over2\hbar T}d_\ldrei^2(\vec q'',\vec q')
                                           -{\i\hbar T\over2m}\bigg]
  \enspace,
  \tag\NUM.\num\endalign$$
and $d_\ldrei(\vec q'',\vec q')$ must be expressed in confocal
ellipsoidal coordinates. Actually, the relation $d<c<\dots$ is only
one possibility to obtain a coordinate system on $\ldrei$. We can
distinguish eighteen further cases, which fall into two classes with
eleven and seven coordinate systems, respectively. In the first class
emerges when ever two of the $\{e_i\}=\{a,b,c,d\}$ are equal. They
represent the negative curvature cases in comparison to the positive
curvature ones as noted in 5.2.6. The second class is obtained by a
similar consideration as for the pseudo-conical coordinate systems in
the two-dimensional case. C.f.~Section 3.1 for the classification into
the four general classes A,B,C,D and [\OLE] for the explicit
construction. Therefore we obtain a total of 34 different coordinate
systems. This concludes the discussion of the separable coordinate
systems on three-dimensional spaces of constant curvature.

\PLUS\glno=0                      
\section{Discussion and Summary}
In this paper I have discussed path integrals in two- and
three-dimensional spaces of constant curvature, i.e.\ the flat
Euclidean spaces $\ezwei$ and $\edrei$, the spheres $\szwei$ and
$\sdrei$, and the  pseudospheres $\lzwei$ and $\ldrei$. In many cases
I could perform the path integration explicitly, thanks to some basic
path integral solutions. In the majority of the cases, in particular on
the  pseudospheres $\lzwei$ and $\ldrei$, only an indirect reasoning
was possible. However, in any case, the {\it explicit\/} expressions of
the propagators and the Green functions are known in terms of the
invariant distance in this space. This is possible for any dimension,
c.f.~[\FEY, \FH] for the Euclidean space, c.f.~[\BJb, \GRSb] for the
spheres, and c.f.~[\BJb, \GRSc, \VENa] for the  pseudospheres, thus
providing coordinate-independent expressions. Knowing them gives rise
to numerous identities connecting the path integral formulations,
explicit solutions in terms of the spectral expansions, and the
coordinate-independent general formul\ae.

There are several generalizations possible one can think of. First, one
can study complex Riemannian manifolds [\KAL, \KAMIb] and try to study
corresponding path integral formulations. At least, the complex
Riemannian manifold of constant negative curvature $\Lambda^{(\bbbc_n)}$
has already attracted some attention in the theory of automorphic
forms, in order to construct a Selberg trace formula [\VENb] similar to
the usual one [\HEJ, \SEL, \VENc]. Here also a path integral evaluation
is possible [\GROn]. However, it is obvious that complexity increases
in these more elaborated cases.

Second, one can introduce potential problems and ask for the separable
ones. There exist some studies of this kind in the literature, e.g.
[\GROj], where the most important one is the Kepler-Coulomb problem
in spaces of constant (positive and negative) curvature [\BIJ,~\GROg].
However, no systematic study has been done until now. In the case of
the Coulomb-problem, one is on the one hand interested in the symmetry
properties and transformations between the bases of the system [\POGOb,
\POGOa, \POGOc, \QUES], where the $\OO(4)$-symmetry of the hydrogen
atom is the best known one and lies at the origin of the separability
of this problem in four coordinate systems. On the other one is
interested in the relation [\HIG, \POGOc] of the Kepler-problem and the
harmonic oscillator in spaces of constant curvature. Whereas in flat
space the transformation which relates the Coulomb problem in $\bbbr^3$
and the isotropic harmonic oscillator in $\bbbr^4$ is the
Kustaanheimo-Stiefel transformation [\DKa, \DKb], such a transformation
is not known for constant (positive and negative) curvature. On $\sdrei$
and $\ldrei$ the Coulomb problem and the harmonic oscillator are only
separable in two coordinate system, the spherical and the
sphero-conical, respectively. the explict form on $\edrei$, $\sdrei$
and $\ldrei$ is given in the following table:
$$\aligned
\vbox{\offinterlineskip
\hrule
\halign{&\vrule#&
  \strut\quad\hfil#\quad\hfil\quad\cr
height2pt&\omit&&\omit&&\omit&\cr
&              &&Kepler problem
               &&Harmonic Oscillator                       &\cr
height2pt&\omit&&\omit&&\omit&\cr
\noalign{\hrule}
\noalign{\hrule}
height2pt&\omit&&\omit&&\omit&\cr
&$\edrei$      &&$-\dsize{\alpha\over r}$
               &&$\dsize{m\over2}\omega^2r^2$              &\cr
height2pt&\omit&&\omit&&\omit&\cr
\noalign{\hrule}
height2pt&\omit&&\omit&&\omit&\cr
&$\sdrei$      &&$-\dsize{\alpha\over R}\cot\theta_1$
               &&$\dsize{m\over2}\omega^2R^2\tan^2\theta_1$&\cr
height2pt&\omit&&\omit&&\omit&\cr
\noalign{\hrule}
height2pt&\omit&&\omit&&\omit&\cr
&$\ldrei$      &&$-\dsize{\alpha\over R}\coth\tau$
               &&$-\dsize{m\over2}\omega^2R^2\tanh^2\tau$  &\cr
height2pt&\omit&&\omit&&\omit&\cr}\hrule}
  \endaligned$$
(where we have reintroduced $R$, and the coordinate notation is as in
Section~5). And as a third, one can introduce constant magnetic fields
on spheres and pseudospheres [\GROb, \GROf]. The corresponding
cases in flat space are well-known, e.g.~[\FEY, \FH, \GRSg, \KOKCAS].

A systematic study for the search of separable potentials in $\bbbr^2$
and $\bbbr^3$ does exist [\EIS, \EWA], a project which has been started
by Smorodinsky, Winternitz et al.\ [\MSVW]. These potentials can be put
into two classes, which are called maximally and minimally
superintegrable, respectively. A Hamiltonian system in three degrees of
freedom is called maximally superintegrable if it admits five globally
defined and single-valued integrals of motion (the Coulomb-problem
and the harmonic oscillator belong to this class). A Hamiltonian system
in three degrees of freedom is called minimally superintegrable if it
admits four globally defined and single-valued integrals of motion
(with the ring-potential as an example [\GROn]). The maximally
superintegrable potentials can be stated explicitly, whereas the
(eight) minimally superintegrable potentials allow in addition to an
explicit expression an arbitrary function of the coordinates according
to either $\propto F(r)$, $\propto F(z)$ or $\propto F(y/x)$,
respectively. The corresponding super-integrable potentials in
$\bbbr^2$ can be obtained in a similar way, and there are a total
number of four independent potentials [\FMSUW]. The five maximally
superintegrable potentials in $\bbbr^3$ are:
\plus$$
   V_1(x,y,z)={m\over2}\omega^2(x^2+y^2+z^2)
   +\hbarm\bigg({k_1\over x^2}+{k_2\over y^2}+{k_3\over z^2}\bigg)
  \enspace;
  \tag\NUM.\num$$
this potential is separable in (with the coordinates systems which
admit explicit path integral evaluation in {\it italic\/}) {\it
cartesian, spherical, circular cylindrical\/}, elliptic cylindrical,
oblate spheroidal, prolate spheroidal, conical, and ellipsoidal
coordinates.
\plus$$
   V_2(x,y,z)=-{\alpha\over\sqrt{x^2+y^2+z^2}}
   +\hbarm\bigg({k_1\over x^2}+{k_2\over y^2}\bigg)\enspace;
  \tag\NUM.\num$$
this potential is separable in {\it spherical, parabolic\/}, and
conical coordinates.
\plus$$
   V_3(x,y,z)=\hbarm
  \bigg({k_1x\over\sqrt{x^2+y^2}}+{k_2\over y^2}
                                 +{k_3\over z^2}\bigg)\enspace;
  \tag\NUM.\num$$
this potential is separable in {\it spherical and parabolic cylindrical
coordinates\/}.
\plus$$
   V_4(x,y,z)=\hbarm
  \bigg({k_1x\over\sqrt{x^2+y^2}}+{k_2\over y^2}\bigg)+k_3z\enspace;
  \tag\NUM.\num$$
this potential is separable in {\it parabolic cylindrical\/} and
parabolic coordinates.
\plus$$
   V_5(x,y,z)={m\over2}\omega^2(x^2+y^2+4z^2)
   +\hbarm\bigg({k_1\over x^2}+{k_2\over y^2}\bigg);
  \tag\NUM.\num$$
this potential is separable in {\it cartesian\/}, parabolic (one
obtains an intractable quadratic plus sextic oscillator), and
elliptic cylindrical coordinates.

In each case it is easy to construct the various path integral
formulations, based on the coordinate systems as listed in Section~5.1,
and obtain therefore the various identities connecting them. The
corresponding path integral evaluations then can be performed by using
the basic path integrals as given in Appendix~1, which is left as an
exercise to the interested reader. Note that in comparison to the
free motion in spaces of constant curvature the statement of the
propagator in a coordinate independent way is not possible.
This concludes the discussion.

\ack
I would like to thank the organizers of the Dubna workshop for the nice
atmosphere and warm hospitality. In particular I would like to thank
V.\ V.\ Belokurov, L.\ S.\ Davtian, A.\ Inomata, G.\ Junker, R.\ M.\
Mir-Kasimov, G.\ S.\ Pogosyan, O.\ G.\ Smolyanov, and S.\ I.\ Vinitsky
for fruitful discussions, in particular G.\ S.\ Pogosyan for drawing
my attention to Refs.~[\EWA, \OLE]. Furthermore I would like to thank
G.\ Holtkamp for providing an english translation of Ref.~[\OLE].


\Chapno=1\glno=0                  
\appendix{Some Important Path Integral Solutions}
In this Appendix we cite some important path integral solutions,
in particular for the radial harmonic oscillator and for the
(modified) P\"oschl-Teller potential, including some special cases.
These three path integral solutions are the most important building
blocks for almost all other path integrals.

\medskip\noindent
{\sl A.1.1. The path integral for the radial harmonic oscillator.}

\noindent
The calculation of the path integral for the radial harmonic oscillator
has first been performed by Peak and Inomata [\PI]. A more general case
is due to Goovaerts [\GOOb] (c.f.\ also [\DURd]). Path integrals related
to the radial harmonic oscillator may be called of Besselian type
[\INOb]. Here we are not going into the subtleties of the Besselian
functional measure due to the Bessel functions which appear in the
lattice approach [\FLM, \GRSb, \STEc] which is actually necessary for
the explicit evaluation of the radial harmonic oscillator path integral
[\DURb, \GOOb, \PI]. One obtains (modulo the mentioned subtleties)
($r>0$)
\plus$$\myalign
  &\int\limits_{r(t')=r'}^{r(t'')=r''}\CD r(t)
  \exp\left\{\ih\int_{t'}^{t''}\bigg[{m\over2}
  \big(\dot r^2-\omega^2r^2)
      -\hbar^2{\lambda^2-\viert\over2mr^2}\bigg]dt\right\}
         \\   &:=
  \int\limits_{r(t')=r'}^{r(t'')=r''}\mu_\lambda[r^2]\CD r(t)
  \exp\left[{\i m\over2\hbar}\int_{t'}^{t''}\big(\dot r^2
           -\omega^2r^2\big)dt\right]
         \\   &=
  {m\omega\sqrt{r'r''}\over\i\hbar\sin\omega T}
  \exp\bigg[-{m\omega\over2\i\hbar}({r'}^2+{r''}^2)\cot\omega T\bigg]
  I_\lambda\bigg({m\omega r'r''\over\i\hbar\sin\omega T}\bigg)
  \tag\AA.\num\endalign$$
with the nontrivial functional weight
$\mu_\lambda[r^2]$:
\plus$$\mu_\lambda[r_jr_{j-1}]=
  \sqrt{2\pi mr_jr_{j-1}\over\i\epsilon\hbar}\,
  \e^{-mr_jr_{j-1}/\i\epsilon\hbar}I_\lambda
  \bigg({mr_jr_{j-1}\over\i\epsilon\hbar}\bigg)\enspace.
  \tag\AA.\num$$
In the special case that $\omega=0$, i.e.\ for the free radial motion,
we obtain
$$\myalign
   \int\limits_{r(t')=r'}^{r(t'')=r''}\mu_\lambda[r^2]&\CD r(t)
  \exp\left({\i m\over2\hbar}\int_{t'}^{t''}\dot r^2dt\right)
         \\   &
  =\sqrt{r'r''}{m\over\i\hbar T}
  \exp\bigg[-{m\over2\i\hbar T}({r'}^2+{r''}^2)\bigg]
  I_\lambda\bigg({mr'r''\over\i\hbar T}\bigg)
  \tag\AA.\num\\    \global\plus
  &=\sqrt{r'r''}\,\int_0^\infty dp\,p \e^{-\i\hbar Tp^2/2m}
    J_\lambda(pr')J_\lambda(pr'')\enspace.
  \tag\AA.\num\endalign$$
\goodbreak

\medskip\noindent
{\sl A.1.2. The modified P\"oschl-Teller Potential.}

\noindent
The path integral solution for the P\"oschl-Teller potential can be
achieved by means of the $\SU(2)$-path integral. We have [\BJb, \DURb,
\INOWI]
\plus$$\myalign
  &\int\limits_{x(t')=x'}^{x(t'')=x''}\CD x(t)
  \exp\left\{\ih\int_{t'}^{t''}\left[{m\over2}\dot x^2
        -\hbarm\bigg(
  {\alpha^2-{1\over4}\over\sin^2x}+{\beta^2-{1\over4}\over\cos^2x}\bigg)
  \right]dt\right\}
  \\   &
  =\sum_{l=0}^\infty
  \exp\bigg[-{\i\hbar T\over2m}(\alpha+\beta+2l+1)^2\bigg]
  \Psi^{(\alpha,\beta)\,*}_l(x')\Psi^{(\alpha,\beta)}_l(x'')\enspace,
  \tag\AA.\num\endalign$$
with the wavefunctions given by
\plus$$\myalign
  \Psi_n^{(\alpha,\beta)}(x)
  &=\bigg[2(\alpha+\beta+2l+1)
  {l!\Gamma(\alpha+\beta+l+1)\over\Gamma(\alpha+l+1)\Gamma(\beta+l+1)}
  \bigg]^{1/2}
         \\   &\qquad\qquad\times
  (\sin x)^{\alpha+1/2}(\cos x)^{\beta+1/2}
  P_n^{(\alpha,\beta)}(\cos2x)\enspace.
  \tag\AA.\num\endalign$$
\goodbreak

\medskip\noindent
{\sl A.1.3. The modified P\"oschl-Teller Potential.}

\noindent
The path integral solution for the modified P\"oschl-Teller
potential can be achieved by means of the $\SU(1,1)$-path integral. We
have [\BJb, \INOWI, \BJa]
\plus$$\myalign
  \int\limits_{r(t')=r'}^{r(t'')=r''}&\CD r(t)
  \exp\left\{\ih\int_{t'}^{t''}\left[{m\over2}\dot r^2
                   -\hbarm
   \bigg({\eta^2-{1\over4}\over\sinh^2r}
   -{\nu^2-{1\over4}\over\cosh^2r}\bigg)\right]dt\right\}
  \\
  &=\sum_{n=0}^{N_M}\Phi_n^{(\eta,\nu)\,*}(r')
  \Phi_n^{(\eta,\nu)}(r'')
  \exp\bigg\{
  -{\i\hbar T\over2m}\Big[2(k_1-k_2-n)-1\Big]^2\bigg\}
         \\   &\qquad\qquad
  +\int_0^\infty dp\,\Phi_p^{(\eta,\nu)\,*}(r')
  \Phi_p^{(\eta,\nu)}(r'') \exp\bigg(-{\i\hbar T\over2m}p^2\bigg)
  \enspace.
  \tag\AA.\num\endalign$$
\goodbreak\noindent
Now introduce the numbers $k_1,k_2$ defined by:
$k_1=\half(1\pm\nu)$, $k_2=\half(1\pm\eta)$, where the correct sign
depends on the boundary-conditions for $r\to0$ and $r\to\infty$,
respectively. In particular for $\eta^2={1\over4}$, i.e.\
$k_2={1\over4},{3\over4}$, we obtain wavefunctions with even and odd
parity, respectively. The number $N_M$ denotes the maximal number of
states with $0,1,\hdots,N_M<k_1-k_2-\half$. The bound state
wavefunctions read as ($\kappa=k_1-k_2-n$)
\plus$$\left.\aligned
  \Phi_n^{(k_1,k_2)}(r)
  &=N_n^{(k_1,k_2)}(\sinh r)^{2k_2-\half}
                    (\cosh r)^{-2k_1+{3\over2 }}
         \\   &\qquad\times
  {_2}F_1(-k_1+k_2+\kappa,-k_1+k_2-\kappa+1;2k_2;-\sinh^2r)
  \\
  N_n^{(k_1,k_2)}
  &={1\over\Gamma(2k_2)}
  \bigg[{2(2\kappa-1)\Gamma(k_1+k_2-\kappa)
                     \Gamma(k_1+k_2+\kappa-1)\over
    \Gamma(k_1-k_2+\kappa)\Gamma(k_1-k_2-\kappa+1)}\bigg]^{1/2}
  \endaligned\qquad\right\}
  \tag\AA.\num$$
The scattering states are given by:
\plus$$\left.\aligned
  \Phi_p^{(k_1,k_2)}(r)
  &=N_p^{(k_1,k_2)}(\cosh r)^{2k_1-\half}(\sinh r)^{2k_2-\half}
         \\   &\qquad\qquad\times
  {_2}F_1(k_1+k_2-\kappa,k_1+k_2+\kappa-1;2k_2;-\sinh^2r)
  \\
  N_p^{(k_1,k_2)}
  &={1\over\Gamma(2k_2)}\sqrt{p\sinh\pi p\over2\pi^2}
  \Big[\Gamma(k_1+k_2-\kappa)\Gamma(-k_1+k_2+\kappa)
         \\   &\qquad\qquad\times
  \Gamma(k_1+k_2+\kappa-1)\Gamma(-k_1+k_2-\kappa+1)\Big]^{1/2},
  \endaligned\qquad\right\}
  \tag\AA.\num$$
[$\kappa=\half(1+\i p)$].
Of course, in the path integral formulation of
the modified P\"oschl-Teller potential a similar functional weight
interpretation must be used as for the P\"oschl-Teller potential in
order to have a proper short-time behaviour, respectively a lattice
regularization [\GROe].

I also cite two special cases, where first only the $1/\sinh^2 r$,
and second where only the $1/\cosh r$ term is present. The special case
$V^{(sh)}(r)=V_0/\sinh^2r$ ($V_0=\hbarm(\lambda^2-\viert)$, $r>0$) is
given by [\GROe, \GRSc, \KLEMUS] (simple Manning-Rosen potential)
\plus$$\myalign
  &\int\limits_{r(t')=r'}^{r(t'')=r''}\CD r(t)
  \exp\left[\ih\int_{t'}^{t''}\bigg({m\over2}\dot r^2-
  \hbarm{\lambda^2-\viert\over\sinh^2r}\bigg)dt\right]
         \\   &
  ={1\over\pi}\sqrt{\sinh r'\sinh r''}
              \int_0^\infty dp\,p\sinh\pi p
  \big\vert\Gamma(\bhalf+\i p-\lambda)\vert^2
         \\   &\qquad\qquad\times
  \big\vert\Gamma(\bhalf+\i p-\lambda)\vert^2
  \CP_{\i p-1/2}^{-\lambda}(\cosh r')
  \CP_{\i p-1/2}^{-\lambda}(\cosh r'')\e^{-\i\hbar p^2T/2m}\enspace.
         \\   &
  \tag\AA.\num\endalign$$
The special case $V^{(ch)}(x)=V_0/\cosh^2x$ ($V_0=\hbarm(k^2+\viert)$,
$x\in\bbbr$) on the other is given by [\GROc, \GROe, \KLEMUS] (simple
Rosen-Morse potential)
\plus$$\myalign
  &\int\limits_{x(t')=x'}^{x(t'')=x''}\CD x(t)
  \exp\left[\ih\int_{t'}^{t''}\bigg({m\over2}\dot x^2-
  \hbarm{k^2+\viert\over\cosh^2x}\bigg)dt\right]
         \\   &
  =\half\int_{\bbbr}\,dp\,p\sinh\pi p
  {P^{-\i p}_{\i k-1/2}(\tanh x')P^{\i p}_{\i k-1/2}(\tanh x'')
   \over\cosh^2\pi k+\sinh^2\pi p}\e^{-\i\hbar p^2T/2m}\enspace.
  \tag\AA.\num\endalign$$
\edef\numAa{\AA.\num}%
Note that for $V_0<0$ also bound state solutions are allowed. These,
however, we do not need.


\PLUS\glno=0                      
\appendix{Discussion of a Dispersion Relation}
We consider the integral representation as given by
Buchholz [\BUC,~p.158]:
\plus$$\myalign
  &{1\over2\pi\i}\int\limits_{-\sigma-\i\infty}^{-\sigma+\i\infty}
  \Gamma\bigg({1+\mu\over2}-s\bigg)\Gamma\bigg({1+\mu\over2}+s\bigg)
  \bigg(\tan{\phi\over2}\bigg)^{2s}
  \\    &\qquad\qquad\times
  \CM_{\chi_1+s-{1+\mu\over2},{\mu\over2}}(-\i x)
  \CM_{\chi_2+s-{1+\mu\over2},{\mu\over2}}(+\i y)ds
  \\   &=
  {\sqrt{xy}\over2}\sin\phi\exp\bigg[
  {\i\over2}(x-y)\cos\phi\bigg]J_\mu(\sqrt{xy}\sin\phi)\enspace,
  \\   &
  \hbox{where}\qquad\bigg\vert\arctan{\phi\over2}\bigg\vert<{\pi\over2}
  \enspace,
  \qquad\vert\sigma\vert={1+\Re(\mu)\over2}\enspace,\qquad
  \Re(\mu)>-1\enspace.\qquad\qquad\qquad\qquad
  \tag\AA.\num\endalign$$
\edef\numCa{\AA.\num}%
In order to apply (\numCa) for the determination of continuous spectra
we perform some manipulations. We replace $x\to2x$, $y\to-2y$,
$s\to+\i p$, $\chi_{1,2}\to(1+\mu)/2$, $\sin\phi\to1/\sin\alpha$, and
$\mu\to2\mu$. This gives
\plus$$\myalign
  &{1\over\sin\alpha}\exp\Big[-(x+y)\cot\alpha\Big]
  I_{2\mu}\bigg({2\sqrt{xy}\over\sin\alpha}\bigg)
  \\   &
  ={1\over2\pi\sqrt{xy}}\int_{\bbbr}
  {\Gamma(\half+\mu+\i p)\Gamma(\half+\mu-\i p)\over\Gamma^2(1+2\mu)}
  \e^{-2\alpha p+\pi p}M_{+\i p,\mu}(-2\i x)M_{-\i p,\mu}(+2\i y)dp
  \enspace,
  \tag\AA.\num\endalign$$
\edef\numCb{\AA.\num}%
where use has been made of some properties of the Whittaker functions
$M_{\mu,\nu}(z)=\Gamma(1+2\mu)\CM_{\mu,\nu}$, and [\BUC, p.11]
$M_{\chi,\mu/2}(z\e^{\pm\i\pi})=\e^{\pm\i\pi(1+\mu)/2}M_{-\chi,\mu/2}(z)
$, respectively. Equation (\numCb) is the desired expansion formula for
the determination of continuous spectra. Note the additional $\e^{\pi
p}$-factor in (\numCb). This representation can also be deduced from
the integral formula [\EMOTb,~p.414; \GRA,~p.884]
\plus$$\multline
  \int_{\bbbr} dx\,\e^{-2\i x\rho}
  \Gamma\big(\bhalf+\nu+\i x\big)\Gamma\big(\bhalf+\nu-\i x\big)
  \CM_{\i x,\nu}(\mu)\CM_{\i x,\nu}(\nu)
  \\
  ={2\pi\sqrt{\mu\nu}\over\cosh\rho}
  \exp\big[-(\mu+\nu)\tanh\rho\big]
  J_{2\nu}\bigg({2\sqrt{\mu\nu}\over\cosh\rho}\bigg)\enspace.
  \endmultline
  \tag\AA.\num$$
We consider the special case of (\numCb), where the Bessel function is
replaced by an exponential, i.e.:
$$I_\mu\bigg({2\sqrt{xy}\over\sin\alpha}\bigg)\to
  \exp\bigg({2\sqrt{xy}\over\sin\alpha}\bigg)\enspace.$$
Together with the properties of the Whittaker functions that for $\mu=
\pm\half$ they can be expressed by parabolic cylinder functions
$E^{(0,1)}_\nu$ we obtain
\plus$$\myalign
  &{1\over\sqrt{2\pi\sin\alpha}}\exp\big[-(x+y)\cot\alpha\big]
  \exp\bigg({2\sqrt{xy}\over\sin\alpha}\bigg)
  \\   &
  ={\sqrt{\half\sqrt{xy}}\over\sin\alpha}
  \exp\big[-(x+y)\cot\alpha\big]
  \bigg[ I_\half\bigg({2\sqrt{xy}\over\sin\alpha}\bigg)
  +      I_{-\half}\bigg({2\sqrt{xy}\over\sin\alpha}\bigg)\bigg]
  \\   &
  ={1\over(2\pi)^2}\int_{\bbbr} dp\,\e^{-2\alpha p+\pi p}
  \\   &\quad\times
  \Bigg[\,\bigg\vert\Gamma\bigg({1\over4}-\i p\bigg)\bigg\vert^2
    E^{(0)}_{-\half+2\i p}\bigg(\e^{-\i\pi/4}2\sqrt{x}\bigg)
    E^{(0)}_{-\half-2\i p}\bigg(\e^{\i\pi/4}2\sqrt{y}\bigg)
   \\   &\qquad
   + \bigg\vert\Gamma\bigg({3\over4}-\i p\bigg)\bigg\vert^2
    E^{(1)}_{-\half+2\i p}\bigg(\e^{-\i\pi/4}2\sqrt{x}\bigg)
    E^{(1)}_{-\half-2\i p}\bigg(\e^{\i\pi/4}2\sqrt{y}\bigg)\,\Bigg]
  \enspace,
  \tag\AA.\num\endalign$$
\edef\numCc{\AA.\num}%
The functions $E^{(0)}_\nu(z)$ and $E^{(1)}_\nu(z)$ are even and odd in
the variable $z$, respectively. We can establish the connection to the
parabolic cylinder functions $D_\nu$ and (\numCc) is equivalent with the
expansion [\GRA,~p.896]:
\plus$$\multline
  \int_{c-\i\infty}^{c+\i\infty}
  \big[D_\nu(x)D_{-\nu-1}(\i y)+D_\nu(-x)D_{-\nu-1}(-\i y)\big]
  {t^{-\nu-1}d\nu\over\sin(-\nu\pi)}
  \\
  =2\i\sqrt{2\pi\over1+t^2}\exp\bigg[
 {1-t^2\over1+t^2}\cdot{x^2+y^2\over4}+{\i txy\over1+t^2}\bigg]\enspace.
  \endmultline
  \tag\AA.\num$$


\PLUS\glno=0                      
\appendix{The $D$-Dimensional Hyperbolic System}
We consider the $D$-dimensional generalization of the hyperbolic
coordinate system:
\plus$$\left.\aligned
  u_0&=\cosh\tau_1\dots\cosh\tau_{D-1}\enspace,
  \\
  u_1&=\sinh\tau_1\enspace,
  \\
  u_2&=\cosh\tau_1\sinh\tau_2\enspace,
  \\
     &\vdots
  \\
  u_{D-1}&=\cosh\tau_1\dots\sinh\tau_{D-1}\enspace.
  \endaligned\qquad\right\}
  \tag\AA.\num$$
Therefore we obtain the path integral formulation by successively
applying (\numAa) ($\tau_1,\dots,\tau_{D-1}\in\bbbr$)
\plus$$\myalign
       &
  \int\limits_{\tau_1(t')=\tau_1'}^{\tau_1(t'')=\tau_1''}
  \cosh^{D-2}\tau_1\CD\tau_1(t)
  \dots
  \int\limits_{\tau_{D-1}(t')=\tau_{D-1}'}
             ^{\tau_{D-1}(t'')=\tau_{D-1}''}\CD\tau_{D-1}(t)
  \\   &\qquad\times
  \exp\Bigg\{\ih\int_{t'}^{t''}\bigg[{m\over2}
     \Big(\dot\tau_1^2+\cosh^2\tau_1\dot\tau_2^2
        +\dots(\cosh^2\tau_1\dots\cosh^2\tau_{D-2})\dot\tau_{D-1}^2\Big)
  \\   &\qquad\qquad\qquad\qquad
    -{\hbar^2\over8m}\bigg((D-2)^2+{1\over\cosh^2\tau_1}
     +\dots+{1\over\cosh^2\tau_1\dots\cosh^2\tau_{D-2}}\bigg)
     \bigg]dt\Bigg\}
  \\   &
  =\Big(\cosh^{D-2}\tau_1'\cosh^{D-2}\tau_1''\dots
        \cosh\tau_{D-2}'\cosh\tau_{D-2}''\Big)^{-1/2}
  \int_{\bbbr}{dp_0\over2\pi}
  \e^{\i p_0(\tau_{D-1}''-\tau_{D-1}')}
  \\   &\qquad\times
  \prod_{j=1}^{D-2}\half\int_{\bbbr}
  \Psi_{\i p_{j-1}}^{-\i p_j}(\tau_{D-1-j}')
  \Psi_{\i p_{j-1}}^{\i p_j}(\tau_{D-1-j}'')
  \exp\bigg[-{\i\hbar T\over2m}
   \bigg(p_{D-2}^2+{(D-2)^2\over4}\bigg)\bigg]
  \enspace.
  \\   &
  \tag\AA.\num\endalign$$
with the wave-functions $\Psi_{\i k}^{\i p}(\tau)$ given by
\plus$$
  \Psi_{\i k}^{\i p}(\tau)
  =\sqrt{p\sinh\pi p\over2(\cosh^2\pi k+\sinh^2\pi p)}\,
  P_{\i k-\half}^{\i p}(\tanh\tau)\enspace.
  \tag\AA.\num$$
This solution represents an alternative path integral solutions
as already outlined for two other coordinate systems in [\GROn]. Note
that irrespective which coordinate system for the path integral
formulation on $\lD$ is chosen (and there are many indeed)
the Green function {\it always\/} has the form [\GROn, \GRSc]
\plus$$\multline
  G^\lD\Big(d_\lD(\vec q'',\vec q');E\Big)
  ={m\over\pi\hbar^2}
  \left({-1\over2\pi\sinh d_\lD(\vec q'',\vec q')}
                                              \right)^{(D-3)/2}
  \\   \times
  \CQ_{-\i\sqrt{2mE/\hbar^2-(D-2)^2/4}-1/2}^{(D-3)/2}
  \Big(\cosh d_\lD(\vec q'',\vec q')\Big)\enspace.
  \endmultline
  \tag\AA.\num$$


\newpage\noindent
{\bf References}
\bigskip
\eightpoint\eightrm
\baselineskip=10pt
\def\refno{\item}
\refno{[\ABS]}
M.Abramowitz and I.A.Stegun (Eds.): Pocketbook of Mathematical
Functions
({\it Verlag Harry Deutsch}, Frankfurt/Main, 1984)
\refno{[\ABHK]}
S.Albeverio, P.Blanchard and R.H\o egh-Krohn:
Some Applications of Functional Integration;
{\it Lecture Notes in Physics} {\bf 153}
({\it Springer-Verlag}, Berlin, 1982)
\refno{[\AHK]}
S.Albeverio and R.J.H\o egh-Krohn:
Mathematical Theory of Feynman Path Integrals;
{\it Lecture Notes in Mathematics} {\bf 523}
({\it Springer-Verlag}, Berlin, 1976)
\refno{[\ART]}
A.M.Arthurs:
Path Integrals in Polar Coordinates;
{\it Proc.Roy.Soc.(London)} {\bf A 313} (1969) 445
\refno{[\AMSS]}
R.Aurich, C.Matthies, M.Sieber and F.Steiner:
Novel Rule for Quantizing Chaos;
{\it Phys.Rev.Lett.}\ {\bf 68} (1992) 1629
\refno{[\ASS]}
R.Aurich, M.Sieber and F.Steiner:
Quantum Chaos of the Hadamard-Gutzwiller Model;
{\it Phys.Rev. Lett.}\ {\bf 61} (1988) 483
\refno{[\AST]}
R.Aurich and F.Steiner:
{}From Classical Periodic Orbits to the Quantization of Chaos;
{\it Proc.Roy.Soc. (Lon\-don)} {\bf  A 437} (1992) 693;
Statistical Properties of Highly Excited Quantum Eigenstates of a
Strongly Chaotic System;
{\it Physica} {\bf D 64} (1993) 185
\refno{[\BV]}
N.L.Balazs and  A.Voros:
Chaos on the Pseudosphere;
{\it Phys.Rep.}\ {\bf 143} (1986) 109
\refno{[\BIJ]}
A.O.Barut, A.Inomata and G.Junker:
Path Integral Treatment of the Hydrogen Atom in a Curved Space of
Constant Curvature;
{\it J.Phys.A: Math.Gen.}\ {\bf 20} (1987) 6271;
Path Integral Treatment of the Hydrogen Atom in a Curved Space of
Constant Curvature II;
{\it J.Phys.A: Math.Gen.}\ {\bf 23} (1990) 1179
\refno{[\BJb]}
M.B\"ohm and G.Junker:
Path Integration Over Compact and Noncompact Rotation Groups;
{\it J.Math.Phys.}\ {\bf 28} (1987) 1978
\refno{[\BJc]}
M.B\"ohm and G.Junker:
Path Integration Over the n-Dimensional Euclidean Group;
{\it J.Math.Phys.}\ {\bf 30} (1989) 1195
\refno{[\BJd]}
M.B\"ohm and  G.Junker:
Group Theoretical Approach to Path Integrations on Spheres;
in ``Path Summation: Achievements and Goals'', Trieste, 1987, p.469,
eds.: S.Lundquist et al.\ ({\it World Scientific}, Singapore, 1988)
\refno{[\BUC]}
H.Buchholz: The Confluent Hypergeometric Function,
Springer Tracts in Natural Philosophy, Volume 15
({\it Springer-Verlag}, Berlin, 1969)
\refno{[\CAST]}
D.P.L.Castrigiano and F.St\"ark:
New Aspects of the Path Integrational Treatment of the Coulomb
Potential;
{\it J.Math.Phys.}\ {\bf 30} (1989) 2785.
\refno{[\CHe]}
L.Chetouani et T.F.Hammann:
Traitement Exact des Syst\`eme Coulombiens, dans le Formalisme des
Int\'egrales de Feynman, en Coordonn\'ees Parabolique;
{\it Nuovo Cimento} {\bf B 98} (1987) 1
\refno{[\DAR]}
G.Darboux:
Le\c cons sur les Syst\'eme Orthogonaux et les Coordon\' ees
Curvilignes (Paris, 1910)
\refno{[\POGOb]}
L.S.Davtyan, G.S.Pogosyan, A.N.Sisakyan, and V.M.Ter-Antonyan:
Transformations Between Parabolic Bases of the Two-Dimensional Hydrogen
Atom in the Continuous Spectrum;
{\it Theor.Math. Phys.}\ {\bf 74} (1988) 157;
\newline
L.G.Mardoyan, G.S.Pogosyan, A.N.Sisakyan, and V.M.Ter-Antonyan:
Hidden Symmetry, Separation of Variables and Interbasis Expansion in
the Two-Dimensional Hydrogen Atom;
{\it J.Phys.A: Math,Gen.}\ {\bf 18} (1985) 455
\refno{[\DEW]}
B.S.DeWitt:
Point Transformations in Quantum Mechanics;
{\it Rev.Mod.Phys.}~{\bf 29} (1957) 377.
\refno{[\REUT]}
W.Dittrich and M.Reuter:
Classical and Quantum Dynamics. From Classical Paths to Path Integrals
({\it Springer-Verlag}, Berlin, 1992)
\refno{[\DOTO]}
J.C.D'Olivio and M.Torres:
The Canonical Formalism and Path Integrals in Curved Spaces;
{\it J.Phys.A: Math.Gen.}~{\bf 21} (1988) 3355;
\refno{[\DURb]}
I.H.Duru:
Path Integrals Over $\SU(2)$ Manifold and Related Potentials;
{\it Phys.Rev.}\ {\bf D 30} (1984) 2121
\refno{[\DURd]}
I.H.Duru:
On the Path Integral for the Potential $V=ar^{-2}+br^2$;
{\it Phys.Lett.}\ {\bf A 112} (1985) 421
\refno{[\DKa]}
I.H.Duru and H.Kleinert:
Solution of the Path Integral for the H-Atom;
{\it Phys.Lett.}~{\bf B 84} (1979) 185
\refno{[\DKb]}
Quantum Mechanics of H-Atoms From Path Integrals;
{\it Fort\-schr.Phys.}\ {\bf 30} (1982) 401
\refno{[\EG]}
S.F.Edwards and Y.V.Gulyaev:
Path Integrals in Polar Co-ordinates;
{\it Proc.Roy.Soc.(London)} {\bf A 279} (1964) 229
\refno{[\EIS]}
L.P.Eisenhart:
Enumeration of Potentials for Which One-Particle Schroedinger
Equations Are Separable;
{\it Phys.Rev.}\ {\bf 74} (1948) 87
\refno{[\EIS]}
J.Elstrod, F.Grunewald and J.Mennicke:
The Selberg Zeta-Function for Cocompact Discrete Subgroups of
$\PSL(2,\bbbc)$;
{\it Elementary Theory of Numbers, Banach Center Publications}
{\bf 17} (1985) 83
\refno{[\EMOTa]}
A.Erd\'elyi, W.Magnus, F.Oberhettinger and F.G.Tricomi (Eds.):
Higher Transcendental Functions, Vol.II
({\it McGraw Hill}, New York, 1985)
\refno{[\EMOTb]}
A.Erd\'elyi, W.Magnus, F.Oberhettinger and F.G.Tricomi (Eds.):
Tables of Integral Transforms, Vol.II
({\it McGraw Hill}, New York, 1954)
\refno{[\EWA]}
N.W.Ewans:
Superintegrability in Classical Mechanics;
{\it Phys.Rev.}\ {\bf A 41} (1990) 5666;
Group Theory of the Smorodinsky-Winternitz System;
{\it J.Math.Phys.}\ {\bf 32} (1991) 3369
\refno{[\FEY]}
R.P.Feynman:
Space-Time Approach to Non-Relativistic Quantum Mechanics;
{\it Rev.Mod.Phys.}~{\bf 20} (1948) 367.
\refno{[\FH]}
R.P.Feynman and A.Hibbs: Quantum Mechanics and Path Integrals
({\it McGraw Hill, New York}, 1965)
\refno{[\FLM]}
W.Fischer, H.Leschke and P.M\"uller:
Changing Dimension and Time: Two Well-Founded and Practical Techniques
for Path Integration in Quantum Physics;
{\it J.Phys.A: Math.Gen.}\ {\bf 25} (1992) 3835;
Path Integration in Quantum Physics by Changing the Drift of the
Underlying Diffusion Process;
{\it Universit\"at Erlangen-N\"urnberg preprint}, February 1992
\refno{[\FMSUW]}
J.Fri\v s, V.Mandrosov, Ya.A.Smorodinsky, M.Uhlir and P.Winternitz:
On Higher Symmetries in Quantum Mechanics;
{\it Phys.Lett.}\ {\bf 16} (1965) 354;
\newline
J.Fri\v s, Ya.A.Smorodinski\v\ii, M.Uhl\'\ii\v r and P.Winternitz:
Symmetry Groups in Classical and Quantum Mechanics;
{\it Sov.J.Nucl.Phys.}\ {\bf 4} (1967) 444
\refno{[\GY]}
I.M.Gelfand and A.M.Jaglom:
Die Integration in Funktionenr\"aumen und ihre Anwendung in der
Quantentheorie;
{\it Fortschr.Phys.}\ {\bf 5} (1957) 517;
\newline
I.M.Gel'fand and A.M.Yaglom:
Integration in Functional Spaces and its Applications in Quantum
Physics;
{\it J.Math.Phys.}\ {\bf 1} (1960) 48
\refno{[\GJ]}
J.-L.Gervais and A.Jevicki:
Point Canonical Transformations in the Path Integral;
{\it Nucl.Phys.}~{\bf B 110} (1976) 93.
\refno{[\GOOb]}
M.J.Goovaerts:
Path-Integral Evaluation of a Nonstationary Calogero Model;
{\it J.Math.Phys.}\ {\bf 16} (1975) 720
\refno{[\GEGR]}
I.M.Gel'fand and M.I.Graev:
Geometry of Homogeneous Spaces, Representations of Groups in Homogeneous
Spaces and Related Questions.\ I;
{\it Amer.Math.Soc.Transl., Ser.2} {\bf 37} (1964) 351
\refno{[\GGV]}
I.M.Gel'fand, M.I.Graev, and N.Ya.Vilenkin:
Generalized Functions, Vol.5
({\it Academic Press}, New York, 1966)
\refno{[\GLJA]}
J.Glimm and A.Jaffe:
Quantum Physics: A Functional Point of View
({\it Springer-Verlag}, Berlin, 1981)
\refno{[\GRA]}
I.S.Gradshteyn and I.M.Ryzhik:
Table of Integrals, Series, and Products,
Academic Press, New York, 1980
\refno{[\GSW]}
M.B.Green, J.H.Schwarz and E.Witten:
Superstring Theory I\&II
({\it Cambridge University Press}, Cambridge, 1988)
\refno{[\GROa]}
C.Grosche:
The Product Form for Path Integrals on Curved Manifolds;
{\it Phys.Lett.}~{\bf A 128} (1988) 113.
\refno{[\GROb]}
C.Grosche:
The Path Integral on the Poincar\'e Upper Half-Plane With
a Magnetic Field and for the Morse Potential;
{\it Ann.Phys.(N.Y.)} {\bf 187} (1988) 110
\refno{[\GROc]}
C.Grosche:
The Path Integral on the Poincar\'e Disc, the Poincar\'e Upper
Half-Plane and on the Hyperbolic Strip;
{\it Fortschr.Phys.}\ {\bf 38} (1990) 531
\refno{[\GROe]}
C.Grosche:
Path Integral Solution of a Class of Potentials Related to the
P\"oschl-Teller Potential;
{\it J.Phys.A: Math.Gen.}\ {\bf 22} (1989) 5073
\refno{[\GROg]}
C.Grosche:
The Path Integral for the Kepler Problem on the Pseudosphere;
{\it Ann.Phys.(N.Y.)} {\bf 204} (1990) 208
\refno{[\GROf]}
C.Grosche:
Path Integration on the Hyperbolic Plane With a Magnetic Field;
{\it Ann.Phys.(N.Y.)} {\bf 201} (1990) 258
\refno{[\GROj]}
C.Grosche:
Separation of Variables in Path Integrals and Path Integral Solution
of Two Potentials on the Poincar\'e Upper Half-Plane;
{\it J.Phys.A: Math.Gen.}\ {\bf 23} (1990) 4885
\refno{[\GROm]}
C.Grosche:
Coulomb Potentials by Path-Integration;
{\it Fortschr.Phys.}\ {\bf 40} (1992) 695
\refno{[\GROn]}
C.Grosche:
Path Integration on Hyperbolic Spaces;
{\it J.Phys.A: Math.Gen.}\ {\bf 25} (1992) 4211
\refno{[\GROq]}
C.Grosche:
Path Integral Solution of Two Potentials Related to the $\SO(2,1)$
Dynamical Algebra;
{\it J.Phys.A: Math.Gen.}\ {\bf 26} (1993) L279
\refno{[\GROw]}
C.Grosche:
Path Integral Solution of a Class of Explicitly Time-Dependent
Potentials;
{\it Trieste preprint}, SISSA/2/93/FM, {\it Phys.Lett.}\ {\bf A}, in
press
\refno{[\GRSa]}
C.Grosche and F.Steiner:
The Path Integral on the Poincar\'e Upper Half
Plane and for Liouville Quantum Mechanics;
{\it Phys.Lett.}\ {\bf A 123} (1987) 319
\refno{[\GRSb]}
C.Grosche and F.Steiner:
Path Integrals on Curved Manifolds;
{\it Zeitschr.Phys.}\ {\bf C 36} (1987) 699
\refno{[\GRSc]}
C.Grosche and F.Steiner:
The Path Integral on the Pseudosphere;
{\it Ann.Phys.(N.Y.)} {\bf 182} (1988) 120
\refno{[\GRSf]}
C.Grosche and F.Steiner:
Classification of Solvable Feynman Path Integrals;
{\it DESY preprint} DESY 92-189, to appear in the {\it Proceedings of
the ``Fourth International Conference on Path Integrals from meV to
MeV'', May 1992, Tutzing, Germany}, World Scientific, Singapore
\refno{[\GRSg]}
C.Grosche and F.Steiner:
Table of Feynman Path Integrals;
to appear in: {\it Springer Tracts in Modern Physics}
\refno{[\GRSh]}
C.Grosche and F.Steiner:
Feynman Path Integrals;
to appear in: {\it Springer Lecture Notes in Physics}
\refno{[\GUTc]}
M.C.Gutzwiller:
The Geometry of Quantum Chaos;
{\it Physica Scripta} {\bf T 9} (1985) 184
\refno{[\GUTd]}
M.C.Gutzwiller:
Chaos in Classical and Quantum Mechanics
({\it Springer-Verlag}, New York, 1990)
\refno{[\HIG]}
P.W.Higgs:
Dynamical Symmetries in a Spherical Geometry;
{\it J.Phys.A: Math.Gen.}\ {\bf 12} (1979) 309
\refno{[\HEJ]}
D.A.Hejhal: The Selberg Trace Formula for $\PSL(2,\bbbr)$, I\&II;
{\it Lecture Notes in Mathematics} {\bf 548, 1001}
({\it Springer-Verlag}, Berlin, 1976)
\refno{[\INOa]}
A.Inomata:
Exact Path-Integration for the Two Dimensional Coulomb Problem;
{\it Phys.Lett.}\ {\bf A 87} (1982) 387
\refno{[\INOb]}
A.Inomata:
Roles of Dynamical Groups in Path Integration;
to appear in the proceedings of the ``International Workshop on
`Symmetry Methods in Physics' in Memory of Prof.\ Ya.\ A.\
Smorodinsky'', Dubna, July 1993
\refno{[\INOWI]}
A.Inomata and R.Wilson:
Path Integral Realization of a Dynamical Group;
{\it Lecture Notes in Physics} {\bf 261}, p.42
({\it Springer}, Berlin-Heidelberg, 1985);
Factorization-Algebraization-Path Integration and Dynamical Groups;
in ``Symmetries in Science II'', eds.: B.Gruber and R.Lenczewski, p.255
({\it Plenum Press}, New York, 1986)
\refno{[\JUNc]}
G.Junker:
Remarks on the Local Time Rescaling in Path Integration;
{\it J.Phys.A: Math.Gen.}\ {\bf 23} (1990) L881
\refno{[\BJa]}
G.Junker and M.B\"ohm:
The $\SU(1,1)$ Propagator as a Path Integral Over Noncompact Groups;
{\it Phys.Lett.}\ {\bf A 117} (1986) 375
\refno{[\KAL]}
E.G.Kalnins:
Separation of Variables for Riemannian Spaces of Constant Curvature
({\it Longman Scientific \&\ Technical}, Essex, 1986)
\refno{[\KAMIa]}
E.G.Kalnins and W.Miller, Jr.:
Separation of Variables on $n$-Dimensional Riemannian Manifolds 1.
The $n$-Sphere and Euclidean $n$-Space;
{\it J.Math.Phys.}\ {\bf 27} (1986) 1721
\refno{[\KAMIb]}
E.G.Kalnins, W.Miller, Jr., and G.J.Reid:
Separation of Variables for Complex Riemannian Spaces of Constant
Curvature 1. Orthogonal Separable Coordinates for $S_{n\bbbc}$ and
$E_{n\bbbc}$;
{\it Proc.Roy.Soc. (London)} {\bf A 394} (1984) 183
\refno{[\KLEh]}
H.Kleinert:
How to do the Time Sliced Path Integral for the H Atom;
{\it Phys.Lett.}\ {\bf A 120} (1987) 361
\refno{[\KLE]}
H.Kleinert:
Path Integrals in Quantum Mechanics, Statistics and Polymer Physics
({\it World Scientific}, Singapore, 1990)
\refno{[\KLEMUS]}
H.Kleinert and I.Mustapic:
Summing the Spectral Representations of P\"oschl-Teller and
Rosen-Morse Fixed-Energy Amplitudes;
{\it J.Math.Phys.}\ {\bf 33} (1992) 643
\refno{[\KOKCAS]}
N.Kokiantonis and D.P.L.Castrigiano:
Propagator for a Charged Anisotropic Oscillator in a Constant
Magnetic Field;
{\it J.Phys.A: Math.Gen.}\ {\bf 18} (1985) 45
\refno{[\KUB]}
R.Kubo:
Path Integration on the Upper Half-Plane;
{\it Prog.Theor.Phys.}\ {\bf 78} (1987) 755;
Geometry, Heat Equation and Path Integrals on the Poincar\'e
Upper Half-Plane;
{\it Prog.Theor.Phys.}\ {\bf 79} (1988) 217
\refno{[\KUZ]}
V.Kuznetsov:
private communication
\refno{[\LL]}
L.D.Landau and E.M.Lifschitz: Lehrbuch der Theoretischen Physik,
Vol.I - III ({\it Akademie Verlag, Berlin,} 1979)
\refno{[\POGOa]}
I.V.Lutsenko, G.S.Pogosyan, A.N.Sisakyan, and V.M.Ter-Antonyan:
Hydrogen Atom as Indicator of Hidden Symmetry of a Ring-Shaped
Potential;
{\it Theor.Math.Phys.}\ {\bf 83} (1990) 633
\refno{[\MCSCH]}
D.W.McLaughlin and L.S.Schulman:
Path Integrals in Curved Spaces;
{\it J.Math.Phys.}~{\bf 12} (1971) 2520
\refno{[\MSVW]}
A.A.Makarov, J.A.Smorodinsky, Kh.Valiev and P.Winternitz:
A Systematic Search for Nonrelativistic Systems with Dynamical
Symmetries;
{\it Nuovo Cimento} {\bf A 52} (1967) 1061
\refno{[\MASA]}
P.T.Matthews and A.Salam:
Covariant Fock Equations;
{\it Proc.Roy.Soc.(London)} {\bf A 221} (1953) 128;
The Green's Function of Quantized Fields;
{\it Nuovo Cimento} {\bf 12} (1954) 563;
\newline
Propagators of Quantized Field;
{\it Nuovo Cimento} {\bf 2} (1955) 120
\refno{[\MADO]}
I.M.Mayes and J.S.Dowker:
Canonical Functional Integrals in General Coordinates;
{\it Proc.Roy.Soc. (London)} {\bf A 327} (1972) 131
\refno{[\MESCH]}
J.Meixner and F.W.Sch\"afke:
Mathieusche Funktionen und Sph\"aroidfunktionen
({\it Springer-Verlag}, Berlin, 1954)
\refno{[\MIZa]}
M.M.Mizrahi:
The Weyl Correspondence and Path Integrals;
{\it J.Math.Phys.}~{\bf 16} (1975) 2201
\refno{[\MDEW]}
C.Morette:
On the Definition and Approximation of Feynman's Path Integrals;
{\it Phys.Rev.}\ {\bf 81} (1951) 848;
Feynman's Path Integral: Definition Without Limiting Procedure;
{\it Commun.Math.Phys.}\ {\bf 28} (1972) 47
\refno{[\MF]}
P.M.Morse and H.Feshbach:
Methods of Theoretical Physics
({\it McGraw-Hill}, New York, 1953)
\refno{[\NEL]}
E.Nelson:
Feynman Integrals and the Schr\"odinger Equation;
{\it J.Math.Phys.}\ {\bf 5} (1964) 332
\refno{[\OMO]}
M.Omote:
Point Canonical Transformations and the Path Integral;
{\it Nucl.Phys.}~{\bf B 120} (1977) 325
\refno{[\OLE]}
N.N.Olevski\v\ii:
Triorthogonal Systems in Spaces of Constant Curvature in which the
Equation $\Delta_2u+\lambda u=0$ Allows the Complete Separation of
Variables;
{\it Math.Sb.}~{\bf 27} (1950) 379 (in russian)
\refno{[\PAKSc]}
N.K.Pak and I.S\"okmen:
General New-Time Formalism in the Path Integral;
{\it Phys.Rev.}\ {\bf A 30} (1984) 1629
\refno{[\PI]}
D.Peak and A.Inomata:
Summation Over Feynman Histories in Polar Coordinates;
{\it J.Math.Phys.}\ {\bf 10} (1969) 1422
\refno{[\PELST]}
A.Pelster and A.Wunderlin:
On the Generalization of the Duru-Kleinert-Propaga\-tor Transformations;
{\it Zeitschr.Phys.}\ {\bf B 89} (1992) 373
\refno{[\PODO]}
B.Podolsky:
Quantum-Mechanically Correct Form of Hamiltonian Function for
Conservative Systems;
{\it Phys.Rev.}~{\bf 32} (1928) 812.
\refno{[\POGOc]}
G.S.Pogosyan, A.N.Sissakian and S.I.Vinitsky:
The Super-Integrable Systems of the Smorodinsky-Winternitz Type
on the Three Dimensional Sphere;
to appear in the proceedings of the ``International Workshop on
`Symmetry Methods in Physics' in Memory of Prof.\ Ya.\ A.\
Smorodinsky'', Dubna, July 1993
\refno{[\POL]}
A.M.Polyakov:
Quantum Geometry of Bosonic Strings;
{\it Phys.Lett.}\ {\bf B 103} (1981) 207
\refno{[\QUES]}
C.Quesne:
A New Ring-Shaped Potential and its Dynamical Invariance Algebra;
{\it J.Phys.A: Math.Gen.}\ {\bf 21} (1988) 3093
\refno{[\ROEP]}
G.Roepstorff:
Pfadintegrale in der Quantenphysik
({\it Vieweg}, Braunschweig, 1992)
\refno{[\SCHUa]}
L.Schulman:
A Path Integral for Spin;
{\it Phys.Rev.}\ {\bf 176} (1968) 1558
\refno{[\SCHU]}
L.S.Schulman:
Techniques and Applications of Path Integration
({\it John Wiley \&\ Sons}, New York, 1981)
\refno{[\SCHW]}
J.Schwinger (ed.): Quantum Electrodynamics (ed.J.Schwinger)
{\it Dover}, New York, 1958
\refno{[\SEL]}
A.Selberg:
Harmonic Analysis and Discontinuous Groups in Weakly Symmetric
Riemannian Spaces with Application to Dirichlet Series;
{\it J.Indian Math.Soc.}\ {\bf 20} (1956) 47
\refno{[\SIMON]}
B.Simon:
Functional Integration and Quantum Physics
({\it Academic Press}, New York, 1979)
\refno{[\STEc]}
F.Steiner:
Path Integrals in Polar Coordinates From eV to GeV;
in ``Bielefeld Encounters in Physics and Mathematics VII; Path
Integrals From meV to MeV'', 1985, p.335;
eds.: M.C.Gutzwiller et al.\ ({\it World Scientific}, Singapore, 1986)
\refno{[\STEP]}
V.V.Stepanov:
The Laplace Equation in Certain Trirectangular Systems;
{\it Math.Sb.}\ {\bf 11} (1942) 204 (in russian, cited after [\OLE])
\refno{[\STORCH]}
S.N.Storchak:
Rheonomic Homogeneous Point Transformation and Reparametrization in the
Path Integral;
{\it Phys.Lett.}\ {\bf A 135} (1989) 77
\refno{[\VENa]}
A.B.Venkov:
Expansion in Automorphic Eigenfunctions of the Laplace Operator
and the Selberg Trace Formula in the Space $\SO_0(n,1)/\SO(n)$;
{\it Soviet Math.Dokl.}~{\bf 12} (1971), 1363.
\refno{[\VENb]}
A.B.Venkov:
Expansion in Automorphic Eigenfunctions of the Laplace-Beltrami
Operator in Classical Symmetric Spaces of Rank One, and the Selberg
Trace Formula;
{\it Proc.Math.Inst.Steklov} {\bf 125} (1973), 6.
\refno{[\VENc]}
A.B.Venkov:
Spectral Theory of Automorphic Functions;
{\it Proc.Math.Inst. Steklov} {\bf 153} (1981) 1
\refno{[\VIL]}
N.Ja.Vilenkin:
Special Functions Connected with Class 1 Representations of Groups of
Motions in Spaces of Constant Curvature;
{\it Trans.Moscow Math.Soc.}\ (1963) 209
\refno{[\VISM]}
N.Ya.Vilenkin and Ya.A.Smorodinsky:
Invariant Expansions of Relativistic Amplitudes;
{\it Sov.Phys. JETP}\ {\bf 19} (1964) 1209
\refno{[\WIE]}
F.W.Wiegel:
Introduction to Path-Integral Methods in Physics and Polymer Science
({\it World Scientific}, Singapore, 1986)
\refno{[\YODEWM]}
A.Young and C.DeWitt-Morette:
Time Substitution in Stochastic Processes as a Tool in Path Integration;
{\it Ann.Phys.}\ {\bf 169} (1986) 140


\enddocument